\documentclass{aa}

\usepackage[varg]{txfonts}
\usepackage[english,english]{babel}
\usepackage{amsmath}
\usepackage{amssymb,amsfonts,textcomp}
\usepackage{array}
\usepackage{supertabular}
\usepackage{hhline}
\usepackage{soul}
\usepackage{hyperref}
\usepackage[usenames,dvipsnames]{color}
\hypersetup{dvips, colorlinks=true, linkcolor=blue, citecolor=blue, filecolor=blue, urlcolor=blue}
\usepackage{graphicx}
\usepackage{orcidlink}
\bibpunct{(}{)}{;}{a}{}{,}

\begin{document}

\title{Galaxies with grains: unraveling dust evolution and extinction curves with hydrodynamical simulations}

\titlerunning{Dust properties and extinction curves of simulated galaxies}
\authorrunning{Dubois et al.}

\author{Yohan Dubois\inst{1}\orcidlink{0000-0003-0225-6387}, Francisco Rodr\'iguez Montero\inst{2}\orcidlink{0000-0001-6535-1766}, Corentin Guerra\inst{3},
Maxime Trebitsch\inst{4}\orcidlink{0000-0002-6849-5375},
San Han\inst{5}\orcidlink{0000-0001-9939-713X},\\
Ricarda Beckmann\inst{6}\orcidlink{0000-0002-2850-0192},
Sukyoung K. Yi\inst{5}\orcidlink{0000-0002-4556-2619},
Joseph Lewis\inst{1}\orcidlink{0000-0001-7917-8474},
and J. K. Jang\inst{5}\orcidlink{0000-0002-0858-5264}}
\institute{Institut d'Astrophysique de Paris, UMR 7095, CNRS, Sorbonne Universit\'e, 98 bis boulevard Arago, 75014 Paris, France \\
\email{dubois@iap.fr}
\and
Sub-department of Astrophysics, University of Oxford, Keble Road, Oxford OX1 3RH, United Kingdom
\and 
Observatoire de Paris, PSL University, Université de Paris, CNRS, LUTH 5 Place Jules Janssen, 92190 Meudon, France
\and
Kapteyn Astronomical Institute, University of Groningen, P.O. Box 800, 9700AV Groningen, The Netherlands
\and
Department of Astronomy and Yonsei University Observatory, Yonsei University, Seoul 03722, Republic of Korea
\and 
Institute of Astronomy and Kavli Institute for Cosmology, University of Cambridge, Madingley Road, Cambridge CB3 0HA, United Kingdom
}

\date{Received 28 February 2024 / Accepted 30 April 2024}

\abstract{Dust in galaxies is an important tracer of galaxy properties and their evolution over time. 
The physical origin of the grain size distribution, the dust chemical composition, and, hence, the associated ultraviolet-to-optical extinctions in diverse galaxies remains elusive.
To address this issue, we introduce a model for dust evolution in the {\sc ramses} code for simulations of galaxies with a resolved multiphase interstellar medium.
Dust is modelled as a fluid transported with the gas component, and is decomposed into two sizes, 5 nm and $0.1 \, \mu\rm m$, and two chemical compositions for carbonaceous and silicate grains.
This dust model includes the growth of dust by accretion of elements from the gas phase and by the release of dust in stellar ejecta, the destruction by thermal sputtering, supernovae, and astration, and the exchange of dust mass between the two main populations of grain sizes by coagulation and shattering.
Using a suite of isolated disc simulations with different masses and metallicities, the simulations can explore the role of these processes in shaping the key properties of dust in galaxies.
The simulated Milky Way analogue reproduces the dust-to-metal mass ratio, depletion factors, size distribution and extinction curves of the Milky Way.
Galaxies with lower metallicities reproduce the observed decrease in the dust-to-metal mass ratio with metallicity at around a few $0.1\,\rm Z_\odot$.
This break in the dust-to-metal ratio corresponds to a galactic gas metallicity threshold that marks the transition from an ejecta-dominated to an accretion-dominated grain growth, and that is different for silicate and carbonaceous grains, with $\simeq0.1\,\rm Z_\odot$ and $\simeq 0.5\,\rm Z_\odot$ respectively.
This leads to more Magellanic Cloud-like extinction curves, i.e. with steeper slopes in the ultraviolet and a weaker bump feature at $2175\, \AA$, in galaxies with lower masses and lower metallicities.
Steeper slopes in these galaxies are caused by the combination of the higher efficiency of gas accretion by silicate relative to carbonaceous grains and by the low rates of coagulation that preserves the amount of small silicate grains.
Weak bumps are due to the overall inefficient accretion growth of carbonaceous dust at low metallicity, whose growth is mostly supported by the release of large grains in SN ejecta.
We also show that the formation of CO molecules is a key component to limit the ability of carbonaceous dust to grow, in particular in low-metallicity gas-rich galaxies.
}


\maketitle

\section{Introduction}

The spectral energy distribution (SED) of galaxies characterises the distance and properties of galaxies~\citep[e.g.][]{fioc97,fioc19,devriendt99,noll09,chevallard16,boquien19}.
As dust efficiently absorbs light in the optical or in the ultraviolet (UV), it re-emits this light in the infrared (IR), which is subsequently absorbed and re-emitted in the IR (producing IR multi-scattering). 
The exact SEDs of galaxies across the entire spectral range are, thus, tied in part to the properties of the dust, and in particular to the size and chemical composition of grains.

Unlike extinction, corresponding to a line-of-sight emitted by a single point source whose light is absorbed and scattered, the attenuation is the result of the collection of multiple sources with different extinctions into a single line-of-sight.
While galaxy attenuation curves offer insights into the dust properties~\citep{calzetti00}, it is crucial to note that they result from the interplay of complex radiative transfer effects within the intricate geometry of the interstellar medium (ISM) and stellar distribution~\citep[e.g.][]{gordon97, witt00, granato00, inoue05, narayanan18}.
A key source of knowledge about dust properties comes from the extinction curves of multiple lines-of-sights of individual stars (for which the complexity of dust absorption is better characterised) in the Milky Way or from the Magellanic Clouds.
The Milky Way extinction curve is characterised by a moderate extinction in the ultraviolet, as opposed to Magellanic Clouds, and by a significant feature in the optical range, called the $2175 \, \AA$ bump, very faint in the Large Magellanic Cloud and almost indistinguishable in the Small Magellanic Cloud~\citep{pei92}.
Thanks to the description with Mie theory of light absorption and scattering by solid spherical bodies, and optical properties of grains, it is possible to interpret the shape of the Milky Way extinction curve by a number density distribution skewed towards small grains (a few nm in size) but dominated by large grains ($0.1\,\mu \rm m$) in mass, as well as a mixed composition of carbonaceous grains and of silicate grains~\citep{weingartner&draine01}.
Although, the exact nature of the carbonaceous grains (amorphous, aromatic, aliphatic) and silicate grains (olivine, pyroxene, iron inclusions) and how they mix up in a single grain body is complex~\citep{jones17}, it is possible to make predictions about the grain properties as a function of the galaxy properties and test them against observables~\citep[e.g.][]{mathis77,draine&lee84,weingartner&draine01,zubko04,draineli07,compiegne11,galliano18,hensley23}, such as the extinction or attenuation laws, or the depletion of the various elements in the gas phase.

Not only is dust key to reconstruct the properties of galaxies from their SED, but it also intervenes directly or indirectly in the thermodynamics of the ISM.
Dust is a catalyst for molecular H$_2$ formation, that is an important tracer of star formation with the so-called Kennicutt-Schmidt relation~\citep{kennicutt98}, and which is a key component for gas cooling by rotational transitions or for heating through H$_2$ formation~\citep{hollenbach79}.
Also dust directly affects the heating of the gas via photo-electric heating, or the cooling of the gas by depletion of elements in the gas phase, by dust cooling at low temperatures in the dark cores of molecular clouds, or by energy exchange through collisions with electrons in X-ray bright gas.
As a contributor to the reprocessing of light in the ISM, dust can have a role in radiative feedback at galactic scales.
In particular, the role of dust is certainly exacerbated in massive starburst galaxies which appear as ultra-luminous IR galaxies (or sub-millimetre galaxies at high redshift), as they can obscure up to 99\% of their UV-visible emission \citep{casey14}.

Observations show that the dust-to-gas ratio (DTG) has a strong correlation with galaxy metallicity~\citep[e.g.][]{lisenfeld98,draine07,galametz11,sandstrom13,remy14}, and that the
dust-to-metal ratio (DTM) shows a weak but positive correlation with the gas phase metallicity of galaxies~\citep{decia13,decia16,wiseman17}.
Evolved galaxies, including the Milky Way, have large values of ${\rm DTM}\simeq 0.4$, while younger galaxies (with larger gas fraction and lower metallicity) have lower DTM.
Large samples of galaxies show that there is also a break in both the DTG- and the DTM-metallicity relation at around $0.1\,\rm Z_\odot$~\citep{remy14,devis19}.
This break predicted by numerical models~\citep{popping17,hou19,li19,parente22} is explained by a transition in the leading mechanism of dust growth in the ISM: from a stellar ejecta-dominated phase at low metallicity to an accretion-dominated phase at high metallicity~\citep{asano13critmet,zhukovska14,popping17,graziani20,li21,lewis23}.

Several dust evolution models in galaxies have been developed which include the dust released by stars, the growth of dust by accretion of elements from the gas phase, and the destruction of dust by astration, supernovae and thermal sputtering.
In the seminal work of~\cite{dwek98}, they developed a one zone analytical model to follow the evolution of carbonaceous and silicate grains separately, and they were able to reproduce the observed amount of dust in Milky Way-like galaxies.
Such one zone models were further refined to track the distribution of grains in sizes including the coagulation and shattering of grains~\citep{asano13size}.
Evolution of dust, followed as a separate component to that of metals from the gas phase, is also modelled in complete hydrodynamical simulations of galaxies~\citep[e.g.][]{bekki13,mckinnon16,aoyama17,gjergo18,li19,li21,graziani20,granato21,trebitsch21,choban22}.
Some of these hydrodynamical simulations make various assumptions about the dust physics, either neglecting the chemical differentiation of grains~\citep[as in e.g.][]{aoyama17} or the grain sizes~\citep[as in e.g.][]{graziani20,choban22}, or both~\citep{mckinnon16,trebitsch21}, or using a unique size distribution for different dust grain species~\citep{li21}, or using too low resolution to separate ab initio the dense gas structures, in which accretion and coagulation efficiently proceed, from the diffuse ISM~\citep{gjergo18,granato21,li21}.

Our goal in this paper is to provide a theoretical understanding of the physical origin of the dust properties and, hence, of the different shapes of extinction curves observed as in the Milky Way, or the Magellanic Clouds.
Therefore, our approach is to implement and test against various observables a model of dust evolution in the adaptive mesh refinement code {\sc ramses}~\citep{teyssier02} that evolves the sizes and the chemical composition of dust.
This model assumes a two-size grain approximation, at $5 \,\rm nm$ and at $0.1\,\rm \mu m$, following separately the evolution of carbonaceous and silicate grains, and is tested in numerical simulations with sufficient resolution to capture the structure of a turbulent multiphase ISM.

The paper is structured as follows.
In section~\ref{section:simulation}, we provide an overview of the galactic physics (cooling, star formation, supernova feedback, chemical enrichment) used in {\sc ramses}, and the setup of initial conditions to simulate isolated disc galaxies.
In section~\ref{section:dustmodel}, we describe the details of the model of dust evolution.
In section~\ref{section:resultsmilkyway}, we test our model in a typical Milky Way-like galaxy including stringent tests on extinction curves.
In section~\ref{section:resultslowmass}, we extend our predictions to various galaxy masses and explain how Magellanic Cloud-like extinction curves can be obtained.
We discuss the limitations of this work in section~\ref{section:caveats} and, in section~\ref{section:conclusion}, we outline our conclusions.

\section{Numerics}
\label{section:simulation}

\subsection{Initial conditions and resolution}

\begin{table*}
    \centering
    \caption{Galaxy initial conditions' identifier and their properties.}
    \begin{tabular}{c c c c c c c c}
         Galaxy & $M_{\rm h}$  & $M_{\rm gas}$ & $M_{\rm \star,d}$ & $M_{\rm \star,b}$ & $f_{\rm gas}$  & $f_{\rm bar}$ & $R_{\rm d}$     \\
         identifier & $(10^{10}\,\rm M_\odot)$ & $(10^{10}\,\rm M_\odot)$ & $(10^{10}\,\rm M_\odot)$ & $(10^{10}\,\rm M_\odot)$ & \% & \% & (kpc) \\
         \hline
         \hline
        G10LG & $100$ & $0.35$ & $3.15$ & $0.35$ & 9 & 3.8 & 3.2  \\
         \hline
        G9LG & $10$ & $0.05$ & $0.3$ & $0.035$ & 13 & 3.8 & 1.5  \\
        G9HG & $10$ & $0.17$ & $0.17$ & $0.035$ & $45$ & $3.8$ & $1.5$  \\
         \hline
        G8HG & $1$ & $0.017$ & $0.017$ & $0.0035$ & $45$ & $3.8$ & $0.7$  \\
        \hline
    \end{tabular}
    \label{tab:ics}
\end{table*}

Our simulations were performed with the adaptive mesh refinement code {\sc ramses}~\citep{teyssier02}.
It solves the hydrodynamics on a cartesian grid with refinement with the Godunov method using the unsplit MUSCL-Hancock scheme with the minmod slope limiter to preserve the total variation diminishing property of the scheme, and the Harten-Lax-van Leer-Contact approximate Riemann solver to obtain the flux at a cell interface.
Dark matter and stars are represented by particles that are evolved with a leap-frog scheme.
The gravitational potential is obtained by a particle-mesh method where particles are projected on the grid with a cloud-in-cell interpolation.

The spatial resolution was varied from  $\Delta x=72 \,\rm pc$ to $9\,\rm pc$ by multiples of 2 with a resolution of $18\,\rm pc$ for our fiducial simulations.
The halo mass was either $M_{\rm h}=10^{10}$, $10^{11}$, or $10^{12}\, \rm M_\odot$, with a total galaxy mass (gas and stars) that represented $f_{\rm bar}=3.8\,\%$ of the halo mass, hence that was respectively $M_{\rm gal}=3.8\times 10^{8}$, $3.8\times 10^{9}$, or $3.8\times 10^{10}\, \rm M_\odot$, for simulations that we named respectively G8, G9 and G10. 
The halo, stellar disc and bulge were initialised with respectively $10^6$, $10^6$ and $10^5$ particles. 
We allowed refinement where the total mass exceeded $8\times 10^3\,\rm M_\odot$ starting at level 6.
In addition to pseudo-Lagrangian mass refinement, we added a refinement criterion on the Jeans length so that it is refined with at least 4 cells down to the maximum level of refinement.
The box sizes of respectively 150, 300, and $600\,\rm pc$ for G8, G9, and G10 were evolved with outflowing boundary conditions.

The dark matter (DM), star and gas distributions were initialised with {\sc makedisk} \citep{springel05}, following the strategy introduced in~\cite{rosdahl17} for {\sc ramses}.
The DM mass distribution followed a~\citet{navarro97} profile with concentration $c=10$ and a spin parameter of $\lambda_{\rm DM}=0.04$~\citep{maccio08}.
The disc distribution was decomposed into a stellar and a gaseous component (with mass $M_{\star,\rm d}$ and $M_{\rm gas}$ resp.), with density profiles of $\rho\propto e^{-R/R_{\rm d}}{\rm sech}^2(z/(2H))$, where $R$ is the cylindrical radius, $R_{\rm d}$ is the disc scale length, $z$ is the disc height, and $H$ is the disc characteristic scale height that we took as $H=0.1R_{\rm d}$.
The stellar bulge component (with mass $M_{\star,\rm b}$) followed a Hernquist density profile with bulge scale length equals to $H$.
The gas disc density profile with a temperature of $10^4\,\rm K$ was used up to a cut radius and a height radius of respectively $R_{\rm c}=5R_{\rm d}$ and $H_{\rm c}=5H$ after which, a uniform density of $10^{-6}\,\rm H\,cm^{-3}$ and a uniform temperature of $10^7\,\rm K$ were imposed.
For the G8, G9, and G10 galaxies, we used respectively a disc scale radius of $0.7$, $1.5$, and $3.2$ kpc.
We varied the gas fraction $f_{\rm gas}=M_{\rm gas}/(M_{\rm gas}+M_{\star,\rm d}+M_{\star,\rm b})$ between two extreme values.
The initial properties of the simulated galaxies and their halos are shown in table~\ref{tab:ics}.

The gas metallicity was initialised with a decreasing radial profile, following
\begin{eqnarray}
    Z(r_{\rm cyl})&=&Z_0 10^{0.5-R/{R_c}}{\rm ,\ if\ } R<R_c {\rm \ and \ } z<H_c, \nonumber \\
    &=&0 {\rm ,\ elsewhere},\nonumber
\end{eqnarray}
where $R_{c}=7.42\,\rm kpc$, and $Z_0$ is the scaling metallicity of the profile, which took values of $Z_0=0.045,0.15,0.45,0.90\,\rm Z_\odot$, with $Z_\odot=0.01345$, which gave respectively a mean metallicity of the gas of $0.1,0.3,1,3\,\rm Z_\odot$.
The various elements followed in this work (H, He, C, N, O, Mg, Fe, Si, and S) were initialised in the gas phase with solar mass fractions \citep{asplund09}. 
The exact mass fraction of each individual element is a function of the metallicity and of star formation history of the galaxy, as can be observed for low metallicity stars in the Milky Way, but we neglected this aspect.
We assumed that 10 per cent of the initial refractory\footnote{Through an intentional terminological adjustment, we will employ the term `refractory' throughout this text to refer to the atomic elements contributing to the composition of the dust grains in our model, i.e. this includes Mg, Fe, Si, O, and C.} elements were trapped in dust grains for the G10LG galaxy and 0.1 per cent for the lower mass galaxies G8 and G9, whose mass were equally distributed in sizes, which corresponded to a ${\rm DTM}=3.6\times 10^{-2}$ and $3.6\times 10^{-4}$ respectively.

For galaxies simulated with the large gas fractions (i.e. galaxy G8HG, G9HG, and G10HG), a special procedure to smoothly relax the initial conditions was required.
Due to the very idealised nature of the disc initial conditions, during the first 10s of Myr, the gas loses pressure support due to rapid cooling and produces a strong burst of star formation rate (SFR).
In turn this burst of SFR creates a massive outflow that can remove very significant amounts of gas from the galaxy and turn a gas-rich galaxy into a gas-poor galaxy.
In order to circumvent this shortcoming, and after some tests, we introduced a perturbation in the initial vertical velocity field with an amplitude of $30\,\rm km\,s^{-1}$ modulated by two sinusoids in the $x$ and $y$ axis of the cartesian box (the $z$ axis is oriented along the spin axis of the disc) with wavelength of $200\,\rm pc$.
In addition to this initial perturbation of the velocity field, we also imposed that the star formation efficiency (see section~\ref{section:starformation}) linearly ramped up with time with a timescale of 100 Myr. 

\subsection{Gas radiative cooling and heating}

Gas was allowed to cool down to $10\, \rm K$ through H and He collisions with a contribution from metals in the gas phase using rates tabulated by~\cite{sutherland&dopita93} above $10^4\,\rm K$ and those from~\cite{dalgarno&mccray72} below $10^4\,\rm K$.
Dust did not directly contribute to the gas cooling rates at low temperature, however, these rates can be significantly reduced in regions where the dust fraction over total metallicity is large, by decreasing the amount of metals in the gas phase available for gas cooling.
The contribution of metals to the overall cooling rates was reduced by the corresponding amount of metals locked into the dust phase.
We note that this approach is not consistent since the cooling curves, as tabulated in this version of {\sc ramses}, are for a total metallicity with a solar composition of elements, while the amount of depletion is not uniform across all elements, and, hence, this should affect differently the cooling curves as a function of temperature.
A more consistent approach will be explored using the {\sc prism} ISM model for {\sc ramses} from~\cite{katz22b} that computes cooling rates from the amount of each individual element and their ionisation state in the gas phase (Rodr\'iguez Montero et al., in prep.).

At high temperatures, fast moving electrons frequently collide with grains, leading to an exchange of internal energy from the gas phase to the dust phase. 
Since dust radiates the stored energy into the IR, this leads to a net cooling for the gas phase.
We used the corresponding cooling rates from~\cite{dwek&werner81}:
\begin{equation}
\Lambda_{\rm D}(a,T)=\mathcal{D}\frac{3m_{\rm H}}{X_{\rm H} 4\pi s a^3}n_{\rm H}n_{\rm e}\mathcal{H}(a,T)\, ,
\end{equation} 
where $a$ is the grain radius, $s$ is the grain material density, $\mathcal{D}$ is the dust-to-gas ratio, $n_{\rm H}$ is the hydrogen number density, $n_{\rm e}$ is the (free) electron number density (for which we assume a gas fully ionised at these temperatures), $X_{\rm H}=0.76$ is the hydrogen mass fraction, $m_{\rm H}$ is the hydrogen mass, and 
\begin{equation}
\frac{\mathcal{H}(a,T)}{\rm erg\,s^{-1}cm^3}=\left\{
    \begin{array}{ll}
    0, x \ge x_{\rm max}\,, \\
	5.38\times 10^{-18}\left(\frac{a}{\mu {\rm m}}\right)^2\left(\frac{T}{\rm K}\right)^{1.5}, 4.5\le x< x_{\rm max}\,,\\
	3.37\times 10^{-13}\left(\frac{a}{\mu {\rm m}}\right)^{2.41}\left(\frac{T}{\rm K}\right)^{0.88}, 1.5\le x <4.5\,,\\
	6.48\times 10^{-6}\left(\frac{a}{\mu {\rm m}}\right)^{3}, x <1.5\,, \nonumber
    \end{array}
\right.
\end{equation} 
where $x=2.71\times10^8 (a/\mu{\rm m})^{2/3}(T/{1\, \rm K})^{-1}$, and $x_{\rm max}=14 000 \left(a/\mu {\rm m} \right)^{2/3}$, which was introduced to impose a sharp cut-off to the dust cooling function at temperatures below $2\times 10^4\,\rm K$ in order to prevent the contribution to this cooling from dust in the cold phase.
While the dust contribution to the gas cooling rate can largely exceed that of the fiducial gas cooling at $T>10^6 \, \rm K$, thermal sputtering from ion collisions efficiently erodes dust grains (see Section~\ref{section:sputtering}), such that grains are destroyed on a timescale shorter than that of the gas cooling from dust~\citep{guillar09}, which reduces the effective contribution to cooling at high temperatures~\citep{montier&giard04, pointecouteau09, vogelsberger19}.  

\subsection{Star formation}
\label{section:starformation}

Star formation followed a Schmidt law: $\dot \rho_\star= \varepsilon_\star {\rho_{\rm g} / t_{\rm ff}}$, where $\dot \rho_\star$ is the star formation rate mass density, $\rho_{\rm g}$ the gas mass density, $t_{\rm ff}$ the local free-fall time of the gas, and $\varepsilon_\star$ is the star formation efficiency.
Star formation occurred in regions with gas number density above $n_0$ which varied with resolution ($n_0=0.6,2.5,10,40 \, \rm H\, cm^{-3}$ for a minimum spatial resolution of respectively $\Delta x=72,36,18,9 \,\rm pc$).
This corresponds to a stellar mass resolution for newly formed stars of respectively $=7.4\times 10^3,3.7\times 10^3,1.9\times 10^3, 0.9 \times 10^3 \,\rm M_\odot$.

As in~\cite{dubois21} (see for details of the implementation and the underlying parameters), the star formation efficiency varied with the virial parameter $\alpha_{\rm vir}=2 E_{\rm turb}/\vert E_{\rm g}\vert<1$, with $E_{\rm turb}$ and $E_{\rm g}$ the turbulent kinetic energy and the gravitational energy of the gas respectively, and with the turbulent Mach number $\mathcal{M}=\sigma_{\rm t}/c_{\rm s}$, where $\sigma_{\rm t}$ is the gas turbulent velocity and $c_{\rm s}$ is the sound speed.
In practice, $\sigma_{\rm t}$ was computed by taking for each cell its own gas velocity and that of its neighbouring cells into account, removing the local bulk (mass-weighted) mean velocity, and constructing $\sigma$ through $\sigma^2={\rm sum} (\nabla \otimes u {\rm d} x)^2$. 
In this theory~\citep[e.g.][]{krumholz&mckee05,padoan&nordlund11,hennebelle&chabrier11,federrath&klessen12}, star formation (SF) is driven by how much gas mass passes a given density threshold.
This amount is controlled by the level of turbulence and how much the molecular cloud is gravitationally bound. 
The unresolved density distribution can be described by a log-normal probability density function (PDF), with standard deviation given by $\mathcal{M}$ and $\alpha_{\rm vir}$: the more turbulent and bound the gas is, the larger the SF efficiency, reaching values as high as $\varepsilon_\star\sim 1$.
The mean SFR between 200 Myr and 400 Myr for the different simulated galaxies are $0.01\,\rm M_\odot\,yr^{-1}$, $0.03\,\rm M_\odot\,yr^{-1}$, $0.5\,\rm M_\odot\,yr^{-1}$, and $0.7\,\rm M_\odot\,yr^{-1}$ for respectively G8HG, G9LG, G9HG, and G10LG.

\subsection{Stellar yields and feedback}
\label{section:stellar_yields}

We limited our simulations to individually track the amount of C, N, O, Mg, Si, Fe, S (which represented $\sim85\%$ of the mass of solar heavy elements), of total metallicity Z, and H, but these can easily be extended to other individual elements.
We assumed a \cite{chabrier05} initial mass function for the distribution of zero age star (ZAS) masses.
Stellar gas plus dust yields for intermediate stars were those of~\cite{karakas10} for ZAS masses $M_{\rm ZAS}<8\,\rm M_\odot$ (asymptotic giant branch, AGB, stars), while yields from~\cite{limongi&chieffi18} for massive stars $M_{\rm ZAS}\ge8\,\rm M_\odot$ were employed.
The yields for massive stars were parametrised by the rotation velocity of the star (either 0, 150 or $300\,\rm km\, s^{-1}$).
We followed the initial distribution of rotational velocities as a function of the initial metallicity of the star from~\cite{prantzos18} (obtained from multiple Milky Way chemical constraints), with the corresponding grid\footnote{The range of available stellar yields for a given value of metallicity and rotational velocity is more limited but we simply bi-linearly interpolated the individual stellar yields from their data, which provided a finer sampled grid of IMF-weighted SSP yields as a function of metallicity. 
In addition, it should be noted that the yields of \cite{limongi&chieffi18} are given as a function of the Fe content in the ZAS. We prefered using the $Z$ metallicity content instead of the Fe content, which is more flexible. We therefore converted their Fe content into $Z$ content by simply assuming solar ratios.} of velocity-to-metallicity of ZAS: $V_{\rm ZAS}=150, 100, 50, 50, 50, 50, 50\,\rm km\, s^{-1}$ for $Z_{\rm ZAS}=10^{-3}, 10^{-2}, 10^{-1}, 10^{-0.6}, 10^{-0.3}, 1, 10^{0.3}\, Z_{\odot}$ (where $Z_\odot=0.01345$ is the solar value of metallicity from~\citealp{asplund09}).
Since the \cite{limongi&chieffi18} stellar yields largely underproduce the amount of Mg, with respect to Si and solar abundances~\citep{prantzos18}, we artificially boosted up by a factor of 2 the amount of Mg released by massive stars.
The mass distribution of stars was assumed to cover the $[0.1,100]\, \rm M_\odot$ mass range, assuming that massive stars only successfully explode in SNII and release their corresponding SN-processed yields for $M_{\rm ZAS}\le 30 \,\rm M_\odot$, while they also release mass through winds over their entire mass range ($[8, 100]\, \rm M_\odot$).
The death of each individual intermediate mass star followed the Padova stellar tracks~\citep{Girardi00} with thermally pulsating AGB~\citep{Vassiliadis&Wood93}.
We further assumed that the mass release of individual AGB stars happens instantaneously at the end of their evolution (pulsating evolution is not individually time-resolved).

For type Ia SN (SNIa), we assumed the delay time distribution of~\cite{maoz12} with a time-integrated number of SNIa of $N_{\rm Ia}(<t)=2.35\times 10^{-4}(\log(t/{\rm Gyr})+3)\,\rm M_{\odot}^{-1}$ (limiting the range of SNIa between 50 Myr and 13.7 Gyr), together with the stellar yields of~\cite{iwamoto99} (using their W70 carbon-deflagration model).
The resulting mass release for each individual elements are shown in Appendix~\ref{appendix:yields}.

We simplified further the feedback from SNII by assuming that all mass and energy are released in one single explosive event after $5\,\rm Myr$ to maximise the impact of SNII feedback.
We assumed a \cite{chabrier05} initial mass function, with a SNII explodability range of $M_{\rm ZAS}=[8,30] \rm M_\odot$ of canonical individual kinetic energy of $10^{51}\rm \, erg$, hence, corresponding to a SNII specific energy of $10^{49}\,\rm erg\, M_\odot^{-1}$.
We included the SN feedback model from~\cite{kimm14}, where the model follows either the energy conserving or momentum conserving phase of the explosion.
If the SN explosion is still in the energy-conserving phase, only internal energy is given to the gas since the code will be able to handle the Sedov explosion and get the correct amount of momentum at the end of the Sedov phase.
However, out of the energy-conserving phase, the correct amount of momentum is given to the gas.
We also included a minimal model for UV radiation from OB young stars, by considering the larger amount of momentum SNII can impart to the gas thanks to the pre-heating by the UV-radiation~\citep{geen15}.

\section{Dust}
\label{section:dustmodel}

The two bin size decomposition of~\cite{hirashita15} (see also the implementations of~\citealp{aoyama17}, \citealp{gjergo18}, or~\citealp{granato21}) was adopted for our subgrid model of dust evolution.
These bins correspond to a small bin and a large bin of dust sizes (radii; $a$) of respectively $a_{\rm S} = 5\,\rm nm$ and $a_{\rm L} = 0.1\,\mu\rm m$.
This decomposition allows to capture the bulk of the grain size distribution~\citep{hirashita15}, while having a negligible impact on the computational footprint (and on the memory load) of hydrodynamical simulations of galaxy formation. 

We separated the composition of dust between carbonaceous grains and silicate grains.
For silicates, we assumed a fixed amorphous olivine composition Mg$_{2-x}$Fe$_x$SiO$_4$ with iron inclusions balancing the amount of Mg ($x=1$), i.e. MgFeSiO$_4$, as it is supposed to make the bulk of silicates~\citep{kemper04,min07}, although there are still large modelling uncertainties plaguing the reconstructed compositions of silicates, particularly on the depletion of oxygen~\citep{jenkins09}. 
Assuming pyroxene with MgFeSi$_2$O$_6$ instead of olivine typically increased the amount of dust mass released as silicates by 20$\%$ for our stellar production models.
The respective grain material density of carbonaceous and silicate grains were respectively $s_{\rm C}=2.2\,\rm g\,cm^{-3}$ (i.e. we assumed a solid structure close to graphite) and $s_{\rm Sil}=3.3\,\rm g\,cm^{-3}$.

The equations of evolution of the dust mass content $D_{\rm i,j}$ of each grain size bin (with subscript $j=\rm S$ and L for respectively small and large grains) with chemical composition from the `key' element $i$, which is C for carbonaceous grains, and Si for silicate grains, are:
\begin{eqnarray}
\frac{{\rm d}D_{i,{\rm S}}}{{\rm d}t}=&&\frac{{\rm d}D_{{\rm acc},i,{\rm S}}}{{\rm d}t}+\frac{{\rm d}D_{{\rm ej},i,{\rm S}}}{{\rm d}t}+\frac{{\rm d}D_{{\rm sha},i,{\rm L}}}{{\rm d}t}\nonumber\\
&-&\frac{{\rm d}D_{{\rm coa},i,{\rm S}}}{{\rm d}t}-\frac{{\rm d}D_{{\rm spu},i,{\rm S}}}{{\rm d}t}-\frac{{\rm d}D_{{\rm SN},i,{\rm S}}}{{\rm d}t}-\frac{{\rm d}D_{{\rm \star},i,{\rm S}}}{{\rm d}t}\, ,\\
\frac{{\rm d}D_{i,{\rm L}}}{{\rm d}t}=&&\frac{{\rm d}D_{{\rm acc},i,{\rm L}}}{{\rm d}t}+\frac{{\rm d}D_{{\rm ej},i,{\rm L}}}{{\rm d}t}+\frac{{\rm d}D_{{\rm coa},i,{\rm S}}}{{\rm d}t}\nonumber\\
&-&\frac{{\rm d}D_{{\rm sha},i,{\rm L}}}{{\rm d}t}-\frac{{\rm d}D_{{\rm spu},i,{\rm L}}}{{\rm d}t}-\frac{{\rm d}D_{{\rm SN},i,{\rm L}}}{{\rm d}t}-\frac{{\rm d}D_{{\rm \star},i,{\rm L}}}{{\rm d}t}\, ,
\end{eqnarray}
where for each equation the first three terms on the right-hand side stand for the increase in dust mass and the last four terms for the decrease in dust mass. The various terms stand for: the accretion of elements from the gas phase to the dust phase (see Section~\ref{section:accretion}); the ejecta mass release from SSP evolution (see Section~\ref{section:SSPdust}); the shattering and coagulation of dust grains which transfer dust from, respectively, large to small sizes (Section~\ref{section:shattering}), and small to large sizes (Section~\ref{section:coagulation}); the thermal sputtering that returns dust elements to the gas phase (see Section~\ref{section:sputtering}); and the destruction of dust by shocks from type II SNe (see Section~\ref{section:SNdust}).

In this work we neglected entirely the differential dynamical motion of dust with respect to gas dynamics, therefore, the dust content of each bin size and chemical composition can be treated as a passive variable that is transported along with the flow of gas in the exact same vein as for the transport of metals in the gas phase.
This is possible due to the relatively short stopping time associated to Epstein aero-dynamical drag in dense gas $t_{\rm st}\simeq 0.5 \, a_{0.1}(n/1\,{\rm cm^{-3}})^{-1}(c_{\rm s}/10\,\rm km\,s^{-1})^{-1}\,\rm Myr$ compared to dynamical time $t_{\rm dyn}\simeq 10 \,  (L/100\,\rm pc)(\sigma_{\rm gr}/10\,\rm km\,s^{-1})^{-1}\,\rm Myr$, where $\sigma_{\rm gr}$ is the grain velocity dispersion. This gives a diffusion length scale of $L\simeq 5\, a_{0.1}(n/1\,{\rm cm^{-3}})^{-1}(\sigma_{\rm gr}/c_{\rm s})\, \rm pc$, which is in general below the resolution scale of this work.
We defer the investigation of the diffusion of dust grains on cloud scales to future work.

\subsection{Stellar production}
\label{section:SSPdust}

The amount of silicates $m_{\rm key}$, traced by its key element (here Si), released by a single star is given by the least abundant element entering its composition
\begin{equation}
m_{\rm key}=\min_{\ell=\rm{Mg,Fe,Si,O}} \left ( \frac{M_\ell}{A_\ell N_\ell} \right ) \, ,
\end{equation}
with $M_\ell$, $A_\ell$, $N_\ell$, respectively, the amount of mass of the $\ell$-th element released by the star, the atomic weight of the $\ell$-th element, and the number of atoms of the $\ell$-th element entering the composition of the silicate.
This available number of atoms of the key element condensates into dust within the ejecta at a given efficiency $\delta_{\rm Sil}$:
\begin{equation}
D_{{\rm ej,sil},\ell}^k=\delta_{\rm Sil}^k m_{\rm key} A_\ell N_\ell\, ,
\end{equation}
which guarantees a constant mass ratio ($A_\ell N_\ell/\sum(A_\ell N_\ell)=0.14,0.33,0.16,0.37$ for resp. $\ell=\rm Mg$, Fe, Si, and O in $x=1$ olivine\footnote{$A_\ell N_\ell/\sum(A_\ell N_\ell)=0.11,0.24,0.24,0.41$ for resp. $\ell=\rm Mg$, Fe, Si, and O in $x=1$ pyroxene.}) between the various elements entering the composition of silicate grains.
The superscript $k$ stands for the different type of mass release by stars with $k=$ AGB, SNII or SNIa,
The corresponding efficiencies were equal to $\delta_{\rm Sil}^{\rm SNII}=0.8$ for SNII following~\cite{dwek98} (and $\delta_{\rm Sil}^{\rm SNIa}=1$ for SNIa\footnote{Although this value of condensation efficiency for SNIa is extreme, SNIa ejecta are completely irrelevant in these simulations, i.e. for galaxies that are evolved for only a few $100\,\rm Myr$.}).
For AGB stars, the condensation efficiency depended on the ratio of C/O: if C/O~$\ge1$ all the oxygen was associated to carbon in CO molecules, and, hence, silicates did not condensate anymore as a result of unavailable O, while with C/O~$<1$ oxygen was still available to condensate into silicates.
We adopted $\delta_{\rm Sil}^{\rm AGB, C/O<1}=0.8$ and $\delta_{\rm Sil}^{\rm AGB, C/O\ge1}=0$ as in~\cite{dwek98}.
For the same reasons, accretion of elements onto silicates from the ISM was limited by the least abundant of the elements in the ISM entering the composition of the grain, replacing the mass $M_\ell$ by the mass of gas of the given element $\ell$ in the cell.

Similarly to silicates, the amount of carbonaceous dust and the condensation efficiency $\delta_{\rm C}$ varied with the nature of the stellar ejecta. 
It was assumed to be $\delta_{\rm C}^{\rm SNII}=0.5$ and $\delta_{\rm C}^{\rm SNIa}=0.5$, while for AGB stars this efficiency depended on the ratio of C/O in the ejecta.
If the ratio of C/O~$<1$, then all the C elements formed into CO molecules that were not available for dust $\delta_{\rm C}^{\rm AGB, C/O<1}=0$, while for C/O~$\ge1$ the efficiency was 1 with a mass released into C dust:
\begin{equation}
D_{\rm ej, C}^{\rm AGB}=\delta_{\rm C}^{\rm AGB, C/O\ge 1} \left (M^{\rm AGB}_{\rm C}-\frac{A_{\rm C}}{A_{\rm O}} M^{\rm AGB}_{\rm O} \right)\, .
\end{equation}

The production of dust in stellar ejecta is balanced between a population of small and large grains, i.e. $\dot D_{{\rm ej},i,{\rm S}}=f_{{\rm ej},i,{\rm S}}\dot D_{{\rm ej},i}$ and $\dot D_{{\rm ej},i,{\rm L}}^k=(1-f_{{\rm ej},i,{\rm S}})\dot D_{{\rm ej},i}$, where $f_{{\rm ej},i,{\rm S}}$ is the fraction of small grains released for a given chemical type $i$ (carbonaceous or silicate grains).
For sake of simplicity, we assumed that all the dust released in the ejecta is condensed into the large grain size population only, thus, $f_{{\rm ej},i,{\rm S}}=0$ (see appendix~\ref{appendix:fejsmall} for the predictions of extinction curves with different values of $f_{{\rm ej},i,{\rm S}}$, and the discussion in section~\ref{section:caveats}).
Finally, it is important to note that the predicted and observed dust condensation efficiencies are very uncertain and can vary by an order of magnitude~\citep[see][]{schneider23}. 

\subsection{Supernova-driven destruction}
\label{section:SNdust}

SNe (type II and Ia) destroy dust already present in the ISM, where the dust destruction is produced by inertial (non-thermal) sputtering and grain collisions~\citep{kirchschlager22}.
The amount of SN-destroyed dust is a fraction of the mass of gas $M_{\rm 100}$ shocked at above $100 \, \rm km \, s^{-1}$ defined as 
\begin{equation}
\Delta M_{{\rm SN},i,j}=\varepsilon_{{\rm SN},i}(a_j) {\rm min}\left ( \frac{M_{\rm 100}}{M_{\rm g}},1\right)M_{{\rm D},i,j}\, ,
\end{equation}
with the size-dependent destruction efficiency of~\cite{aoyama20}
\begin{equation}
\varepsilon_{{\rm SN},i}(a_j)=1-\exp\left(-\frac{\delta_{{\rm SN},i}}{a_{0.1,j}}\right)\, ,
\end{equation}
where $M_{\rm g}$ is the local gas mass, $M_{{\rm d},i,j}$ is the local dust mass, $a_{0.1,j}$ is the dust grain size in units of $0.1 \,\rm \mu m$, and $\delta_{{\rm SN},i}$ is the destruction efficiency.
Small grains are more efficiently decelerated by drag forces, trapping them near the shock region where thermal sputtering can quickly destroy them~\citep{nozawa06}.
This functional form of the destruction efficiency has been constructed to capture this behaviour with grain size $a$.
The mass of shocked gas is provided by the Sedov solution in a medium of homogeneous gas density, i.e. $M_{\rm 100}=6800 E_{\rm SN, 51}\, \rm M_\odot$, where $E_{\rm SN, 51}$ is the energy of the SN explosion in units of $10^{51}\, \rm erg$.
Multiple SNe were released at once over one time step $\Delta t$ (each individual SN engulfing the gas swept up by the previous explosion), following \citep[see][]{hou17}:
\begin{equation}
\Delta M_{{\rm SNe},i,j}=\left [ 1-\left ( 1-\frac{\Delta M_{{\rm SN},i,j}}{M_{{\rm D},i,j}}\right)^{N_{\rm SN}} \right ] M_{{\rm D},i,j}\, ,
\end{equation}
with $N_{\rm SN}$ the number of SNe.
We stress that this destruction step by the SN blast solution does explicitly destroy the returned ejecta from the stellar production, and destroys only the background dust material, hence, the dust condensation efficiency implicitly takes this term into account.
We used a different value of the dust destruction efficiency for carbonaceous and silicate grains with respectively $\delta_{\rm SN,C}=0.10$ and $\delta_{\rm SN,Sil}=0.15$ following the qualitative behaviour of~\cite{hu19}, where silicate grains are approximately destroyed 50 per cent more than carbonaceous grains.

\subsection{Thermal sputtering}
\label{section:sputtering}

Dust is also destroyed by thermal sputtering, the ejection of atoms from grains by the transfer of kinetic energy from gas particles at temperatures high enough to overcome the energy barrier of the binding energy.
We adopted the fitting form of~\cite{hu19} to the thermal sputtering yields $Y_{\rm th}$ calculated by~\cite{nozawa06}, which relates to the thermal sputtering timescale as (recall that $t_{{\rm spu},i,j}=m_{{\rm gr},i,j}/\vert \dot m_{{\rm gr},i,j} \vert=a_{i,j}/(3\vert \dot a_{i,j} \vert)$ with $m_{{\rm gr},i,j}$ the mass of a single dust grain):
\begin{equation}
t_{{\rm spu},i,j}=\frac{a_i}{3 n_{\rm H} Y_{{\rm th},j}(T)}\, .
\end{equation}
We note that sputtering yields are approximately 3 times lower at $T \gtrsim 10^6\,\rm K$ for carbonaceous grains than for silicate grains~\citep[see also][]{draine&salpeter79,tielens94}.
Hence, they differ from the widely adopted yields of~\cite{tsai&mathews95} giving a unique sputtering timescale based on the intermediate values for silicates and carbonaceous grains of~\cite{draine&salpeter79} and \cite{tielens94} (see Appendix~\ref{appendix:sputtering}).
Using cosmological hydrodynamical simulations of galaxy clusters, \cite{gjergo18} and~\cite{vogelsberger19} show that the dust mass content in observed galaxy clusters can be reproduced only for ten times lower sputtering yields than the canonical values. 
We defer this investigation to future work, but we immediately stress that thermal sputtering is already a sub-dominant destruction mechanism in our set of simulations compared to direct destruction by SNe.
We performed a G10LG simulation with sputtering time artificially increased by a factor of 10: there is no noticeable difference in any of the dust properties investigated in section~\ref{section:resultsmilkyway}.

\begin{table*}
    \centering
    \caption{Simulation names with their corresponding initial gas-phase metallicity, spatial resolution, and dust physics.}
    \begin{tabular}{c c c c c c c }
         Simulation & $Z_{\rm g,0}$ & Resolution & Accretion & SN destruction & Coagulation & Shattering \\
         name & ($Z_\odot$) & (pc) & model & model & model & model \\
         \hline
        \hline
        G10LG      & 1 & 18 & Fid. & Fid. & Fid. & Fid.  \\
        G10LG\_HR      & 1 & 9 & Fid. & Fid. & Fid. & Fid.  \\
        G10LG\_LR  & 1 & 36 & Fid. & Fid. & Fid. & Fid.  \\
        G10LG\_VLR & 1 & 72 & Fid. & Fid. & Fid. & Fid. \\
        G10LG\_LB\_MR  & 1 & 18 & LB12 & Fid. & Fid. & Fid.  \\
        G10LG\_LB\_LR  & 1 & 36 & LB12 & Fid. & Fid. & Fid.  \\
        G10LG\_LB\_VLR & 1 & 72 & LB12 & Fid. & Fid. & Fid.  \\
        G10LG\_NCO & 1 & 18 & No CO & Fid. & Fid. & Fid.  \\
        G10LG\_sn      & 1 & 18 & Fid. & Reduced & Fid. & Fid.  \\
        G10LG\_cF   & 1 & 18 & Fid. & Fid. & $3\times$ Fast & Fid.  \\
        G10LG\_cS   & 1 & 18 & Fid. & Fid. & $3\times$ Slow & Fid.  \\
        G10LG\_sF   & 1 & 18 & Fid. & Fid. & Fid. & $3\times$ Fast  \\
        G10LG\_sS   & 1 & 18 & Fid. & Fid. & Fid. & $3\times$ Slow  \\
        G10LG\_csF   & 1 & 18 & Fid. & Fid. & $3\times$ Fast & $3\times$ Fast  \\
        G10LG\_HZ  & 2 & 18 & Fid. & Fid. & Fid. & Fid.  \\
        G10LG\_LZ  & 0.3 & 18 & Fid. & Fid. & Fid. & Fid.  \\
        G10LG\_VLZ  & 0.1 & 18 & Fid. & Fid. & Fid. & Fid.  \\
        G10LG\_NCO & 1 & 18 & No CO & Fid. & Fid. & Fid.  \\
        \hline
        G9HG\_LZ  & 0.3 & 18 & Fid. & Fid. & Fid. & Fid.  \\
        G9HG\_VLZ & 0.1 & 18 & Fid. & Fid. & Fid. & Fid.  \\
        G9HG\_VVLZ  & 0.03 & 18 & Fid. & Fid. & Fid. & Fid.  \\
        G9HG\_VVVLZ & 0.01 & 18 & Fid. & Fid. & Fid. & Fid.  \\
        G9HG\_VLZ\_NCO  & 0.1 & 18 & No CO & Fid. & Fid. & Fid.  \\
        G9HG\_VVLZ\_NCO  & 0.03 & 18 & No CO & Fid. & Fid. & Fid.  \\
        \hline
        G9LG\_LZ     & 0.3 & 18 & Fid. & Fid. & Fid. & Fid.  \\
        G9LG\_VLZ    & 0.1 & 18 & Fid. & Fid. & Fid. & Fid.  \\
        G9LG\_VVLZ   & 0.03 & 18 & Fid. & Fid. & Fid. & Fid.  \\
        G9LG\_VVVLZ  & 0.01 & 18 & Fid. & Fid. & Fid. & Fid.  \\
        G9LG\_VLZ\_NCO    & 0.1 & 18 & No CO & Fid. & Fid. & Fid.  \\
        G9LG\_VVLZ\_NCO   & 0.03 & 18 & No CO & Fid. & Fid. & Fid.  \\
        \hline
        G8HG\_LZ  & 0.3 & 18 & Fid. & Fid. & Fid. & Fid.  \\
        G8HG\_VLZ & 0.1 & 18 & Fid. & Fid. & Fid. & Fid.  \\
        G8HG\_VVLZ & 0.03 & 18 & Fid. & Fid. & Fid. & Fid.  \\
        G8HG\_VVVLZ & 0.01 & 18 & Fid. & Fid. & Fid. & Fid.  \\
        G8HG\_VLZ\_NCO & 0.1 & 18 & No CO & Fid. & Fid. & Fid.  \\
        G8HG\_VVLZ\_NCO & 0.03 & 18 & No CO & Fid. & Fid. & Fid.  \\
        \hline
    \end{tabular}
    \label{tab:simulations}
\end{table*}

\subsection{Gas accretion}
\label{section:accretion}

The dust mass content can finally grow by accretion through the metals in the gas phase, following~\cite{dwek98} 
\begin{equation}
\dot M_{{\rm acc},i,j}=\left( 1-\frac{M_{{\rm D},i,j}}{M_{ Z,i}} \right) \frac{M_{{\rm D},i,j}}{t_{{\rm acc},i,j}}\, ,
\end{equation}
where $M_{Z,i}$ is the mass of metals (gas + dust) of the key element, $t_{{\rm acc},i,j}$ is the accretion timescale 
\begin{equation}
t_{{\rm acc},i,j}^{-1}= \alpha \frac{\pi \bar a_j^2}{f_{X}m_{{\rm gr},i,j}} \rho_{X} u_{{\rm th},X}\, ,
\end{equation}
where $m_{{\rm gr},i,j}=s_{i} a_j^3 4\pi /3$ is the grain mass, $s_{i}$ is the grain material density, $f_{X}$ is the mass fraction of the limiting element in the chemical composition of the grain, $u_{{\rm th},X}=(8 k_{\rm B} T/(\pi m_{X}))^{1/2}$ is the gas thermal velocity, $m_{X}=A_{X}m_{\rm H}$ is the atomic mass, and $\rho_{X}$ is the total (gas plus dust phase) mass density of the limiting element.
$\pi \bar a_{i,j}^2$ is the surface of the grain with Coulomb enhancement factor $E_i(a_j)$ due to electrostatic effects caused by ionised gas interacting with charged grains~\citep{weingartner&draine99}, which is integrated over the whole distribution of the grain sizes. 
In our case, where the size distribution was discretised over two bins and assuming a Dirac distribution, we simply got $\bar a_{i,j}=E_i(a_j) a_j$.
For carbonaceous grains, $X=i=\rm C$, however, for silicate grains, the limiting element depends on the grain composition, i.e. 
\begin{equation}
\frac{\rho_X u_{{\rm th,}X}}{f_{X}}={\rm min}\left [ \frac{\rho_\ell u_{{\rm th,}\ell}}{f_\ell}\right ]_{\ell={\rm Mg, Fe, Si, O}} \, .
\end{equation}
The accretion timescale can be rewritten as
\begin{equation}
t_{{\rm acc},i,j}= 0.28\alpha^{-1} a_{{\rm 0.005},j}  s_{3,i}  \frac{A_{X}^\frac{1}{2}f_{X}}{E_i(a_j) Z_{X}}\left(\frac{n}{1\,\rm H\, cm^{-3}}\right)^{-1} \left(\frac{T}{50\,\rm K}\right) ^{-\frac{1}{2}} \, \rm Myr\, , 
\end{equation}
where $a_{{\rm 0.005},j}=a_j/5\,\rm nm$, $s_{3,i}=s_i/3\,\rm g\,cm^{-3}$, $\alpha$ is the sticking coefficient of gas particles onto dust, and  $Z_{X}=\rho_{X}/\rho_{\rm g}$ is the mass fraction of the limiting element. 
\cite{weingartner&draine99} gave values of $E(a)$ for a range of sizes of carbonaceous and silicate grains for three typical phases of the ISM. 
To mimic their complex dependencies, we simply adopted a value of $E=1$ for all kind of grains everywhere, except for large carbonaceous and small silicate grains in the cold neutral medium with $T<2\times10^4\,\rm K$ and $n>10\,\rm H\,cm^{-3}$, for which we adopted $E_{\rm C}(0.1\,\mu\rm m)=0$ and $E_{\rm Sil}(5\,\rm nm)=10$.

Sub-grid accretion can be estimated assuming that the gas density with mean value $\bar n$ (the value of the gas density in a given resolution element) follows a log-normal PDF at unresolved scales, which is shaped by the turbulence:
\begin{equation}
p(s)=\frac{1}{\sqrt{2\pi} \sigma_s} \exp{\left(-\frac{(s-s_0)^2}{2\sigma_s^2}\right)}\, , 
\end{equation}
where $s=\log(n/\bar n)$, $s_0=-\sigma_s^2/2$ and $\sigma_s=\log(1+b^2\mathcal{M}^2)$, with $b$ the compression ratio (we take $b=0.4$ for mixed turbulence which is the same value used in the gravo-turbulent model for SF efficiency) and $\mathcal{M}$ the turbulent Mach number.
The effective timescale $t_{\rm acc,eff}$ was obtained by integrating the mass accretion rate over the log-normal PDF for a given set of mean density, temperature, and metallicity ($\bar n, \bar T, \bar Z$), hence:
\begin{equation}
\frac{t_{\rm acc}(\bar n, \bar T, \bar Z)}{t_{\rm acc,eff}}=\int_{-\infty}^{s_{\rm max}} \frac{t_{\rm acc}(\bar n, \bar T, \bar Z)}{t_{\rm acc}(n, T, Z)} \frac{n}{\bar n} p(s)ds\, ,
\end{equation}
up to a maximum density $s_{\rm max}=\log(n_{\rm max}/\bar n)$, where the maximum density is $n_{\rm max}=10^4 \, \rm H\, cm ^{-3}$ where grains starts to be significantly coated with water ice mantle ($n\ge 10^3 \,\rm H\, cm^{-3}$, \citealp{cuppen&herbst07,hollenbach2009}) suppressing the accretion of refractory material onto the grain surface.
We used the value of the gas density and metallicity of a given cell for $\bar n$ and $\bar Z$, and we assumed that the gas temperature is always $\bar T=100\,\rm K$ for gas that fulfils the conditions to trigger this unresolved log-normal PDF of density (see below) for simplicity.
In addition, we assumed that $T=\bar T$ and $Z=\bar Z$ for the sampled values of $n$ by the log-normal PDF. 
This led to 
\begin{eqnarray}
\frac{t_{\rm acc}(\bar n,\bar T,\bar Z)}{t_{\rm acc,eff}}&=&\int_{-\infty}^{s_{\rm max}} \exp(2s) p(s)ds\, \nonumber \\
&=&\frac{e^{\sigma_s^2}}{2}{\rm Erfc}\left(\frac{3\sigma_{s}^2/2-s_{\rm max}}{\sqrt{2}\sigma_s} \right)  \, ,
\end{eqnarray}
where $\rm Erfc$ is the complementary error function.
This unresolved effective accretion timescale was estimated only for cells with gas density larger than $0.1\,\rm H\,cm^{-3}$, temperature lower than $10^4\,\rm K$, and Jeans length smaller than 4 times the local cell size, assuming a sticking coefficient of $\alpha_{\rm eff}=1/3$, which value is representative of the sticking coefficient obtained at temperatures of $T=100-1000\,\rm K$ (see Fig.~\ref{fig:stickcoef}).
Otherwise the accretion time was that obtained for $\bar n$ (the local gas density) with the~\cite{lebourlot12} sticking coefficient (see Appendix~\ref{appendix:sticking_coefficient} for a comparison of sticking coefficients).

\begin{figure}
\centering \includegraphics[width=0.48\textwidth]{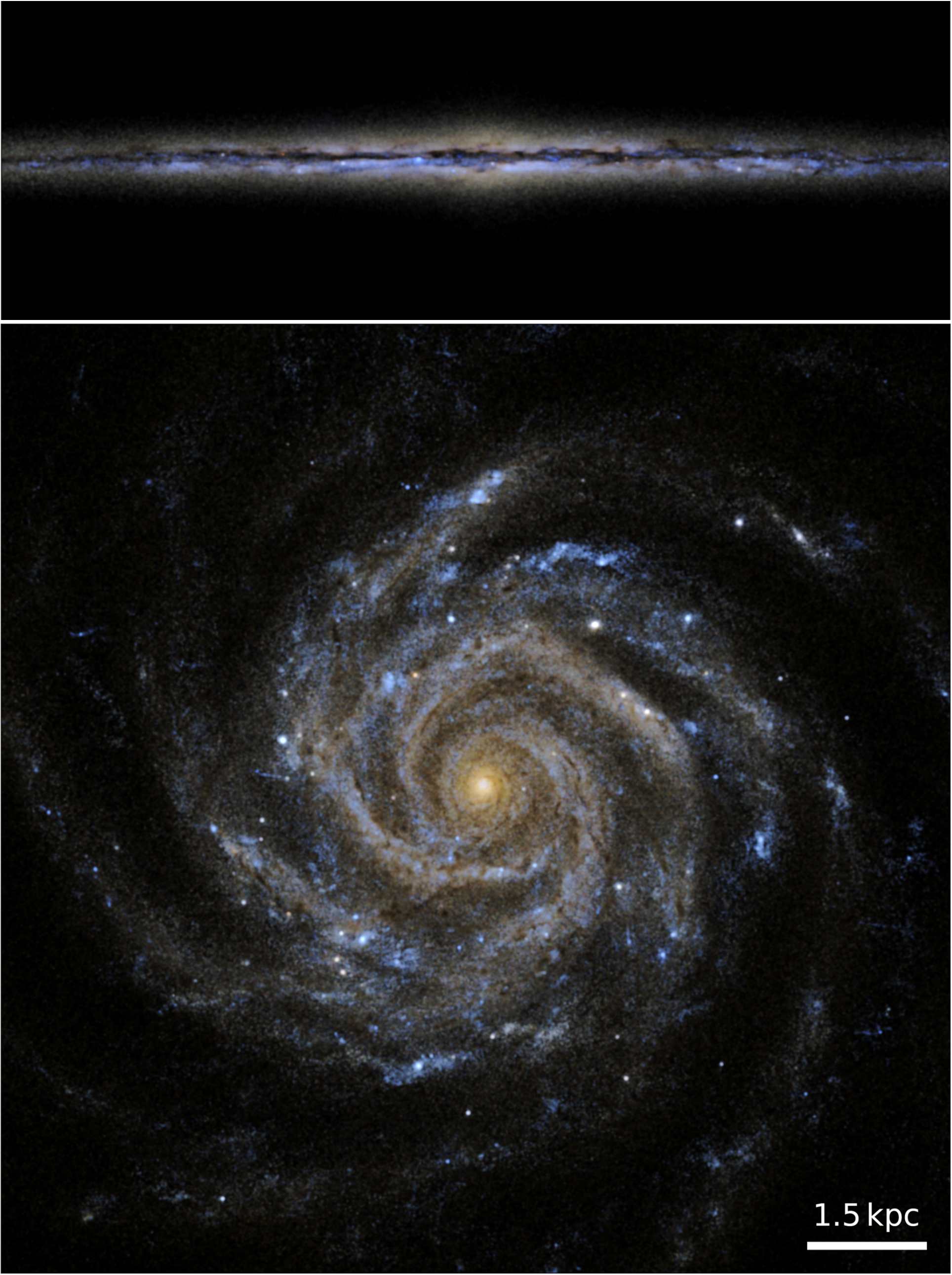}
\caption{Mock image of the fiducial G10LG simulation seen edge-on (top) or face-on (bottom) in SDSS g-r-i filter bands at $t=400\,\rm Myr$.}
\label{fig:skirt}
\end{figure}

For carbonaceous grains, we further assumed that carbon in the gas phase is fully molecular (CO) above a gas density of $n>10^3 \, \rm H\,cm ^{-3}$.
Indeed, simulations~\citep[e.g.][]{safranek17,clark19,hu21,katz22b} give a transition to full CO gas at $n=2\times10^3-10^4 \, \rm H\,cm ^{-3}$ depending on the cosmic ray ionisation rate.
We note that while our value might seem fairly low (or corresponding to a very weak cosmic ray ionisation rate), our subgrid gas distribution neglects the fact that gas collapses further at smaller unresolved scales (higher density) and extends the log-normal PDF with a power-law tail~\citep{vallini18}.

We will see in Section~\ref{section:resolution} that this turbulence-driven subgrid accretion model allows for better converged amounts of dust with respect to resolution (as opposed to ignoring the unresolved density structure of the gas).
Other work in the literature had to make similar, though ad hoc, assumptions on how much gas concentrates into typical molecular cloud densities~\citep[$10^3\,\rm H\,cm^{-3}$,][]{aoyama17,granato21}, or using a Sobolev-like shielding length to infer the amount of unresolved dense gas~\citep{choban22}.

\begin{figure*}
\centering \includegraphics[width=1.05\textwidth]{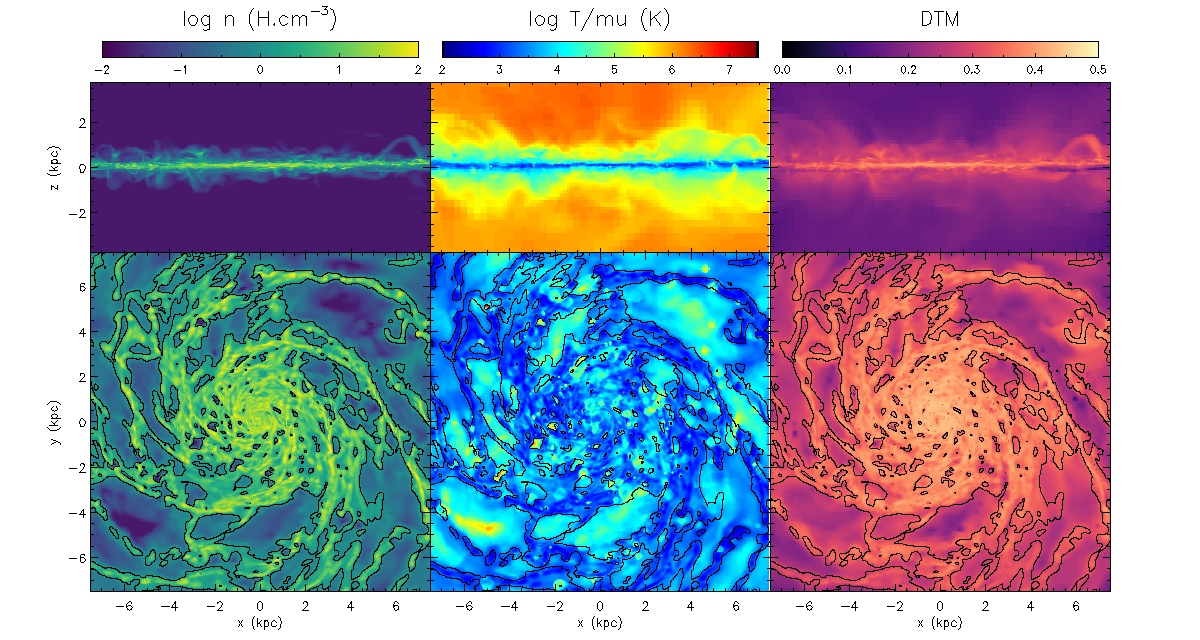}
\centering \includegraphics[width=1.05\textwidth]{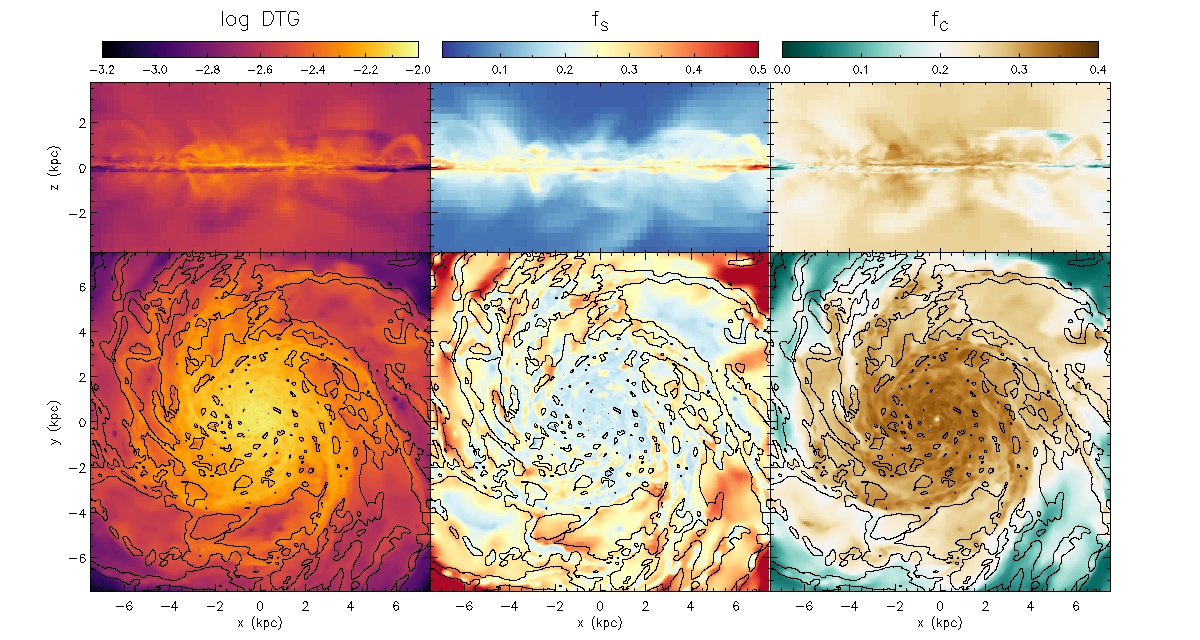}
\caption{Mass-weighted projections of the gas density (top two left panels), temperature (top two middle panels), dust-to-metal mass ratio (top two right panels), and dust-to-gas mass ratio (bottom two left panels), small grain fraction (bottom two middle panels), and carbonaceous grain fraction (bottom two right panels), viewed edge-on (top panel with a depth of $30\,\rm kpc$) or face-on (bottom panel with a depth of $600\,\rm pc$) for the high resolution version of G10LG galaxy with the fiducial physics at $t=400 \,\rm Myr$.
In black are shown isocontours of density with $n=1\,\rm H\,cm^{-3}$.}
\label{fig:illustrative_example_g10lg}
\end{figure*}

\subsection{Shattering}
\label{section:shattering}

Shattering is the process by which grains with sufficiently large velocities collide and fragment into smaller size grains.
Therefore, it is a transfer of mass from the large to the small grain population that conserves the total dust mass.
The corresponding timescale is obtained by considering the timescale for grain collisions $\tau_{\rm col}=(\sigma_{\rm D} \tilde \sigma n_{\rm D})^{-1}$, where $\tilde \sigma$ is the grain cross section, $\sigma_{\rm D}$ the dust grain velocity dispersion, and $n_{\rm D}$ the dust number density, which boils down to~\citep{aoyama17}:
\begin{equation}
t_{\rm sha,i}= 54 a_{\rm 0.1,L}  s_{3,i} \left(\frac{\mathcal{D}_{\rm L}}{0.01}\right)^{-1}\left(\frac{n}{1\,\rm H\,cm^{-3}}\right)^{-1}\left(\frac{\sigma_{\rm D,L}}{10\,\rm km\,s^{-1}}\right)^{-1}\, \rm Myr\, ,
\end{equation}
where $\mathcal{D}_{\rm L}$ and $\sigma_{\rm D,L}$ are, respectively, the dust-to-gas ratio and velocity dispersion of the large grains population.
Typical velocities required to shatter grains are above a few $\rm km\,s^{-1}$~\citep{jones96}, which limit the effect of shattering to the diffuse phase of the ISM.
\cite{yan04} provided grain-gas relative velocities for various phases of the ISM and showed that large grains ($a_{\rm L}=0.1\,\rm \mu m$) reach the turbulent velocity of the forcing scale, i.e. $\sigma_{\rm D,L}\simeq 10 \,\rm km\,s^{-1}$ in the warm ionised medium, $\simeq 1 \,\rm km\,s^{-1}$ for the warm or cold neutral medium, and $\simeq 0.1-1 \,\rm km\,s^{-1}$ for molecular clouds.
We followed~\cite{granato21} and adopted the following functional form:
\begin{equation}
t_{\rm sha}= 54 a_{\rm 0.1}  s_3 \left(\frac{\mathcal{D}_{\rm L}}{0.01}\right)^{-1}\left(\frac{n}{1\,\rm H\,cm^{-3}}\right)^{-p_{\rm sh}}\, \rm Myr\, ,
\label{eq:sha_time}
\end{equation}
with $p_{\rm sh}=1$ for $n<1\,\rm H\, cm^{-3}$ and $p_{\rm sh}=1/3$ for $1<n<10^3\,\rm H\, cm^{-3}$, which is equivalent to having a dispersion velocity varying from $10\,\rm km\,s^{-1}$ to $1\,\rm km\,s^{-1}$ from the warm ionised phase ($1\,\rm H\,cm^{-3}$) to the molecular cloud density\footnote{We note that considering the gas-drag timescale and a Kolmogorov turbulence, the velocity dispersion of grains follow~\citep{ormel09, hirashita&aoyama19}: $\sigma_{\rm gr}= 1.2\mathcal{M}^{3/2} a_{0.1}^{1/2}  s_3^{1/2} T_4^{1/4} n_1^{-1/4}\, \rm km\,s^{-1}$.
 Hence, with a temperature-density scaling relation of $T\propto n^{-2/3}$, which well approximates the obtained relation in our simulations for the ISM gas below $T<10^4\,\rm K$, one gets $p_{\rm sh}=7/12$, steeper than the power of $1/3$ adopted here.} ($10^3\,\rm H\,cm^{-3}$).

\subsection{Coagulation}
\label{section:coagulation}

When dust grains are embedded in gas that is dense and cold, they have small velocity dispersions.
This leads to the coagulation of grains by their direct collisions, thereby, transferring mass from small grains to large grains~\citep{yan04}.
We set the timescale for grain coagulation to that obtained by considering the typical timescales obtained previously for shattering but now for the coagulation of small grains into large grains, i.e.
\begin{equation}
t_{\rm coa}= 0.27 \frac{a_{\rm 0.005} s_3}{F} \left(\frac{\mathcal{D}_{\rm S}}{0.01}\right)^{-1}\left(\frac{n}{10^3\,\rm H\, cm^{-3}}\right)^{-1} \left( \frac{\sigma_{\rm D,S}}{0.1\, {\rm km\, s^{-1}}} \right)^{-1}\, \rm Myr\, ,
\end{equation}
where $F$ is a fudge factor.
The density to achieve grain velocity dispersions that are low enough to pass below the coagulation threshold velocity of $\lesssim 0.1-1\,\rm km\, s^{-1}$ for grains of size $a=0.005\,\rm \mu m$~\citep{chokshi93,poppe&blum97} is of the order $n\sim 10^2-10^3\,\rm H\, cm^{-3}$, which is barely resolved at resolutions of $\Delta x\simeq 20 \,\rm pc$ as in this work.

Since the velocity dispersion of small grains has complex dependencies with the gas turbulent velocity dispersion at the forcing scale, the magnetic field strength, the grain charge, and the ionisation state of the gas~\citep{yan04}, we greatly simplified the problem (following~\citealp{aoyama17}) by assuming that half ($F=0.5$) the gas mass, which Jeans length is unresolved and with gas density above $0.1\, \rm H\, cm^{-3}$ and temperature below $10^4\,\rm K$, has an actually larger gas density of $10^3\,\rm H\,cm^{-3}$ with a small grain turbulent velocity dispersion of $0.1\,\rm km\,s^{-1}$~\citep{yan04}.
We also performed two simulations with a coagulation model where $F=1$ and where $n$ is sampled by the log-normal PDF shaped by turbulence as for the subgrid model for dust growth by accretion.
In this model, we tried two different density cuts, $n_{\rm max,coa}=10^5$ or $10^7\,\rm H\, cm^{-3}$, and their results show very negligible differences with respect to the fiducial coagulation model for G10LG (they produce respectively $\sim 0.2$ and $3\,\%$ less small grains than the fiducial G10LG simulation at final time).

\begin{figure}
\centering \includegraphics[width=0.45\textwidth]{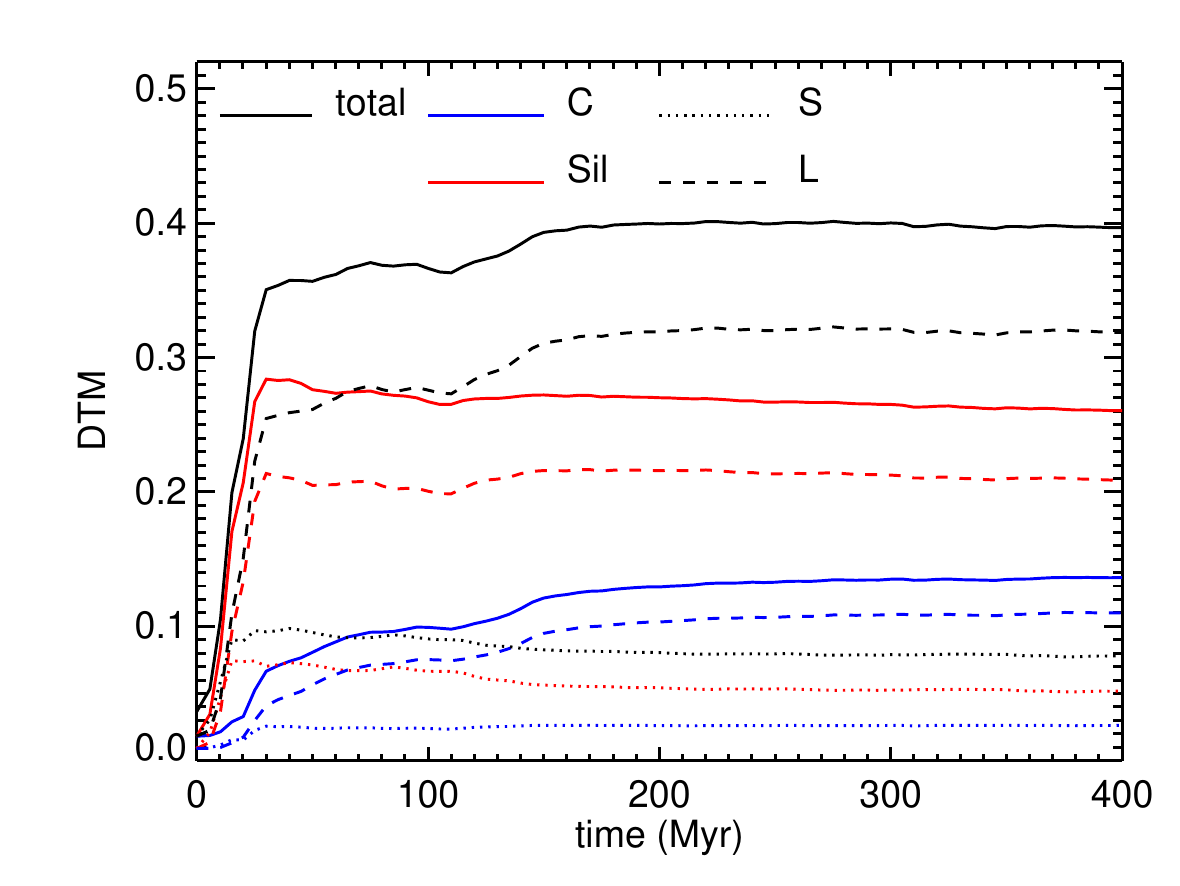}\\
\caption{Dust-to-metal mass ratio (DTM) as a function of time for the fiducial G10LG galaxy. The DTM is further decomposed into the carbonaceous ('C' in blue) and silicate ('Sil' in red) grains, and into small $5\,\rm nm$ ('S' in dotted) and large $0.1 \,\mu\rm m$ ('L' in dashed). The DTM is close to the canonical value of 0.4 in the Milky Way, where the bulk of the mass is composed of large silicate grains.}
\label{fig:dtm_fiducial}
\end{figure}

\begin{figure}
\centering \includegraphics[width=0.45\textwidth]{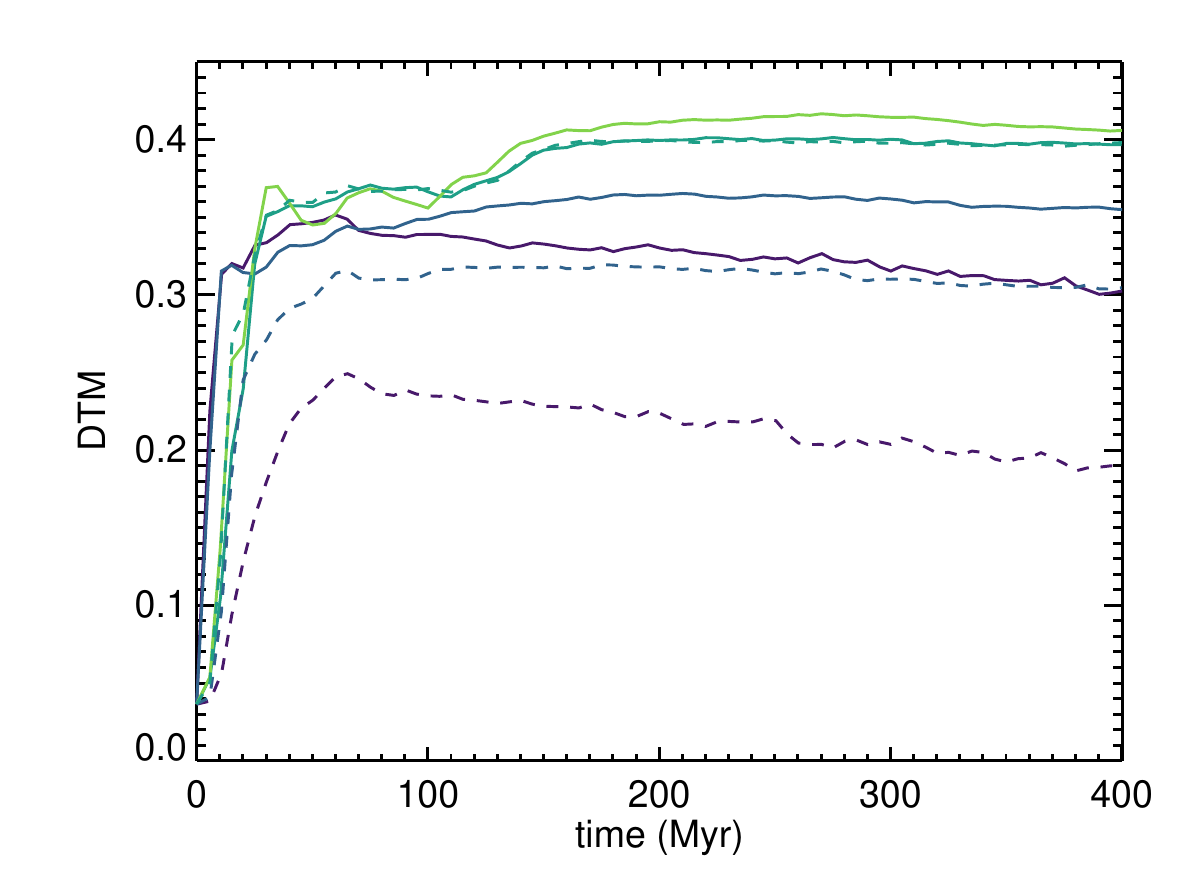}\\
\caption{Dust-to-metal mass ratio (DTM) as a function of time for the different resolutions of the G10LG galaxy using either the fiducial model for dust accretion with turbulence-driven for unresolved densities (solid lines) or the model without (dashed lines).
From bottom to top are the simulations with $\Delta x=72$, 36, 18, and $9\,\rm pc$ resolution (the $9\,\rm pc$ resolution simulation is for the dust fiducial model only). 
The subgrid turbulent model for accretion significantly improves the numerical convergence of the DTM.}
\label{fig:dtm_rescomp_lb12}
\end{figure}

\section{Results for the Milky Way-like galaxy}
\label{section:resultsmilkyway}

The entire set of simulations with variations in resolution, metallicity, or dust physics are summarised in table~\ref{tab:simulations}. 
We will first focus on the fiducial model of our Milky Way-like galaxy G10LG to validate the model of dust physics against available observations for the Milky Way.

\subsection{Visual inspection}

In Fig.~\ref{fig:skirt}, we show a mock image of the fiducial G10LG simulation at $t=400\,\rm Myr$ obtained with the radiative transfer code {\sc skirt}~\citep{camps15} in the g-r-i filter bands.
For the purpose of this image, the stars set up from initial conditions have an age with a cylindrical radial dependency of ${\rm age}=(9-R_{\rm cyl}/{2\,\rm kpc)}\,\rm Gyr\pm 3\,\rm Gyr$ and have solar metallicities.
One can see that the obtained light distribution shows a diffuse old stellar component structured spiral arms, together with more clumpy  blue regions of star formation, with absorption features from dusty gas.

Figure~\ref{fig:illustrative_example_g10lg} gives the visual representation of various gas quantities in the galaxy of the fiducial run, as mass-weighted edge-on (30 kpc in depth) and face-on (0.6 kpc in depth) projections at time $t=400\,\rm Myr$.
The gas is clearly multiphase and strongly clustered into regions of neutral gas with high densities ($n>10-100\,\rm H\,cm^{-3}$) and low temperatures ($T\simeq100 \,\rm K$) embedded in a more diffuse ionised medium at intermediate density ($n\simeq 0.1-1\,\rm H\, cm^{-3}$) and at warm temperatures ($T\simeq10^4\,\rm K$), and SN-driven hot pockets of ultra-diffuse gas ($n<0.1\,\rm H\,cm^{-3}$ and $T\gtrsim 10^6 \,\rm K$).
The galactic wind exhibits only a two-phase structure with the warm and the hot phases from the ejected ISM.

The dust-to-metal mass ratio ${\rm DTM}=\rho_{\rm D}/\rho_{\rm Z}$ and the dust-to-gas mass ratio ${\rm DTG}=\rho_{\rm D}/\rho$, where $\rho_{\rm D}$ is the total dust mass density and $\rho_{\rm Z}$ is the total metal (dust+gas) density, show significant spatial variations in the ISM. 
The DTM is higher in regions of higher densities due to the increase in accretion rates of refractory elements on dust with gas density, and to the higher destruction rates by thermal sputtering at higher temperatures.
Dense gas also corresponds to regions where the mass fraction of small grains $f_{\rm S}$ is lower and the mass fraction of carbonaceous grains $f_{\rm C}$ is higher compared to the more diffuse ISM.

We will ignore this aspect in this work, but it can already be noted that the circum-galactic medium has an appreciable level of dust, with the DTM reaching at $0.1-0.2$ composed mostly of large $0.1\,\mu\rm m$ grains.
Due to the high values of temperature ($T\gtrsim 10^6\,\rm K$) reached in the galactic outflow, the smallest populations of dust grains are faster destroyed by thermal sputtering.

\subsection{Dust-to-metal ratio}
\label{section:resolution}

The evolution over time of the DTM for the fiducial G10LG galaxy is shown in Fig.~\ref{fig:dtm_fiducial}.
This quantity is computed for all the gas and dust contained in a cylinder of $r=4\,\rm kpc$ cylindrical radius centred on the box and $h/2=200\,\rm pc$ above and below the disc plane.
The DTM rises steeply in less than $50\,\rm Myr$ to its approximately steady-state value at nearly ${\rm DTM}=0.4$ comparable to the canonical ratio in present-day galaxies with solar-like metallicities~\citep[e.g.][]{remy14,devis19} or in the Milky Way~\citep[e.g.][]{jenkins09}.
The DTM is decomposed into the four simulated dust bins, i.e. into small (dotted) and large (dashed) grains, and into carbonaceous (blue) and silicate (red) grains.
The bulk of the DTM is from large grains ($80\,\%$) and from silicates ($66\,\%$), and in particular from large silicate grains ($53\,\%$).

Since the accretion time scales inversely with the density of refractory elements in the gas phase, the value of the DTM is sensitive to the ability of the simulation to resolve the dense gas where most of the accretion and growth of dust proceeds.
This motivates the use of a subgrid model (as introduced in section~\ref{section:accretion}) for accretion if the typical molecular cloud densities are not captured.
This is illustrated in Fig.~\ref{fig:dtm_rescomp_lb12} where we compare the G10LG galaxy with the fiducial dust physics run at different spatial resolutions (solid lines) to the same simulated galaxies where the turbulence-driven subgrid dust accretion is turned off (dashed lines for the series of G10LG\_LB simulated galaxies).
The numerical convergence to ${\rm DTM}=0.4$ is obtained at lower resolution for the fiducial model for unresolved molecular cloud densities than in the naive model, although both give similar values of DTM when sufficiently resolved (few per cent relative difference at a resolution of $9 \,\rm pc$).

\subsection{Dust size distribution}

\begin{figure}
\centering \includegraphics[width=0.45\textwidth]{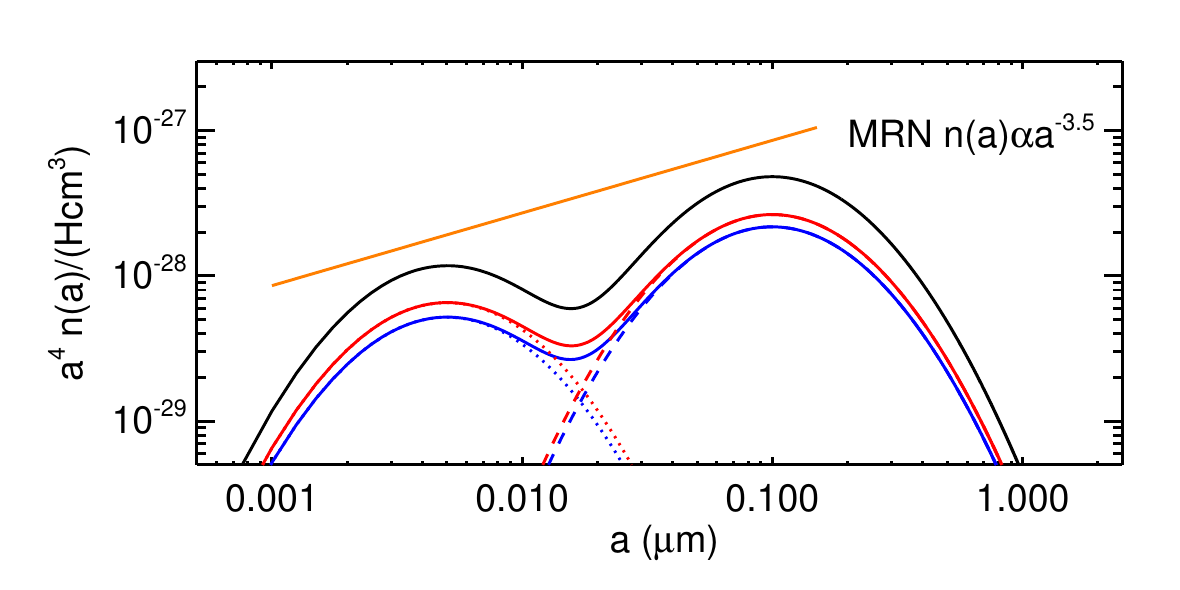}\vspace{-0.5cm}
\centering \includegraphics[width=0.45\textwidth]{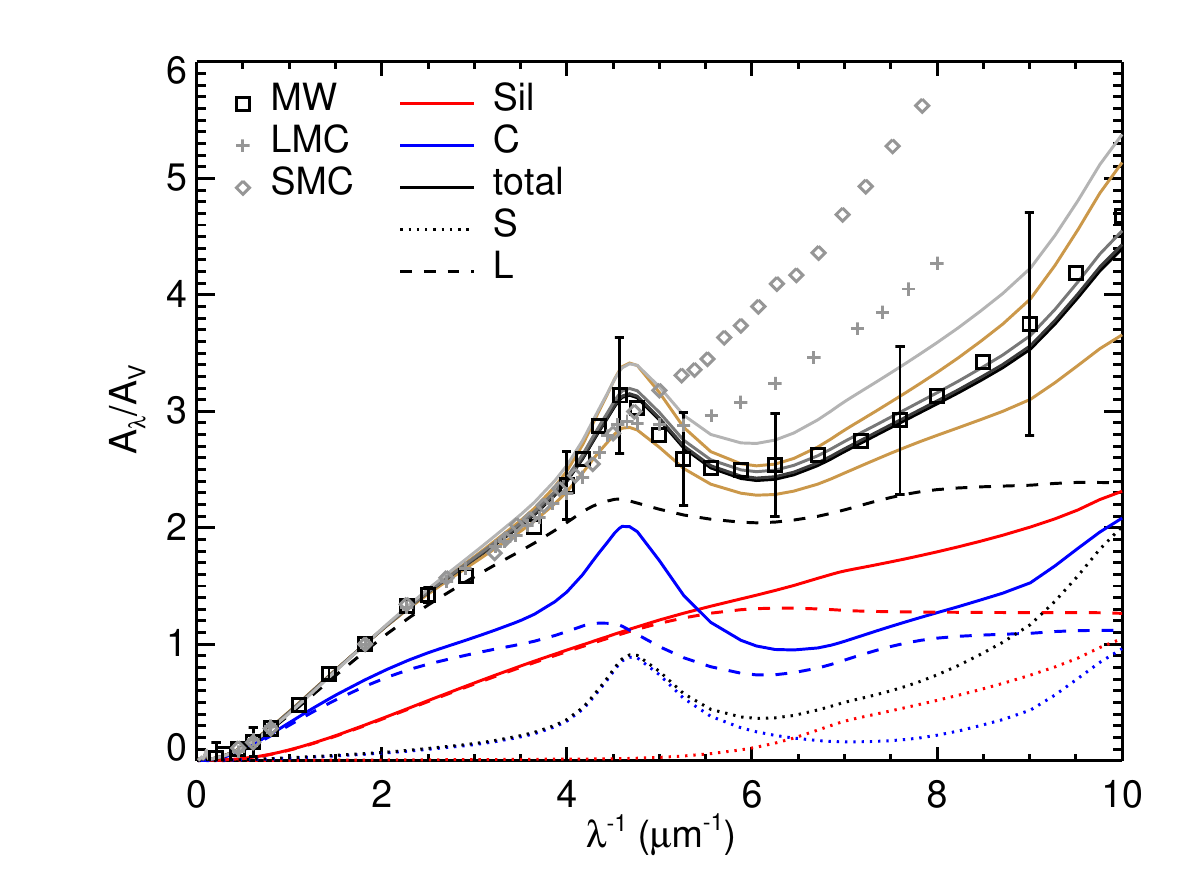}
\caption{Respectively top and bottom panels: Dust size distribution and extinction curve obtained from the fiducial G10LG simulation at $t=400\,\rm Myr$ for the total dust content (black solid), and the contribution from carbonaceous (blue, 'C'), silicate (red, 'Sil') grains, and small (dotted, 'S') and large grains (dashed, 'L'). The extinction curves at different times $t=100$, 200, 300, and $400 \,\rm Myr$ are shown from light to dark grey scales. The scatter of the extinction curves from the simulation are the brown solid lines. The dust size distribution is compared to that of the~\cite{mathis77} (MRN) size distribution in the Milky Way (in orange). The extinction from the Milky Way (MW), Large Magellanic Cloud (LMC), and Small Magellanic Cloud (SMC) from ~\cite{pei92} are shown as labelled in the corresponding panel, with the typical scatter estimated from the data of~\cite{fitzpatrick07}. Our simulated MW analogue is in remarkably good agreement with the MRN size distribution and the MW extinction curve.}
\label{fig:size_ext_fiducial}
\end{figure}

\begin{figure}
\centering \includegraphics[width=0.45\textwidth]{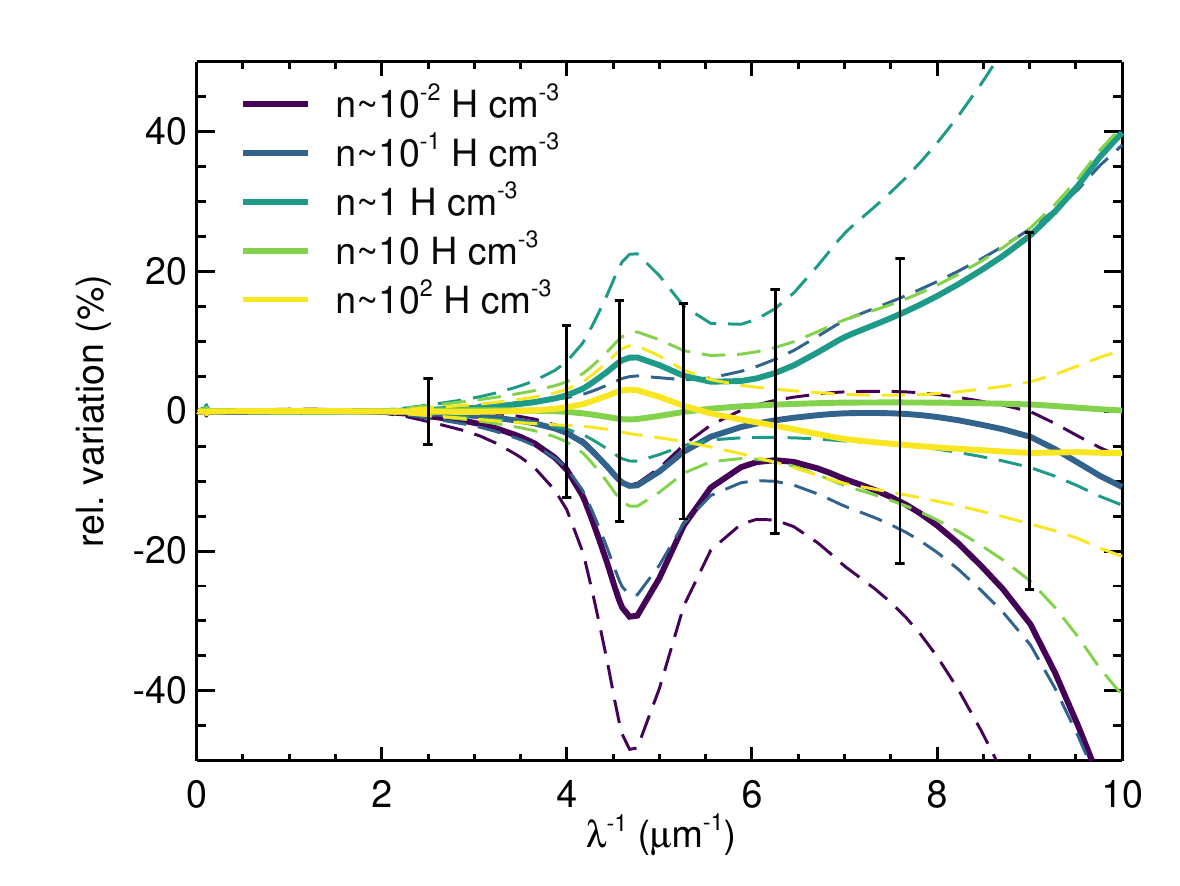}
\caption{Relative variations of the mean extinction curve $\delta(A_\lambda/A_{\rm V})\vert_n$ (solid line) at a given gas density (color-coded as indicated in the panel) with respect to the galactic extinction curve in the G10LG simulation at $t=400\,\rm Myr$. The dashed lines indicate the 1-$\sigma$ scatter for each sampled density. The error bars stand for the scatter of the observed extinction curves of the Milky Way from~\cite{fitzpatrick07} relative to the mean extinction curve from~\cite{pei92}.}
\label{fig:rel_ext_ngas}
\end{figure}

The size distribution of grains in the Milky Way, $n^{\rm D}(a)$ the number density of grains at a given size $a$, can be well represented by a power law with slope $n^{\rm D}\propto a^{-3.5}$. This is the so-called MRN size distribution from~\cite{mathis77}.

The top panel of Fig.~\ref{fig:size_ext_fiducial} shows the reconstructed size distribution of grains (in black solid line) from our four populations with size $a_j$ ($j$ standing for S or L) and chemical composition $i$ (for carbonaceous or silicate grains), assuming that each population is sampled with a modified log-normal PDF (as in~\citealp{hirashita15} and \citealp{hou17}) of 
\begin{equation}
    n^{\rm D}_{i,j}(a)=\frac{C_{i,j}}{a^4}\exp \left( -\frac{ \left[ \ln(a/a_{j})\right]^2 }{2\sigma_{i,j}^2} \right)\, ,
\label{eq:sizes}
\end{equation}
where $\sigma_{i,j}=0.75$ is the standard deviation of the log-normal part and $C_{i,j}$ is the normalisation of the distribution. The normalisation is given by:
\begin{equation}
    \mu m_{\rm p}\mathcal{D}_{i,j}=\int_0^\infty \frac{4}{3}\pi a^3 s_{i} n^{\rm D}_{i,j}(a) {\rm d}a \, ,
\end{equation}
where $\mu=1.22$ is the mean molecular weight of the gas (assuming, for simplicity, that the dusty gas mainly contributing to the extinction curve is fully neutral).
The size distribution of grains measured by $\mathcal{D}_{i,j}=\sum_\Omega D_{i,j}/ \sum_\Omega M_{\rm g}$, over a control volume $\Omega$ that is a cylinder of $r=4\,\rm kpc$ radius and $h/2=0.2\,\rm kpc$ semi-height around the centre of the galaxy, is in good agreement with the expected MRN distribution of grains in the Milky Way, also shown in Fig.~\ref{fig:size_ext_fiducial}.
It shows that the size distribution of carbonaceous and silicate grains are extremely similar, and that the amount of carbonaceous grains is lower than that of silicate grains (as expected from the decomposition of DTM in Fig.~\ref{fig:dtm_fiducial}).

\subsection{Extinction curve}

Extinction curve is a key quantity for nearby galaxies including that of the Milky Way~\citep{pei92}.
It is well known~\citep{mathis77,draine&lee84,weingartner&draine01} that extinction curves are direct probes of the grain size distribution and of the grain chemical composition.
In the particular case of the Milky Way extinction curve, it exhibits in the optical range a pronounced bump at $\lambda=2175\,\rm \AA$ attributed to small carbonaceous grains\footnote{This population of small carbonaceous grains is likely further divided into aliphatic ($5\,\rm nm$) grains and smaller ($1-10\,\AA$) polycyclic aromatic hydrocarbon molecules~\citep[e.g.][]{leger84,weingartner&draine01,zubko04}.}, and a moderate slope in the ultraviolet, compared to that of the Magellanic Clouds, for which their slopes are steeper and the bumps are erased.
Extinction curves, or grain properties, are also central to understanding the emission of distant galaxies, shaped by their attenuation laws. Attenuation curve (not to be confused with the extinction curve) is the result of the combined radiative transfer effects with inhomogeneous stellar emission and dust distribution~\citep{inoue05,narayanan18}.
However, we will keep the investigation of attenuation curves for future work.

The extinction curve $A_{\lambda}$ is obtained from the contribution $A_{\lambda,i,j}$ from each grain population, which is given by 
\begin{equation}
    \frac{A_{\lambda,i,j}}{N_{\rm H}}=2.5\log {\rm e} \int_0^\infty \pi a^2 n^{\rm D}_{i,j}(a) Q_{{\rm ext},i}(a,\lambda){\rm d}a\, ,
\end{equation}
where $N_{\rm H}$ is the hydrogen column number density and $Q_{{\rm ext},i}(a,\lambda)$ is the extinction coefficient that is a function of grain size $a$, wavelength $\lambda$ and grain composition $i$. 
The size distribution is obtained from the log-normal distribution\footnote{We note that the extinction curve is not sensitive to the exact spread of the log-normal reconstruction for the small bin size, but is very sensitive to the one for the large bin (and in particular to that of carbonaceous grains), which should have a sharp cut-off above $a=0.1\,\mu\rm m$~\citep{weingartner&draine01}.} of equation~(\ref{eq:sizes}).
$Q_{{\rm ext},i}(a,\lambda)$ is obtained by using the Mie theory~\citep{bohren83} with the optical constants for carbonaceous and silicate grains from~\cite{weingartner&draine01}.
Bottom panel of Fig.~\ref{fig:size_ext_fiducial} shows the extinction curve $A_\lambda$ normalised by the extinction $A_{\rm V}$ in the V-band at $\lambda_{\rm V}=0.55\,\rm \mu m$, obtained for the fiducial G10LG simulation at time $t=100$, 200, 300, and $400 \,\rm Myr$ for all the dust in the same region (within a cylinder of $r=4\,\rm kpc$ and $h/2=0.2\,\rm kpc$) as used previously for measuring the dust size distribution (top panel of Fig.~\ref{fig:size_ext_fiducial}).
We also show the scatter of the extinction curve at $t=400 \,\rm Myr$ by sampling the corresponding extinction curves for each individual cells within the same region.
The extinction curve from the Milky Way analogue G10LG shows an excellent agreement with that observed for the Milky Way~\citep{pei92}: it shows a similar slope at short wavelengths and the characteristic bump feature at $\lambda=2175\,\rm \AA$ with similar strength.
This shape is clearly distinct from the extinction curves from Magellanic Clouds.
As expected, the $2175\,\rm \AA$ bump is obtained from the small carbonaceous grains (blue dotted line) due to their optical properties, while the UV-to-optical slope is the result of the small silicate grains (red dotted line).
It follows immediately that to obtain steeper slopes and a shallower $2175\,\rm \AA$ bump (as in the Magellanic Clouds extinction curves), the fraction of small carbonaceous grains must be reduced and the fraction of small silicate grains must be enhanced. 
This will be investigated further in section~\ref{section:resultslowmass}.

Diversity in the extinction curves is also a significant feature of Milky Way extinction curves with typical relative variations across lines-of-sights of $A_\lambda/A_{\rm V}$ of $\sim 20\,\%$ at $\lambda^{-1}=8\,\mu\rm m^{-1}$~\citep{cardelli89,fitzpatrick07}, as can be seen in Fig.~\ref{fig:size_ext_fiducial}.
Although, extreme care must be taken when comparing directly with observational data (line-of-sight effects), we investigate what are the different extinction curves as a function of the various gas phases of the ISM to get a better indication of what are variations of the dust properties across them.
In Fig.~\ref{fig:rel_ext_ngas}, we show the extinction curves at $t=400\,\rm Myr$ in G10LG, as relative variations of $\delta(A_\lambda/A_{\rm V})\vert_n=(A_\lambda/A_{\rm V})\vert_n/(A_\lambda/A_{\rm V})-1$ at a given $n$ with respect to the global extinction curve $A_\lambda/A_{\rm V}$ (as shown in Fig.~\ref{fig:size_ext_fiducial}), for different mean gas densities $n\sim 10^{-2},10^{-1},1,10$, and $10^{2}\,\rm H\,cm^{-3}$ 
(sampled in narrow bins of $\pm 0.1 \,\rm dex$ around the mean)
collected within the same cylinder of $r=4\,\rm kpc$ and $h/2=0.2\,\rm kpc$.
Gas at densities of $n\sim 10$ and $10^{2}\,\rm H\,cm^{-3}$ exhibits extinction curves close to the global value. 
The variation of extinction curves is not monotonic with gas density. 
The extremely low value of gas density $n\sim 10^{-2}\,\rm H \,cm^{-3}$ corresponds to the larger decrement in the extinction curves: lower bump strength and shallower UV-to-optical slope.
This is understood as the consequence of small grain destruction by SN explosions and thermal sputtering.
At intermediate densities $n\sim 10^{-1}$ and $1\,\rm H\,cm^{-3}$, the extinction curves approach their global values and stand above the global expectation at $1\,\rm H\,cm^{-3}$ as a signature of shattering.
Shattering transfers mass from large grains to small grains preferentially in the diffuse ISM, reinforcing both the bump strength and the UV slope.
The highest gas densities are the closest to the global value as this is where most of the dust mass is concentrated ($50\,\%$ of the total dust mass at $n\sim10\,\rm H\,cm^{-3}$, and $30\,\%$ at $n\sim10^2\,\rm H\,cm^{-3}$).
They have lower extinction curves relative to the lower ISM gas density of $n\sim 1 \,\rm H\, cm^{-3}$ due to the efficient coagulation of small grains into large grains.
Finally the different sampled densities all have large scatter of the order of the mean values, which is indicative of the intrinsic diversity of extinction curves at a given gas density.

\begin{figure}
\centering \includegraphics[width=0.45\textwidth]{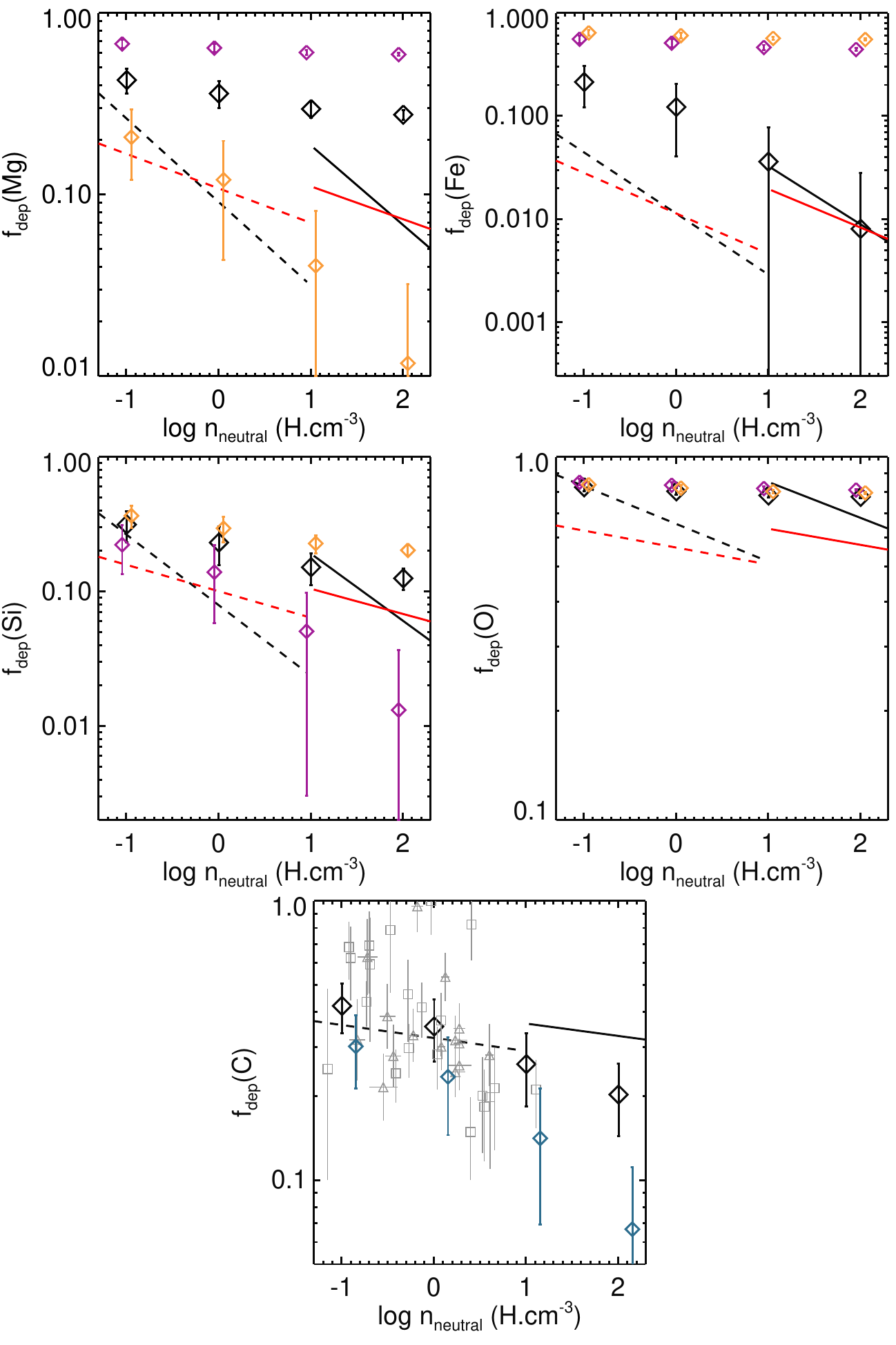}
\caption{Depletion factors as a function of gas density for the G10LG simulation at $400\,\rm Myr$ (black diamonds with standard deviation) compared to the results from~\cite{jenkins09} (black dashed lines) and from~\cite{decia16} (red dashed lines). The solid lines are the rescaled observational fits following~\cite{zhukovska16}. The purple and orange diamonds are the depletion factors for the G10LG simulation using a pyroxene compound (MgFeSi$_2$O$_6$), and an iron-poor olivine compound (Mg$_{1.5}$Fe$_{0.5}$SiO$_4$), respectively, instead of the default olivine compound (MgFeSiO$_4$) for silicates. The blue diamonds stand for the G10LG simulation without CO formation. For C depletion, we also show the data points from~\cite{jenkins09} (triangles) and ~\cite{parvathi12} (squares).}
\label{fig:depletion}
\end{figure}

\subsection{Depletion factors}
\label{section:depletion}

The presence of dust in the ISM can be characterised by the depletion of the corresponding atomic elements in the gas phase $f_{\rm dep}(X)=1-M_{\rm D}(X)/M_{\rm Z}(X)$, where $M_{\rm D}(X)$ and $M_{\rm Z}(X)$ are respectively the dust mass and the total (dust+gas) mass of the element $X$.
We show in Fig.~\ref{fig:depletion} the depletion factor of the various elements as a function of the neutral gas density $n_{\rm neutral}$ (assuming that gas with temperature below $T\le 10^4\,\rm K$ is neutral and fully ionised otherwise) compared to observations.
The observational relations are shown with the raw value of~\cite{jenkins09} as the black dashed lines, and with the recommended renormalisation of their data in black solid line by~\citealp{zhukovska16} (a correction also adopted in~\citealp{choban22}) that compensates for the fact that the observed densities are under-estimated since they are averaged along the line-of-sight.
We also show the fit provided by~\cite{richings22} of the data from~\cite{decia16}, which extends the~\cite{jenkins09} data to low density lines-of-sight (with or without the rescaling in solid and dashed red lines respectively).
For C depletion, we followed~\cite{choban22}, and we reduced the depletion values of~\cite{jenkins09} by a factor 2 (in that case, we did not rescale in density because the relation is sufficiently shallow) to account for the apparent over-estimate of the C gas-phase abundance from weak compared to strong CII lines~\citep{sofia11,parvathi12}.
We also show the data points from~\cite{jenkins09} (rescaled by a factor of 2) and the data points from~\cite{parvathi12} to underline the large uncertainties associated to the estimate of the depletion of C.

As expected, there is more depletion of individual elements in the gas phase ($f_{\rm dep}$ decreases) with increasing density as a result of larger accretion in the dense gas and of destroyed grains in the diffuse phase.
Given the large uncertainties in the reconstruction of the observed gas densities, our depletion values are in broad agreement with the data, except for the Mg which clearly stands out of the data, while the other compounds of silicate grains (Fe, Si, and O) are in the observational ballpark.
Iron is the most depleted of the four elements entering the composition of silicates, and, thus, is the limiting element in the accretion for the formation of olivine with that type of iron inclusions (i.e. MgFeSiO$_4$).
Depletion factors of the silicate-bearing elements are naturally extremely sensitive to the exact composition of the silicate.
We performed additional simulations where we have assumed a different silicate compound (MgFeSi$_2$O$_6$ or Mg$_{1.5}$Fe$_{0.5}$SiO$_4$).
In the pyroxene compound MgFeSi$_2$O$_6$, the mass fraction of Si is $50\, \%$ larger than in the olivine and Fe is $35\,\%$ lower, and, thus, Si becomes the limiting element.
Indeed, Fig.~\ref{fig:depletion} shows that $f_{\rm dep}({\rm Si})$ is lower for a G10LG simulation that is performed assuming pyroxene compound instead of olivine, and that Si, then, becomes the lowest of the silicate-bearing elements.
Similarly, decreasing the amount of iron inclusions in olivine (or pyroxene) changes the picture:
a G10LG simulation assuming iron-poor olivine Mg$_{1.5}$Fe$_{0.5}$SiO$_4$ instead of iron-rich olivine MgFeSiO$_4$ leads to lower values of $f_{\rm dep}({\rm Mg})$ that makes Mg the new limiting element of silicate grains.

For C depletion, our resulting values are well within the range of observations although the observational data exhibits a large scatter. 
We also highlight the role of CO formation at high gas densities in limiting the depletion of C elements by grains~\citep[see also][]{choban22}.
If we do not account for CO formation in the limitation of carbonaceous dust growth from ISM accretion, i.e. as for silicate grains, carbonaceous grain growth is only limited by ice mantle coating starting above $n=10^4\,\rm H\,cm^{-3}$ (G10LG\_NCO simulation), then carbonaceous grains accrete much more efficiently from the ISM, leading to too much depletion of C. 
Therefore, CO formation indirectly allows for more moderate values of $f_{\rm dep}({\rm C})$ in better agreement with the data.

\subsection{Which process dominates the dust growth?}
\label{section:dominantdustmechanism}

\begin{figure}
\centering \includegraphics[width=0.45\textwidth]{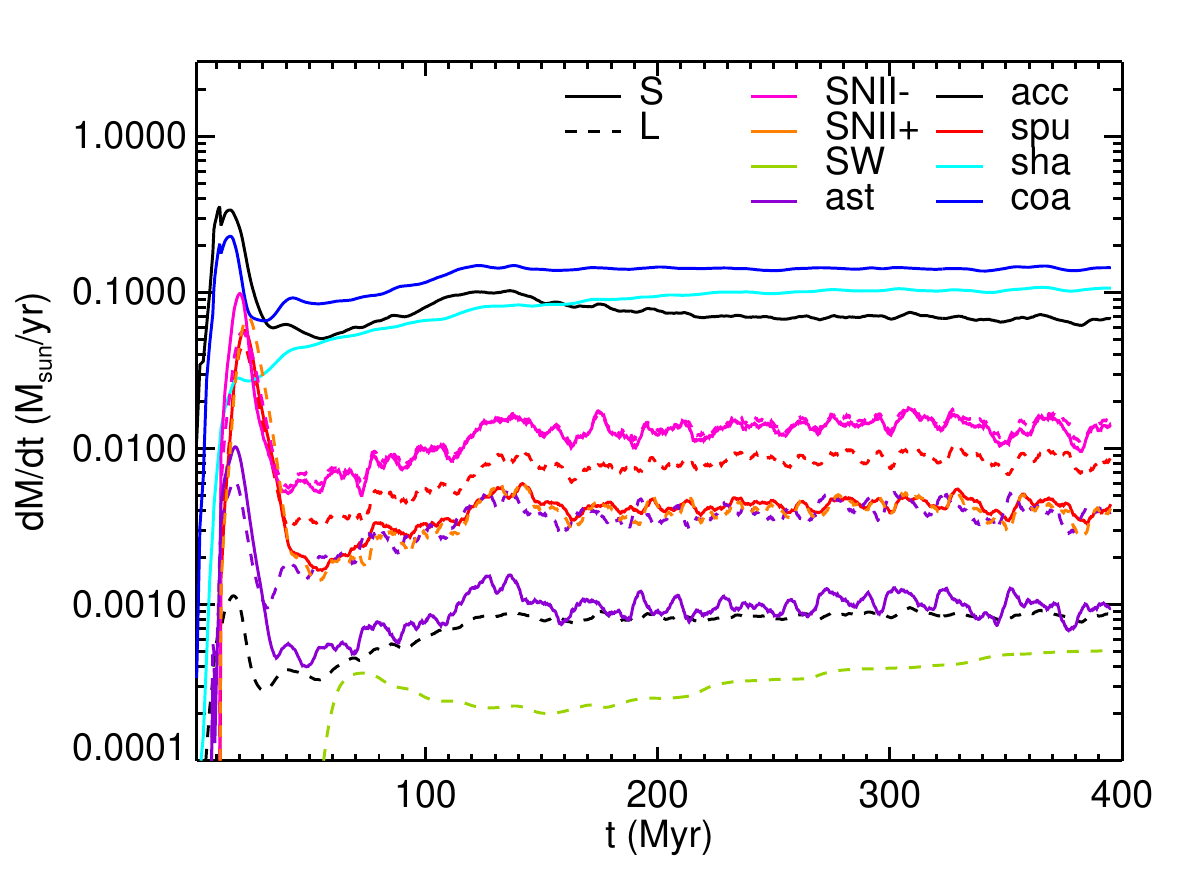}
\caption{Dust mass variation rates as a function of time in the fiducial G10LG simulation for various dust processes: dust growth by gas accretion (`acc', black), dust destruction by thermal sputtering (`spu', red), by supernovae (`SNII$-$', pink), or by astration (`ast', purple), dust released by supernovae (`SNII$+$', orange) or stellar winds (`SW', green), and dust mass transfer between small and large grains by coagulation (`coa', blue) and vice versa by shattering (`sha', cyan).}
\label{fig:dmdt_fiducial}
\end{figure}

\begin{figure*}
\centering \includegraphics[width=0.35\textwidth]{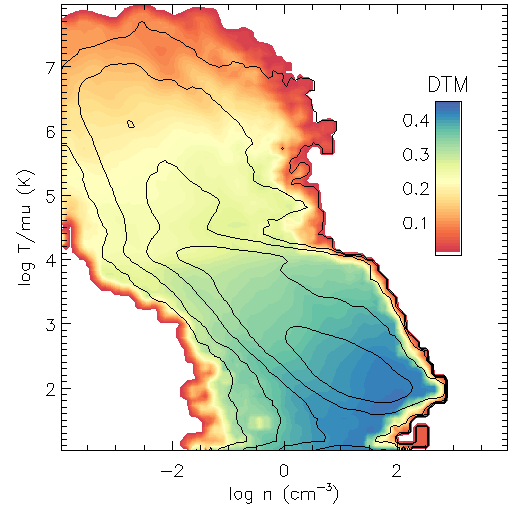}
\centering \includegraphics[width=0.35\textwidth]{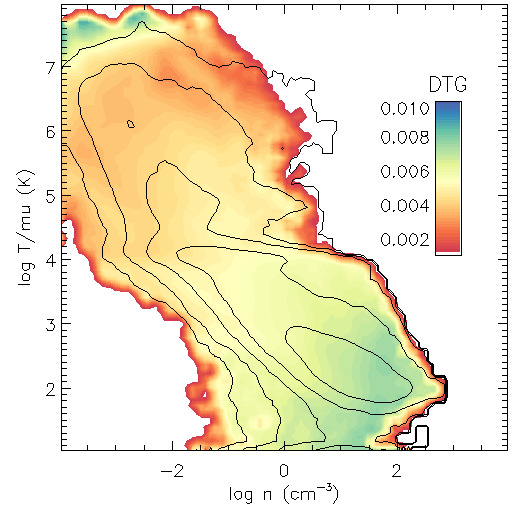}\\
\centering \includegraphics[width=0.35\textwidth]{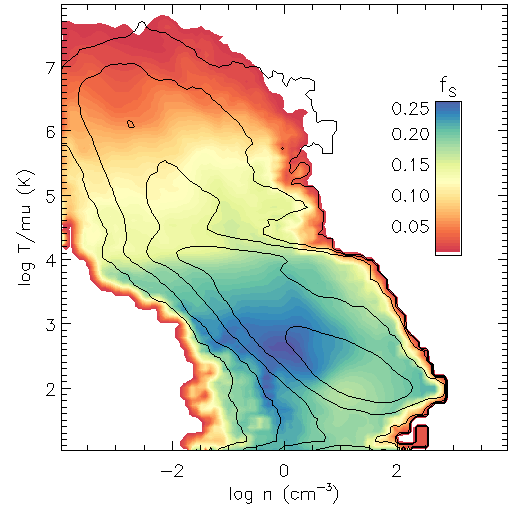}
\centering \includegraphics[width=0.35\textwidth]{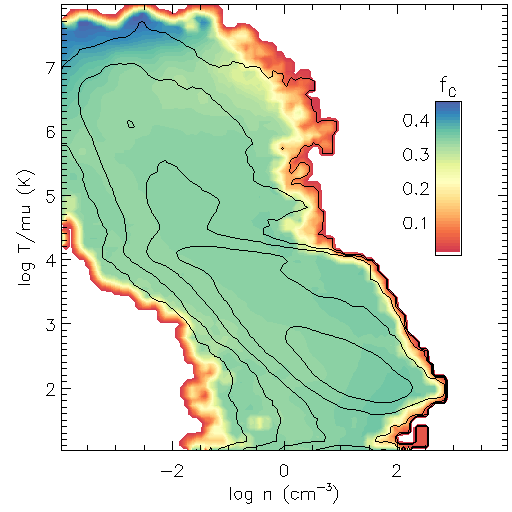}
\caption{Dust-to-metal ratio (DTM, top left), dust-to-gas ratio (DTG, top right), fraction of small grains $f_{\rm S}$ (bottom left), and fraction of carbonaceous grains $f_{\rm C}$ (bottom right) as a function of density $n$ and temperature $T$ for the G10LG galaxy at time $t=400\,\rm Myr$. Black contours are for the mass distribution of gas.}
\label{fig:histo_dust_g10lg}
\end{figure*}

\begin{figure}
\centering \includegraphics[width=0.45\textwidth]{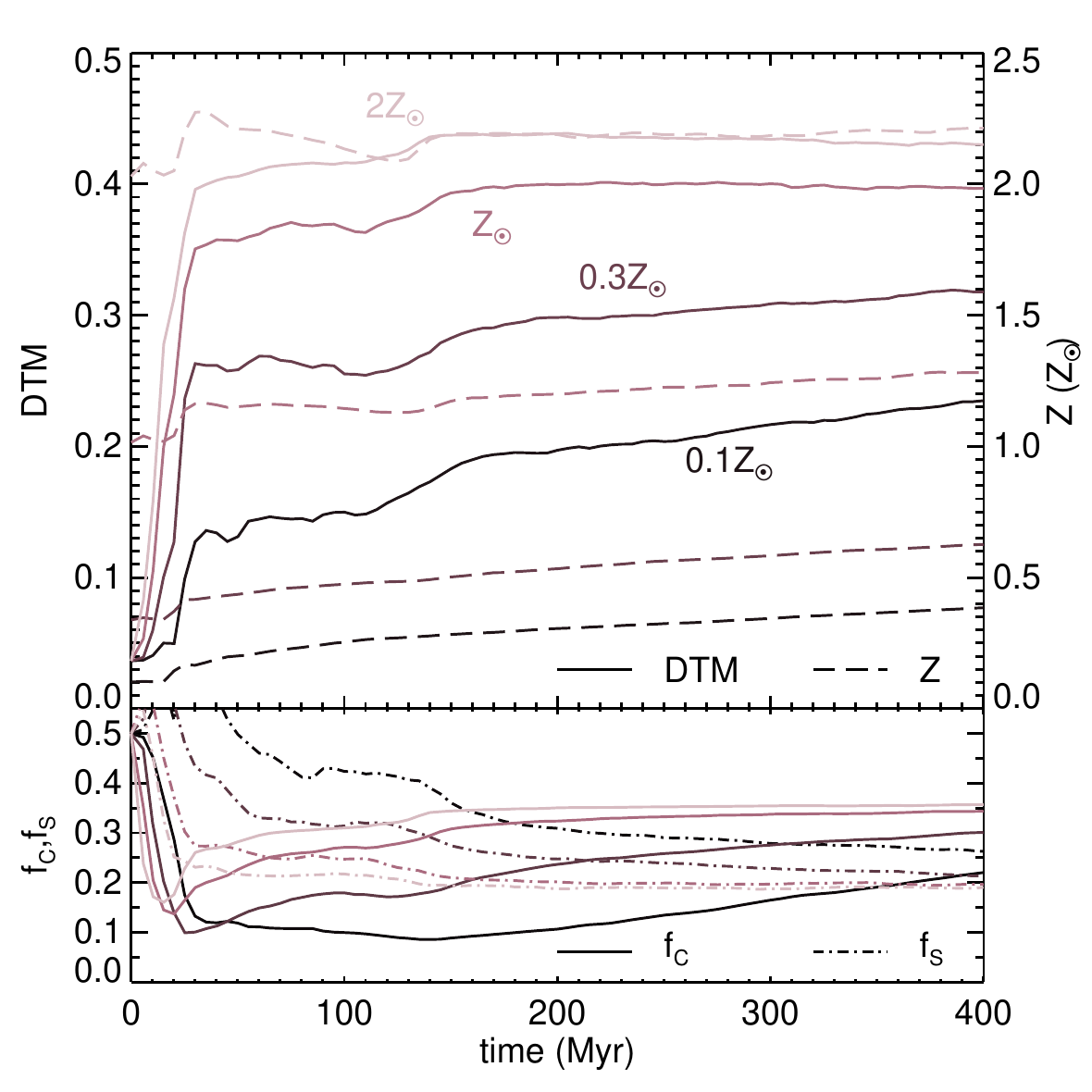}
\caption{Top panel: dust-to-metal ratio (solid lines, left axis) and gas metallicity (dashed lines, right axis) as a function of time for the 4 different initial metallicities of the G10LG galaxy: $0.1$, $0.3$, $1$, and $2\,\rm Z_\odot$ (from bottom to top).
Bottom panel: fraction of carbonaceous grains (solid lines) and of small grains (dot-dashed lines).
The DTM increases with metallicity, the fraction of carbonaceous grains increases with metallicity and the fraction of small grains decreases with metallicity.}
\label{fig:dtm_metcomp}
\end{figure}

\begin{figure}
\centering \includegraphics[width=0.45\textwidth]{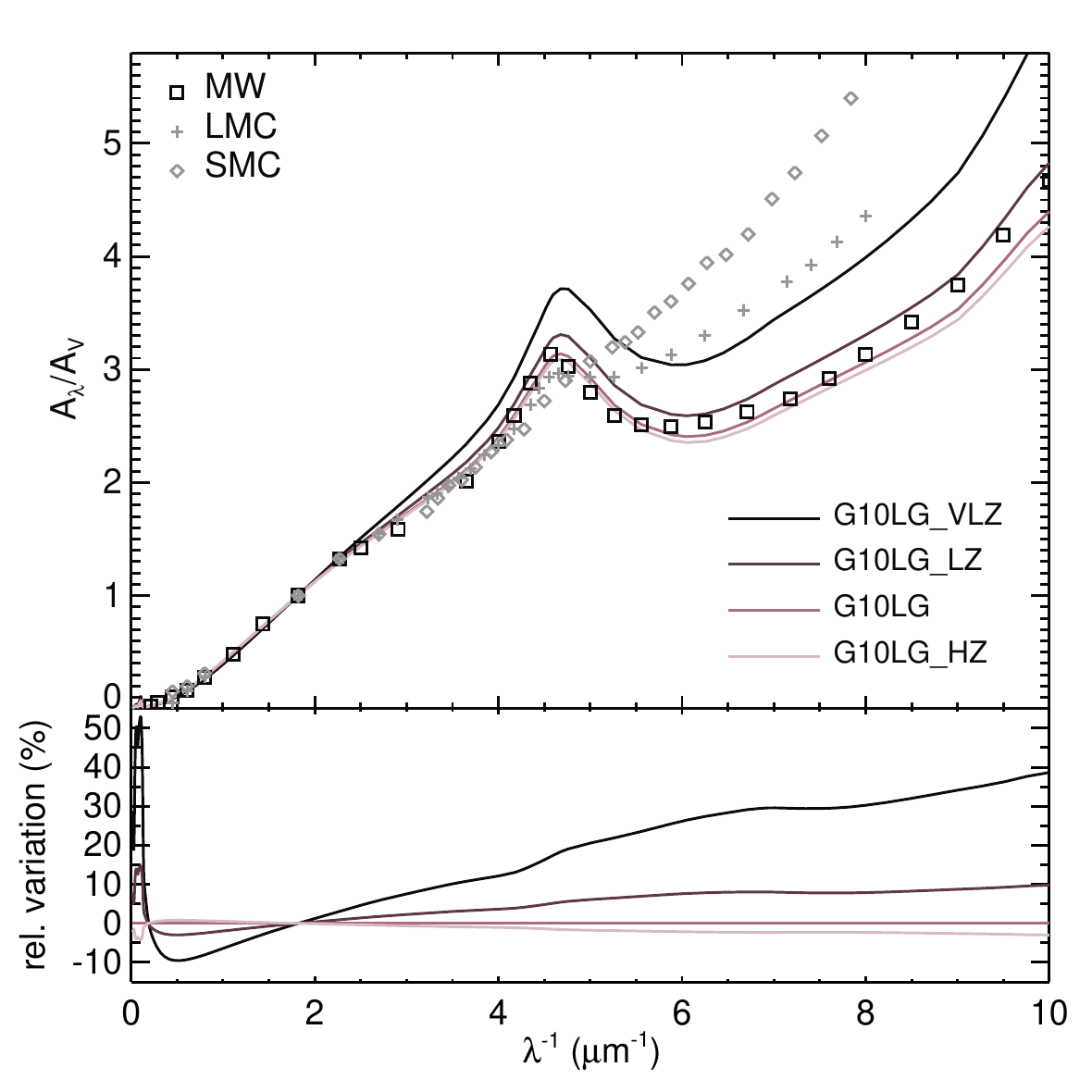}
\caption{Comparison of extinction curves in the G10LG galaxy at time $t=400 \,\rm Myr$ for different initial metallicities ($Z_{\rm g,0}=0.1$, $0.3$, $1$, and $2\,\rm Z_\odot$ for G10LG\_VLZ, G10LG\_LZ, G10LG, and G10LG\_HZ respectively). The top panel shows the extinction curves and the bottom panel shows the relative variation of the extinction curve with respect to the fiducial simulation G10LG. Galaxies with lower metallicities produce steeper UV-to-optical slopes due lower accretion efficiencies in the dense gas where coagulation proceeds, hence, the fraction of small grains is larger.}
\label{fig:ext_curve_metcomp}
\end{figure}

\begin{figure}
\centering \includegraphics[width=0.45\textwidth]{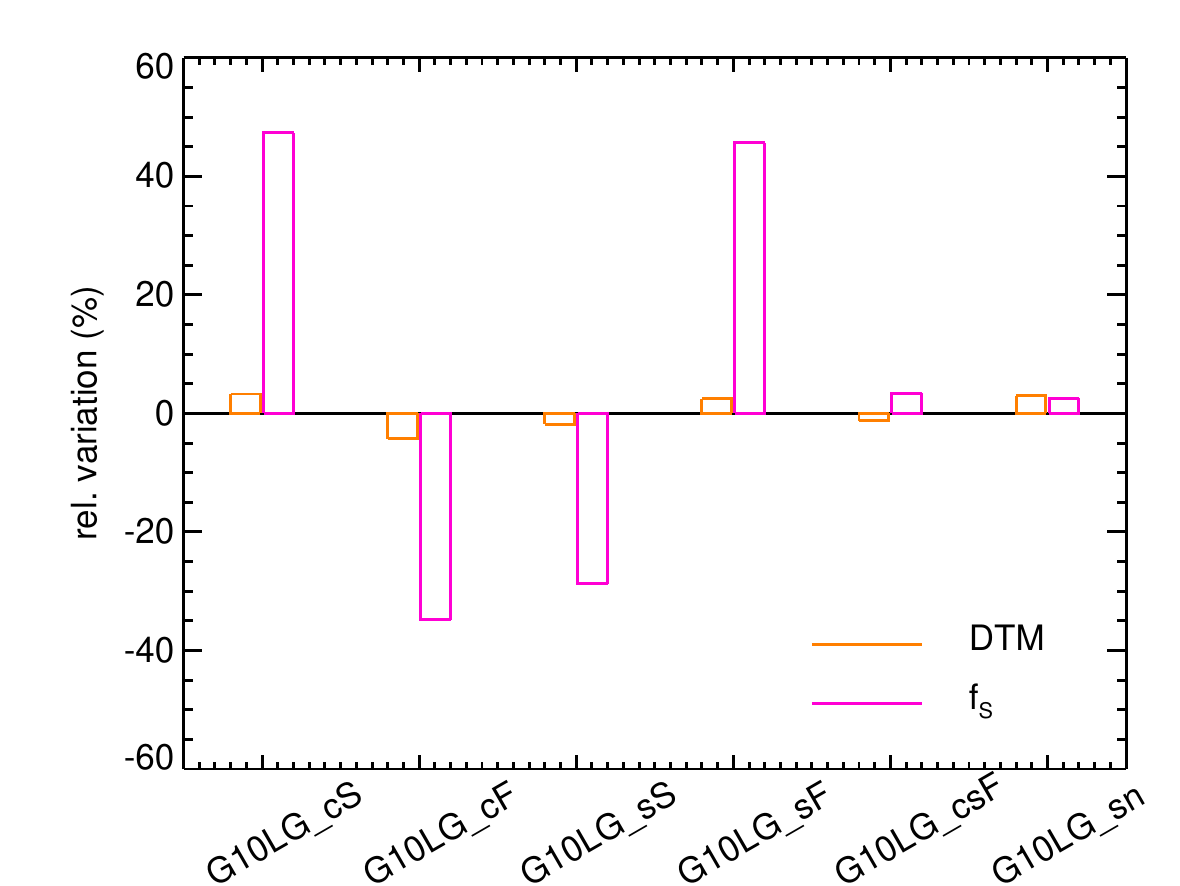}
\caption{Relative variations of the DTM (orange) and of the fraction of small grains $f_{\rm s}$ (magenta) for the given run as labelled on the x-axis with respect to the G10LG simulation, decreasing or increasing the coagulation rate (resp. G10LG\_cS and G10LG\_cF), decreasing or increasing the shattering rate (resp. G10LG\_sS and G10LG\_sF), increasing both the coagulation and the shattering rates (G10LG\_csF), and reducing the SN dust destruction rates (G10LG\_sn). Quantities are averaged from 300 to 400 Myr. Variations of the rates of these processes have little impact on the DTM, however, variations of coagulation and shattering rates can significantly affect the fraction of small grains.}
\label{fig:rel_variation_coa_sha_sn}
\end{figure}

\begin{figure}
\centering \includegraphics[width=0.45\textwidth]{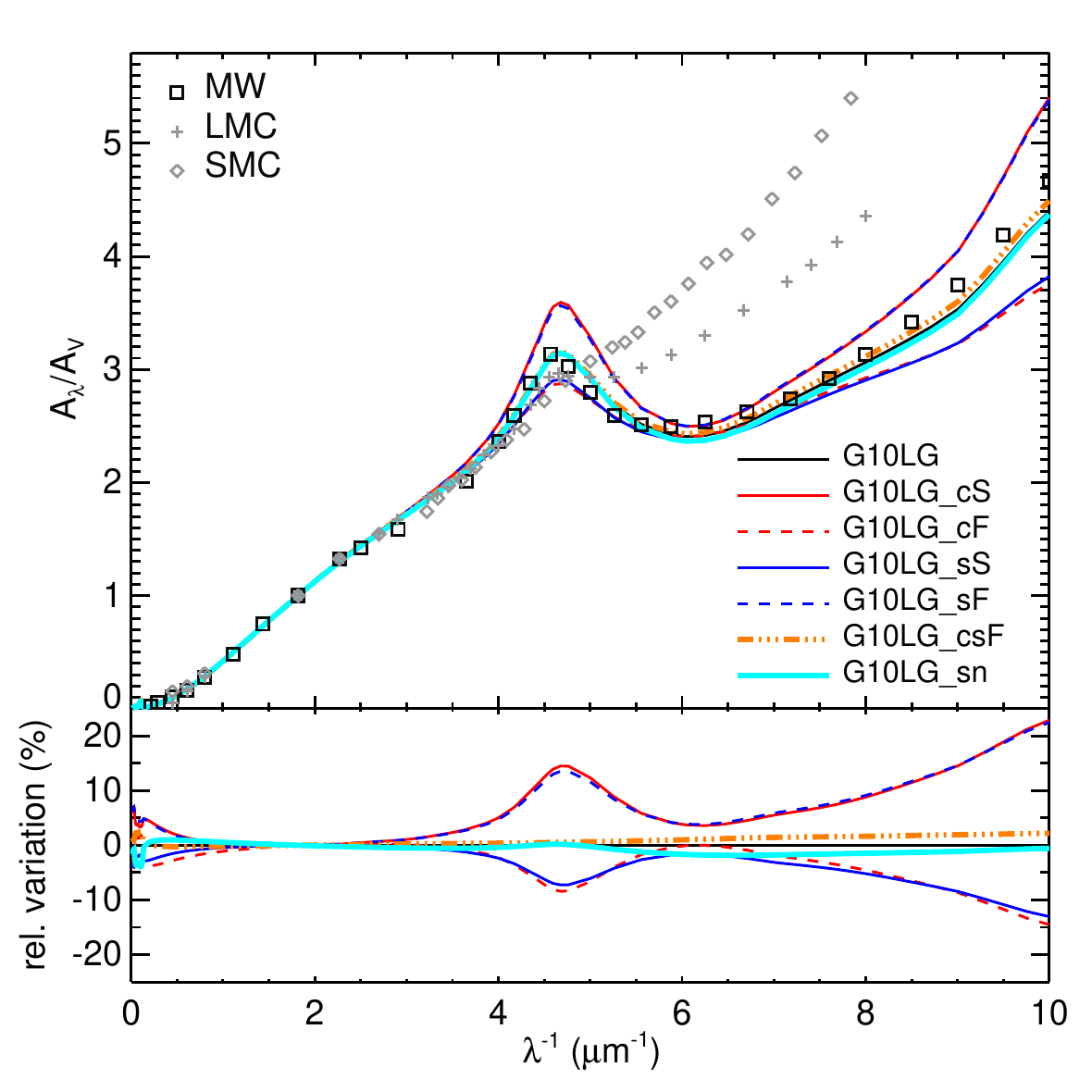}
\caption{Comparison of extinction curves in the G10LG galaxy at time $t=400 \,\rm Myr$ for variations in coagulation (red), shattering rates (blue), and in SN dust destruction efficiency for small grains (cyan). An increase (decrease) in the corresponding rate is depicted by a dashed (respectively solid) line. The triple-dot-dashed orange line stands for an increased in both the shattering and the coagulation rates. Top panel shows the extinction curves and the bottom panel shows the relative variation of the extinction curve with respect to the fiducial simulation.}
\label{fig:ext_curve_coa_sha_sn}
\end{figure}

\begin{figure}
\centering \includegraphics[width=0.45\textwidth]{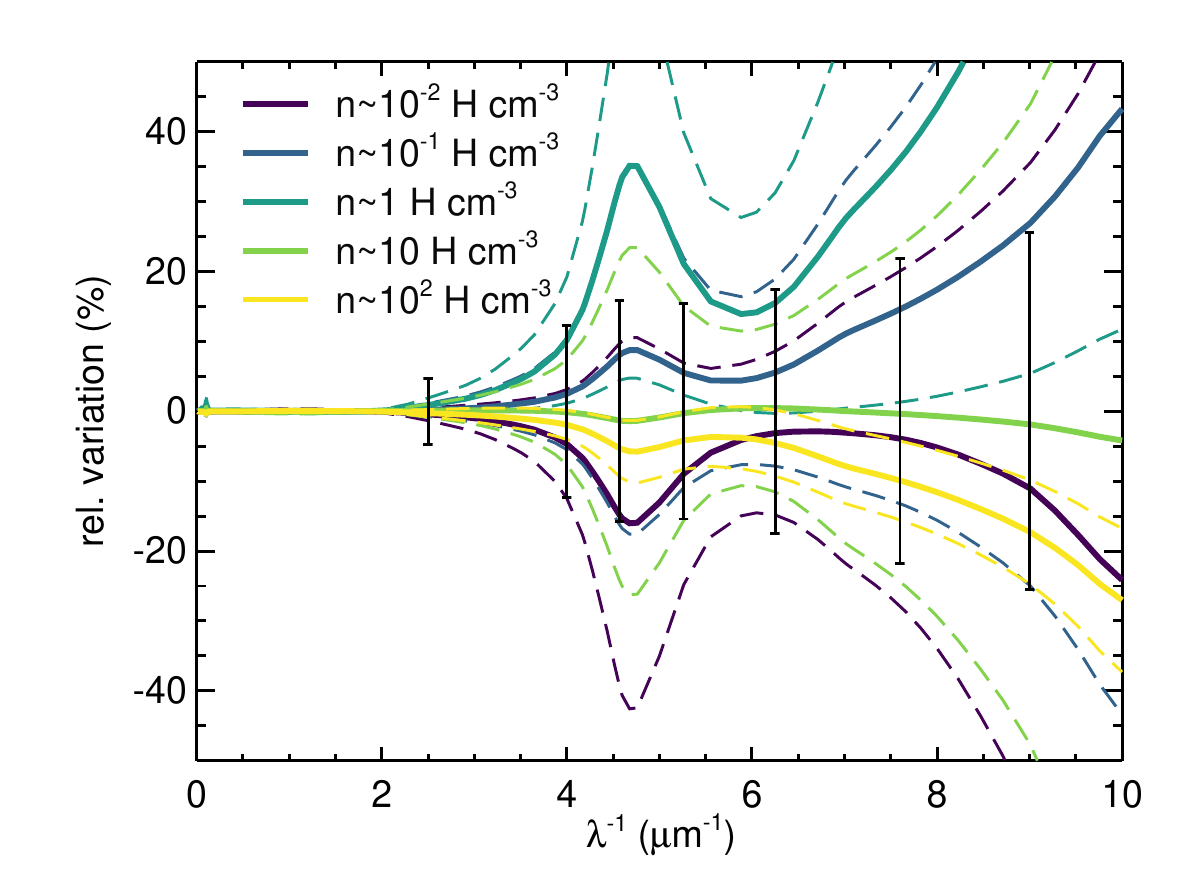}
\caption{As Fig.~\ref{fig:rel_ext_ngas} for the G10LG\_csf simulation that has faster coagulation and shattering rates with respect to the fiducial G10LG simulation. There is more diversity in extinction curves across gas density compared to the fiducial run. The error bars stand for the scatter of the observed extinction curves of the Milky Way from~\cite{fitzpatrick07} relative to the mean extinction curve from~\cite{pei92}.}
\label{fig:rel_ext_ngas_csf}
\end{figure}

The evolution of the dust mass in the simulated galaxies obeys different growth and destruction processes.
Figure~\ref{fig:dmdt_fiducial} shows the different dust mass variation rates for different processes involved in the dust evolution of the fiducial G10LG run: accretion, thermal sputtering, SN shock destruction, dust released by stars through stellar winds in SN ejecta, astration, coagulation and shattering.
All rates are obtained within the whole simulation box, and curves are smoothed over a 5 Myr timescale for sake of readability.
Apart from the first initial peak at a few $10\,\rm  Myr$ after the start of the simulation, all curves show little variation with time due to the low SFR and no significant evolution in gas-phase metallicity (which could affect the accretion time since $t_{\rm acc}\propto Z^{-1}$).
Accretion of the available refractory elements from the ISM onto the small grains is by far the dominant process of dust mass growth compared to dust released in the ejecta of SNe or of stellar winds~\citep[as expected for sufficiently metal-rich galaxies, e.g.][]{popping17,aoyama17,vijayan19,choban22}, or by accretion onto large grains.
Destruction by SNe is a factor of 3 below the mass growth by accretion and is the dominant mechanism for dust mass removal.
Thermal sputtering, which is the mechanism by which the dust is destroyed by pockets of hot ($T>10^6\,\rm K$) and diffuse ($n<0.01\,\rm H\,cm^{-3}$) gas, is a factor 2 below the destruction rate by SNe.
Finally destruction of the dust by astration is lower by a factor 4 compared to destruction by SNe.
Interestingly, the thermal sputtering removes more mass on large grains than on small grains, and SNII destruction removes an equal mass of small and large grains, while their physical timescale for dust destruction is an increasing function of the grain size.
Hence, one would naively expect the sputtering to remove more dust mass from small grains.
However, since the dust mass is 4 times larger for large grains than small grains (see Fig.~\ref{fig:dtm_fiducial}), this effect is compensated by the available reservoir of dust mass in each bin of grain size.

The coagulation rate is slightly larger than the shattering rate and they both transfer more mass between their two respective dust grain sizes than accretion from the gas phase to the dust phase. 
Therefore, coagulation and shattering, and their exact balance, play the critical role in the dust size distribution rather than accretion balanced by destruction effects.

At a solar abundance, for which Fe is the limiting element to the growth by accretion of the silicate with olivine compound, the typical accretion times for silicate  and carbonaceous grains for gas with $n=30 \,\rm H\,cm^{-3}$, $\mathcal{M}=3$ (corresponding to the average mass-weighted gas density and turbulent Mach number of the gas in G10LG) are $t_{\rm acc,Sil}\simeq 1.8\,\rm Myr$ and $t_{\rm acc,C}\simeq 9.5\,\rm Myr$.
Compared with the typical free-fall time at this density, that is $t_{\rm ff}\simeq 8.2 \,\rm Myr$ as well as the time for the first SNe to explode ($5\,\rm Myr$), it shows that both silicate and carbonaceous grains can grow efficiently by accretion, and why there are larger amount of silicate grains with respect to carbonaceous grains ($t_{\rm acc,Sil}<t_{\rm acc,C}$).

\subsection{Dust grain distribution in the multiphase interstellar medium}

As shown in Fig.~\ref{fig:illustrative_example_g10lg}, the dust distribution shows a significant structure in the ISM.
We investigate further how dust distributes in the ISM by measuring the DTM, the fraction of small grains and the fraction of carbonaceous grains as a function of the gas density and temperature.
This is shown in Fig.~\ref{fig:histo_dust_g10lg} together with the distribution of gas mass (black contours).

As could naively be expected by the scaling of the dust accretion time with $n$ and $T^{-1/2}$, the DTM and the DTGs increase with increasing density and decreasing temperature, with the largest values corresponding to densities of $n\simeq 100 \,\rm cm^{-3}$ and temperatures of $100\,\rm K$.
We note that the quoted densities are the raw gas density values extracted from the simulation although dust accretion proceeds from unresolved gas densities at larger values driven by turbulence. It is interesting to see that appreciable levels of dust even in the hot component $T>10^4\,\rm K$ and in particular in the galactic wind component at $T\sim 10^6 \,\rm K$ stays at DTM of $\sim 0.1$.

The size distribution of grains shows a more complex structure.
The fraction of small grains $f_{\rm S}$ is the largest in the cold neutral medium at $n\simeq 1\,\rm cm^{-3}$ and $T\simeq$ a few $100\,\rm K$.
This corresponds to densities and temperatures where shattering is efficient at reducing the size of large grains, originated by coagulation in the densest star-forming regions.
The fraction of small grains decreases in denser and colder regions as the consequence of efficient coagulation of grains in dense gas.
Furthermore, the fraction of small grains diminishes as temperature increases from approximately $1\, 000\,\rm K$.
This reduction occurs because of the enhanced efficiency of thermal sputtering of grains and the fact that high temperatures are remnants of regions propelled by SNe, which obliterate small grains.

Finally, the fraction of carbon $f_{\rm C}$ in Fig.~\ref{fig:histo_dust_g10lg} shows very little variation in the gas-temperature diagram as it is typically of $\sim 10\,\%$, except at the largest temperatures ($T\gtrsim 10^7\,\rm K$), where $f_{\rm C}$ increases more significantly.
This reflects the more efficient sputtering yields of silicate grains with respect to carbonaceous grains.

\subsection{Effect of the metallicity}
\label{section:metallicity}

We now explore how changing the metallicity of the gas affects the dust mass content in the galaxy.
Figure~\ref{fig:dtm_metcomp} (top panel) shows the evolution of the DTM in the G10LG galaxy with different initial gas metallicities, sampling from $0.1, 0.3, 1$ and $2\,\rm Z_\odot$.
There is no significant evolution of the gas metallicity for the two highest metallicity bins, while there is an increase to, respectively, 0.3 and 0.5 $\rm Z_\odot$ for the two lowest metallicity bins.
The DTM shows a similar behaviour in all cases: a steep rise in 50-100 Myr that saturates to a steady state value that is larger for larger values of gas metallicity, which is in good agreement with observations~\citep{remy14,devis19} as shown in section~\ref{section:dtmallgal}.
This behaviour is naturally obtained by the scaling of dust accretion time with the inverse of the metallicity.
However, by comparing the final value of the DTM with the final value of the gas phase metallicity, it is immediate to see that the relation of DTM with $Z$ is not linear.
The reason stems in the saturation of the depletion of refractory elements when metallicity is increasing.
In particular, silicates are closer to saturation  earlier on with respect to carbonaceous grains: there is extreme depletion of iron elements at solar metallicities, which is the limiting element of silicates for olivine, while there are still significant amounts of C in the gas phase (see Fig.~\ref{fig:depletion}).
Consequently, the fraction of carbonaceous grains (bottom panel of Fig~\ref{fig:dtm_metcomp}) increases with metallicity.

We compare in Fig.~\ref{fig:ext_curve_metcomp} the resulting extinction curves for the different metallicities at $t=400\,\rm Myr$.
Simulations with lower metallicities have extinction curves with a steeper UV-to-optical slope.
With lower metallicity of the gas, accretion onto grains is less efficient given that $t_{\rm acc}\propto Z^{-1}$.
As a result, there is less dust in the dense phase, which also reduces the dust coagulation efficiency as it is directly proportional to the dust density of small grains.
Therefore, the dust size distribution becomes further biased towards small grains (see bottom panel of Fig~\ref{fig:dtm_metcomp}).
Combined to the lower fraction of carbonaceous grains, low metallicity leads to an enhancement of the UV part of the extinction curve since there is an increasing fraction of small silicate grains (see in Fig.~\ref{fig:size_ext_fiducial} how small silicate grains contribute to the extinction curve in the UV part).

\subsection{Changing the coagulation and shattering rates}

Our model for grain size evolution relies on several aspects that we explore further: coagulation in dense star forming gas and shattering in diffuse medium, which control the shape of the extinction curve by establishing the balance between small and large grains.
To explore the effect of coagulation and shattering, we vary their baseline velocity dispersion -- or timescale -- by a factor of 3. 
Figure~\ref{fig:rel_variation_coa_sha_sn} shows the relative variation of the DTM and of the fraction of small grains with respect to the values from the fiducial G10LG simulation from their values averaged between time $t=350$ and $400\,\rm Myr$.
As expected, faster (resp. slower) coagulation rates lead to less (resp. more) small grains, and vice versa for the shattering rates.
The relative variations in DTM are less sensitive to variation in coagulation of shattering rates compared to variations in the fraction of small grains.

Extinction curves could eventually be used to discuss the validity of the different coagulation and shattering models.
Since the coagulation and shattering rates change the balance between the amount of small and large grains, the extinction curve should change its UV-to-optical slope and its bump strength.
Figure~\ref{fig:ext_curve_coa_sha_sn} shows the extinction curves for these simulations compared to the fiducial model for coagulation and shattering.
As expected the simulations that produce larger fractions of small grains also exhibit a reinforced bump and a steeper UV-to-optical slope (simulations G10LG\_cF and G10LG\_sS), that seems to be significantly out of the characteristic Milky Way extinction curve.
Simulations with larger fractions of large grains (simulations G10LG\_cS and G10LG\_sF) have a shallower UV-to-optical slope and a weaker bump feature. 
Although, for these two cases, the bump signature is acceptable compared to the Milky Way signal, the extinction in the UV seems slightly too low.

Finally, we investigate the run with both faster coagulation and shattering rates (with an increase by a factor of 3 in their rates). Figures~\ref{fig:rel_variation_coa_sha_sn} and~\ref{fig:ext_curve_coa_sha_sn} show that the DTM, size distributions and extinction curves remain largely unaffected by this simultaneous change in coagulation and shattering rates.
As can be seen in Fig.~\ref{fig:rel_ext_ngas_csf}, the increased coagulation and shattering rates produce larger variations in the extinction curves as a function of density compared to our fiducial model (compare with Fig.~\ref{fig:rel_ext_ngas}), which is reflective of the larger variety in the grain size distribution in the ISM.
In particular, it is in the diffuse ISM ($n\lesssim 1\,\rm H\, cm^{-3}$) that the difference in the fraction of small grains is the largest.
Indeed, at $n= 1\,\rm H\, cm^{-3}$, the free fall time is $t_{\rm ff}\simeq 60\,\rm Myr$, which is comparable to the canonical value of the shattering time (see equation~\ref{eq:sha_time}) adopted for the fiducial run $t_{\rm sha}\simeq50\,\rm Myr$.
Increasing the shattering rates by a factor of 3, therefore, allows the diffuse ISM to efficiently shatter the big grains produced in dense clouds before this diffuse phase collapses into dense star-forming clouds.

\subsection{Changing the dust destruction efficiency by supernovae}

Since SN explosions are the leading destruction mechanism for dust (compared to thermal sputtering or astration), it is valuable to investigate how changing the SN dust destruction efficiency $\varepsilon_{\rm SN}$ affects the predicted dust distribution in the simulated galaxy.
We recall that the SN dust destruction efficiency is stronger for smaller grains (see Section~\ref{section:SNdust}), and is $\varepsilon_{\rm SN,Si}(5{\rm nm})\simeq 0.95$ and $\varepsilon_{\rm SN,C}(5{\rm nm})\simeq0.86$ for respectively small silicate and carbonaceous grains and $\varepsilon_{\rm SN,Si}(0.1\mu{\rm m})\simeq0.14$ and $\varepsilon_{\rm SN,C}(0.1\mu{\rm m})\simeq0.095$ for respectively large silicate and carbonaceous grains.
We now impose the same value for the destruction efficiency of small and large grains by reducing the SN dust destruction efficiency of small grains to that of the large grains (simulation `G10LG\_sn').

\begin{figure*}
\centering \includegraphics[width=0.90\textwidth]{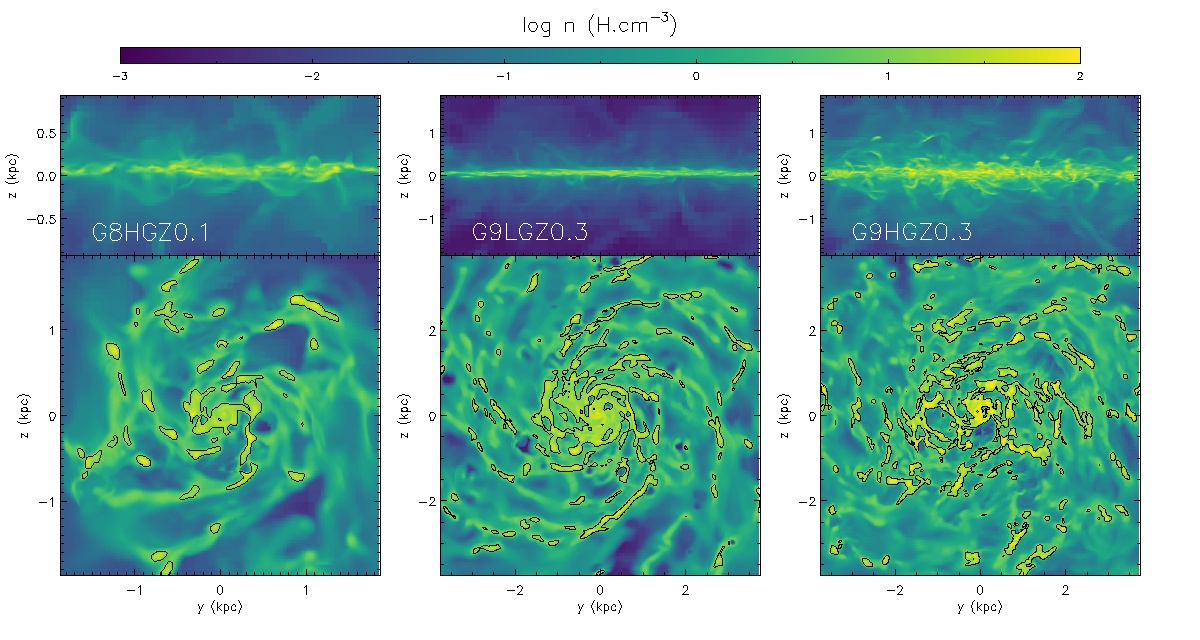}\vspace{-0.25cm}\\
\centering \includegraphics[width=0.90\textwidth]{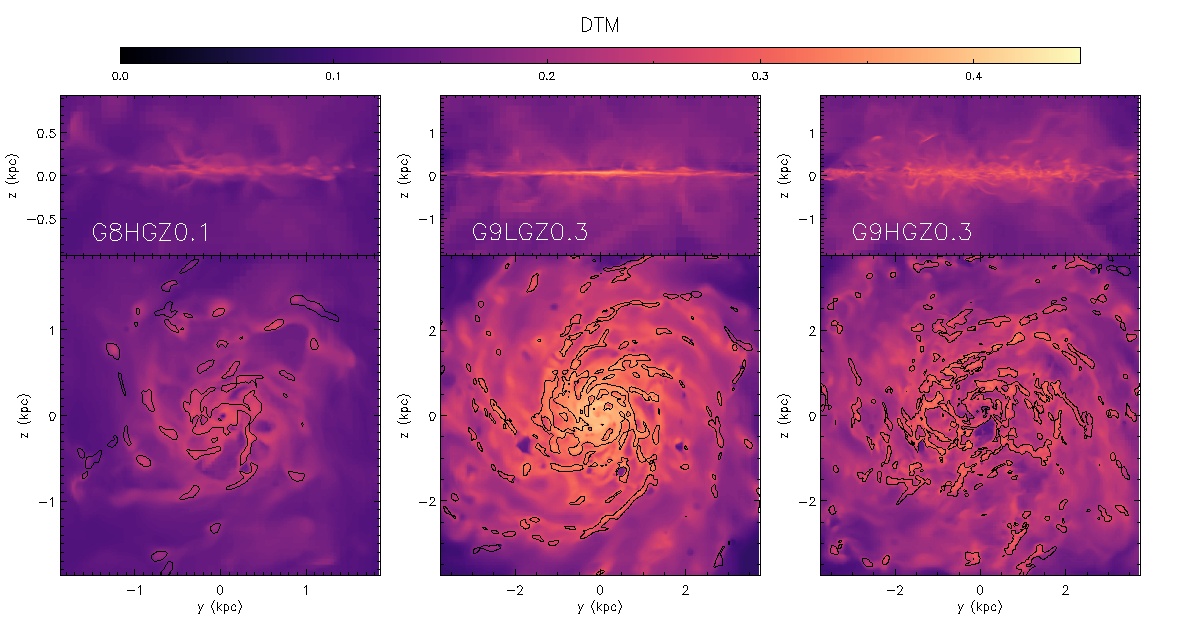}
\caption{Density-weighted projections of the gas density (top two rows), and of the dust-to-metal ratio (bottom two rows) viewed edge-on (first and third rows) or face-on (second and fourth rows) for the G8HG, G9HG and G10LG galaxies from left to right columns. Overplotted black contours encapsulate regions of gas density larger than $1\,\rm H\,cm^{-3}$.}
\label{fig:illustrative_examples}
\end{figure*}

Reducing the SN destruction efficiency for small grains by an order of magnitude still has a limited impact (\% increase compared to G10LG) on both the final DTM and the fraction of small grains as can be seen in Fig.~\ref{fig:rel_variation_coa_sha_sn}.
This is the result of having an efficient accretion of refractory elements on grains that is mostly limited by the depletion of gas.
Consequently, the corresponding extinction curve is not markedly different to that from the fiducial G10LG simulation, with relative variations of less than 5\% (see Fig.~\ref{fig:ext_curve_coa_sha_sn}).

\section{Low mass galaxies}
\label{section:resultslowmass}
\begin{figure}
\centering \includegraphics[width=0.5\textwidth]{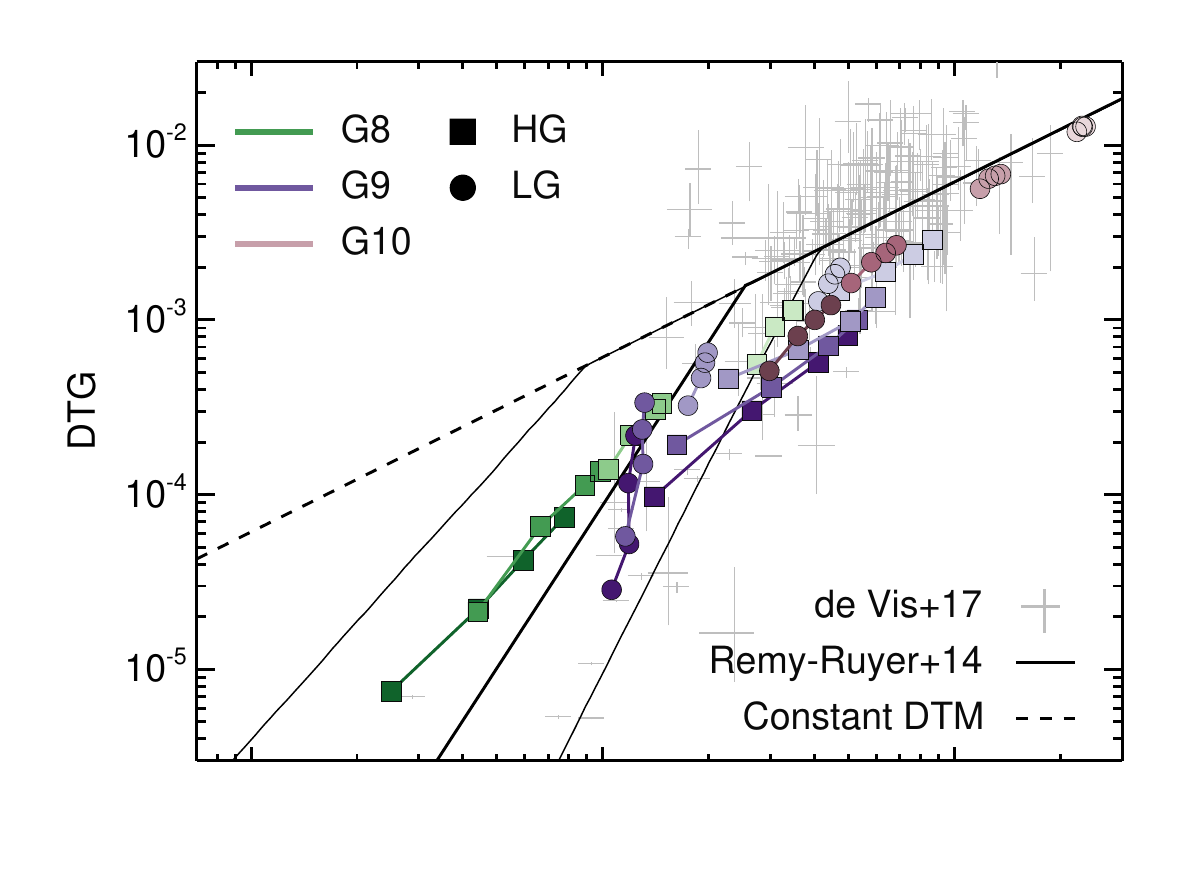}\vspace{-1.5cm}
\centering \includegraphics[width=0.5\textwidth]{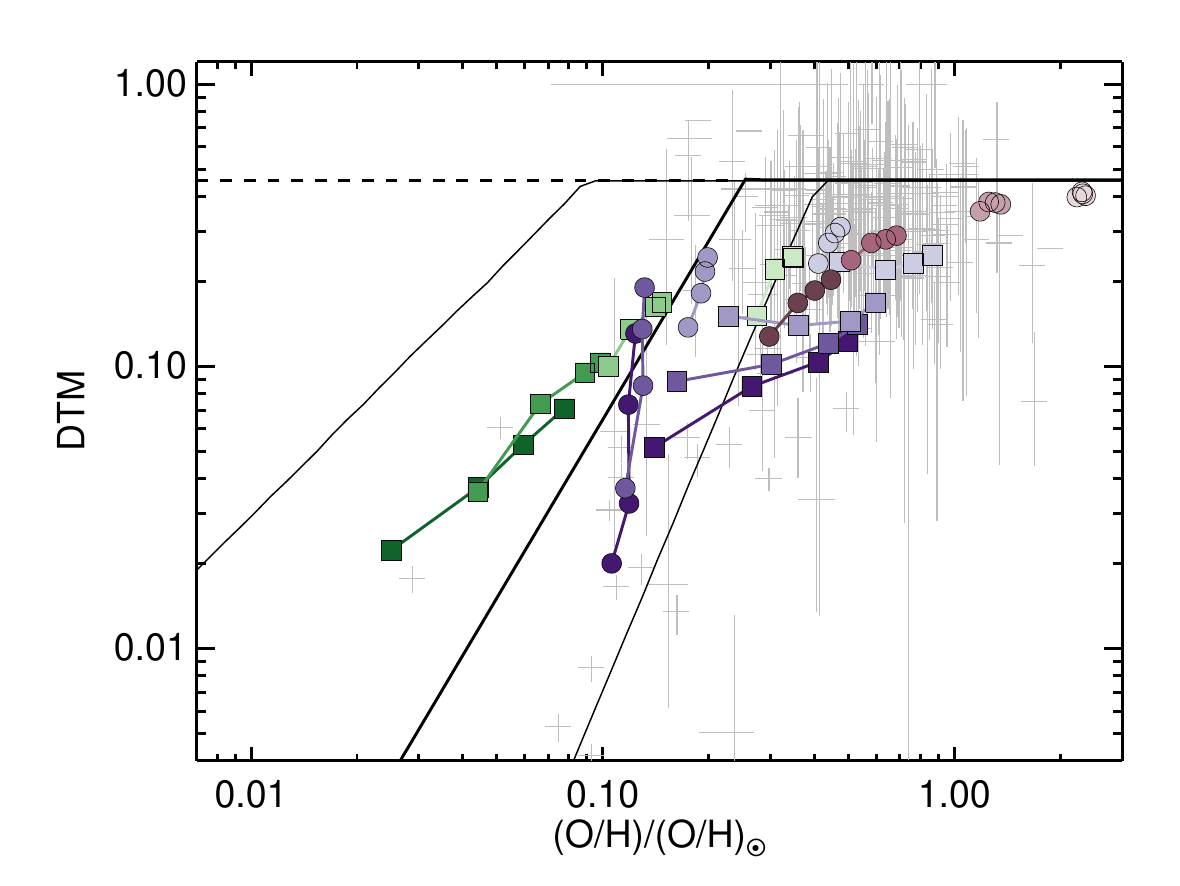}
\caption{Dust-to-gas ratio (DTG, top panel) and dust-to-metal (DTM, bottom panel) as a function of the gas phase metallicity for different simulated galaxies (varying galaxy mass and initial gas metallicity) using the fiducial model for dust evolution. Each symbol with connected lines represent a simulation at different times (100, 200, 300 and $400\,\rm Myr$ with increasing metallicities). Squares are for galaxies with high gas (HG) fractions, and circles for galaxies with low gas (LG) fractions. The dashed black line is the linear relation $\rm DTG\propto [O/H]$, and the solid black lines are the broken power-law from observations of~\cite{remy14} for the mean (black thick solid) and the first and third quartiles (thin solid lines). The grey crosses are the observational data from~\cite{devis17} with error bars.}
\label{fig:dtgvsooverh}
\end{figure}

\begin{figure}
\centering \includegraphics[width=0.45\textwidth]{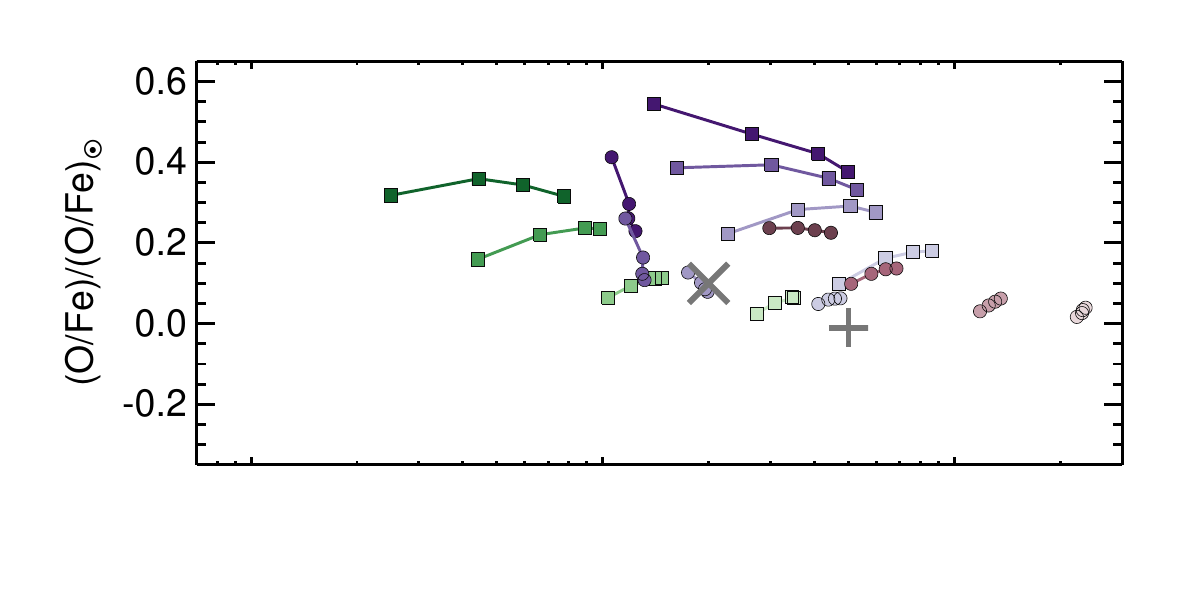}\vspace{-1.35cm}
\centering \includegraphics[width=0.45\textwidth]{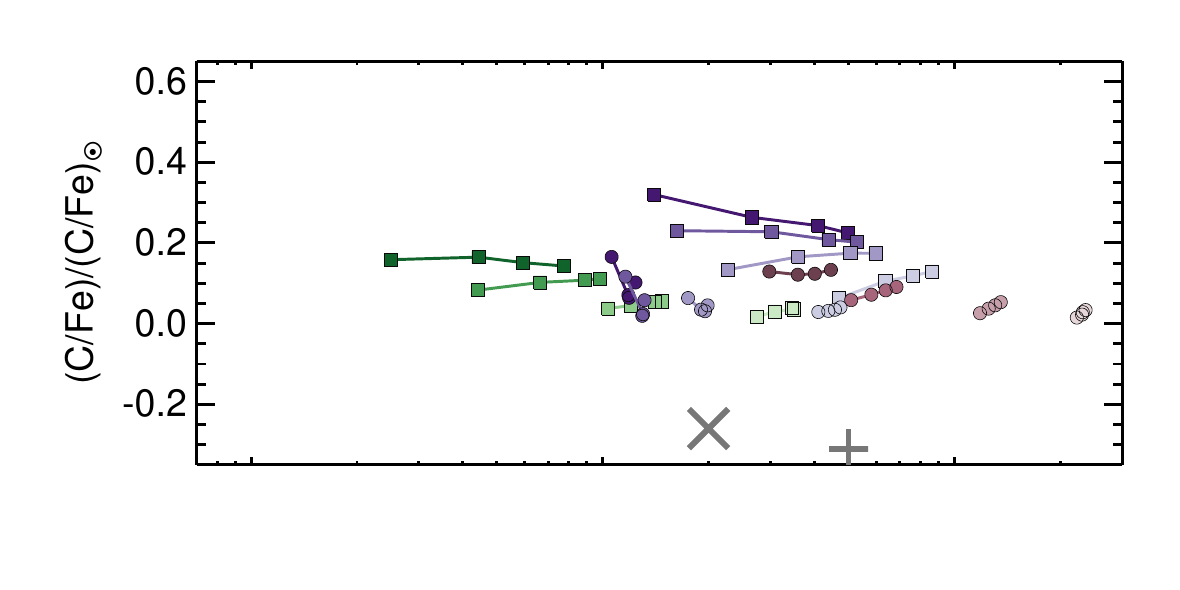}\vspace{-1.35cm}
\centering \includegraphics[width=0.45\textwidth]{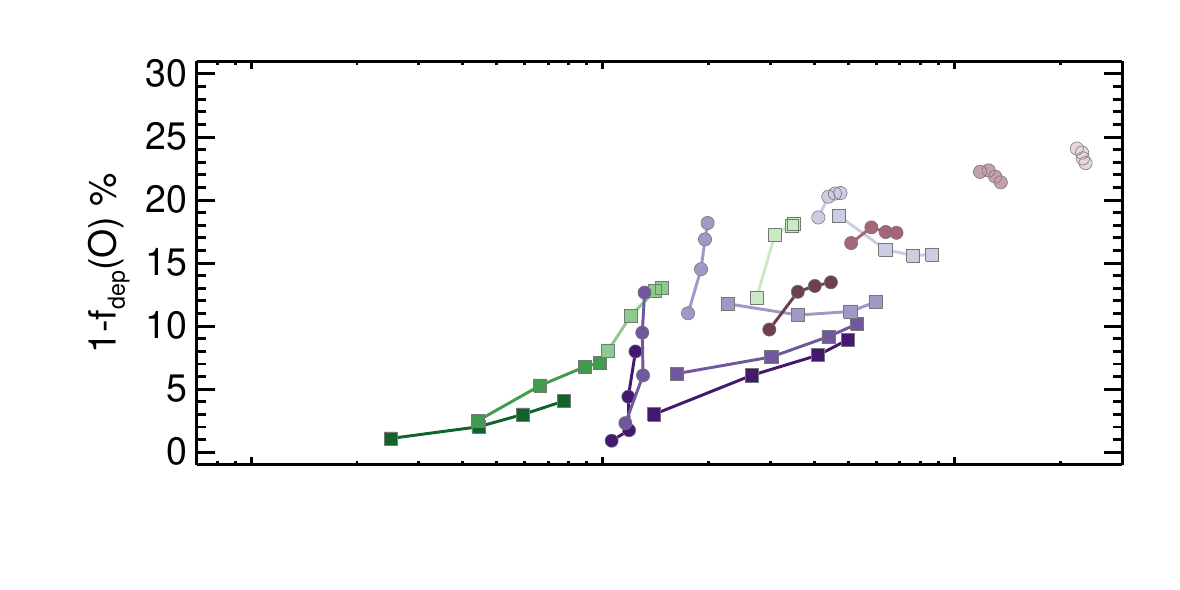}\vspace{-1.35cm}
\centering \includegraphics[width=0.45\textwidth]{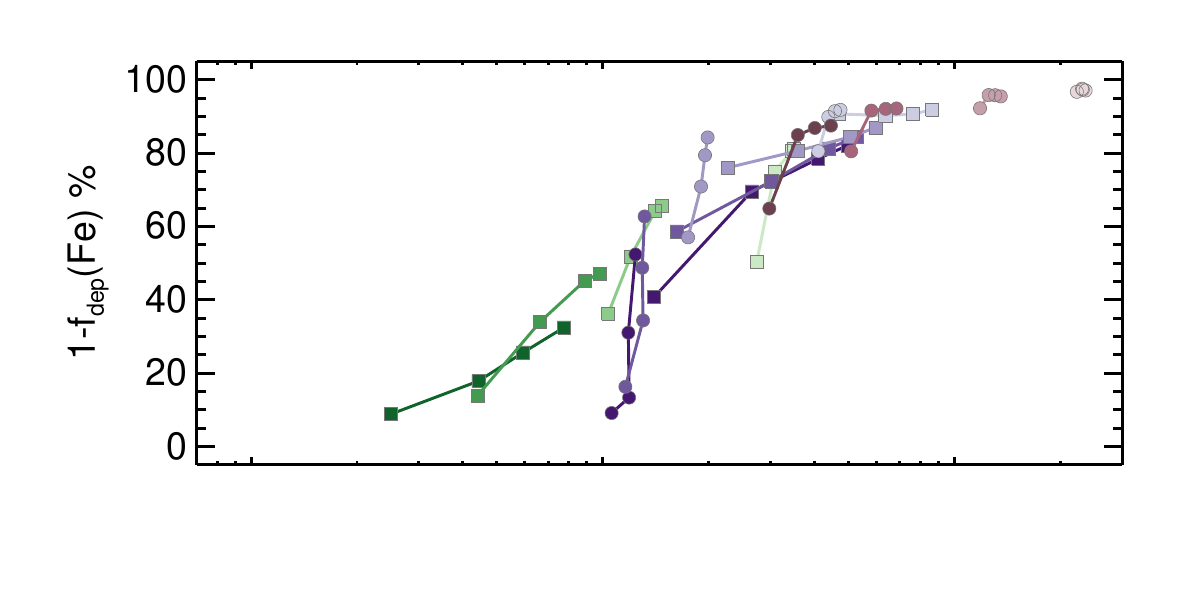}\vspace{-1.35cm}
\centering \includegraphics[width=0.45\textwidth]{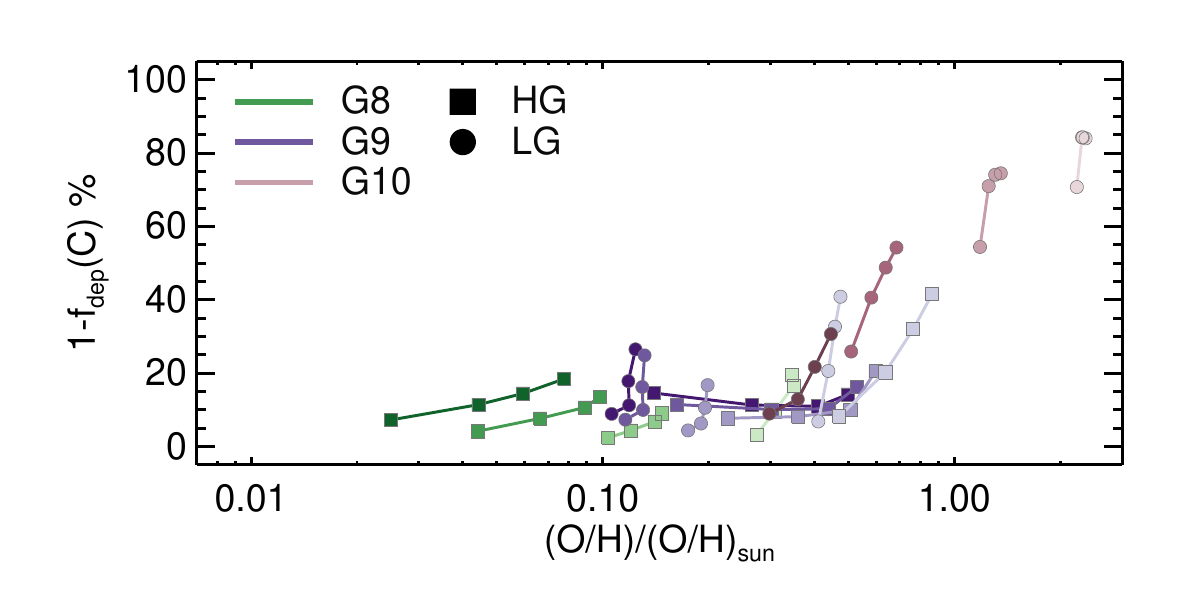}
\caption{Depletion factors for O (top panel), Fe (middle panel) and C (bottom panel) for the different simulations at different times. Similar symbol and color coding as in Fig.~\ref{fig:dtgvsooverh}. Values of abundances and metallicities in SMC and LMC are taken from the compilation of data in~\cite{roman22}.}
\label{fig:depvsooverh}
\end{figure}

We now investigate the dust content and the properties of extinctions curves in ten and hundred times lower mass galaxies (respectively called G8 and G9) than the fiducial Milky Way-like galaxy G10. 

\subsection{Dust mass content}
\label{section:dtmallgal}

Low mass galaxies are generally younger and more metal-poor than more massive galaxies, typically decreasing by 0.3 dex in gas metallicity per 1 dex in stellar mass, albeit with very large dispersion in metallicities from one galaxy to another at any given mass~\citep[e.g.][]{sanchez19,curti20}.
Therefore, for the lower mass simulations, we focus on lower gas phase metallicities.
Figure~\ref{fig:illustrative_examples} gives the visual representation of the gas density and the dust-to-metal (DTM) ratio for the low mass gas-rich galaxy, G8HG\_VLZ, and for the intermediate mass galaxy, either gas poor, G9LG\_LZ, or gas rich, G9HG\_LZ.
For the gas-rich cases, the gas is clearly multi-phase and strongly clustered into regions of neutral gas with high densities ($n>10-100\,\rm H\,cm^{-3}$) corresponding to low temperatures ($T\simeq100 \,\rm K$) surrounded by a more diffuse ionised medium with intermediate density ($n\simeq 0.1-1\,\rm H\, cm^{-3}$) that is warm ($T\simeq10^4\,\rm K$), and SN-driven hot pockets of ultra-diffuse gas ($n<0.1\,\rm H\,cm^{-3}$ and $T>10^6 \,\rm K$).
Compared to the gas density distribution in the gas poor galaxies G9LG\_LZ and G10LG, the gas and the corresponding DTM distributions in the gas rich cases appear to be more flocculent and spiral arms pattern are less well ordered.
Also it immediately appears that for the gas rich G9HG\_LZ case the DTM is lower than for the gas poor G9LG\_LZ case.

\begin{figure*}
\centering \includegraphics[width=0.55\textwidth]{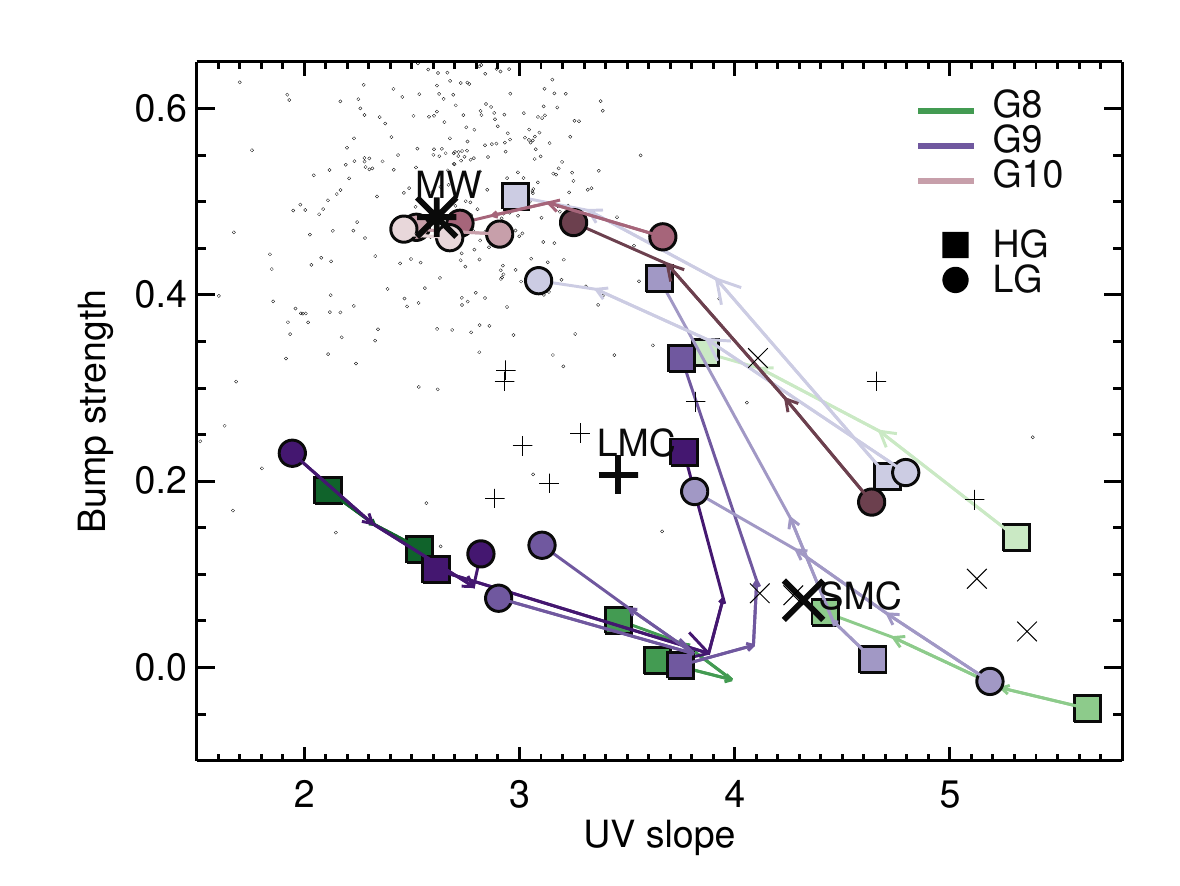}\hspace{-2.27cm}
\centering \includegraphics[width=0.55\textwidth]{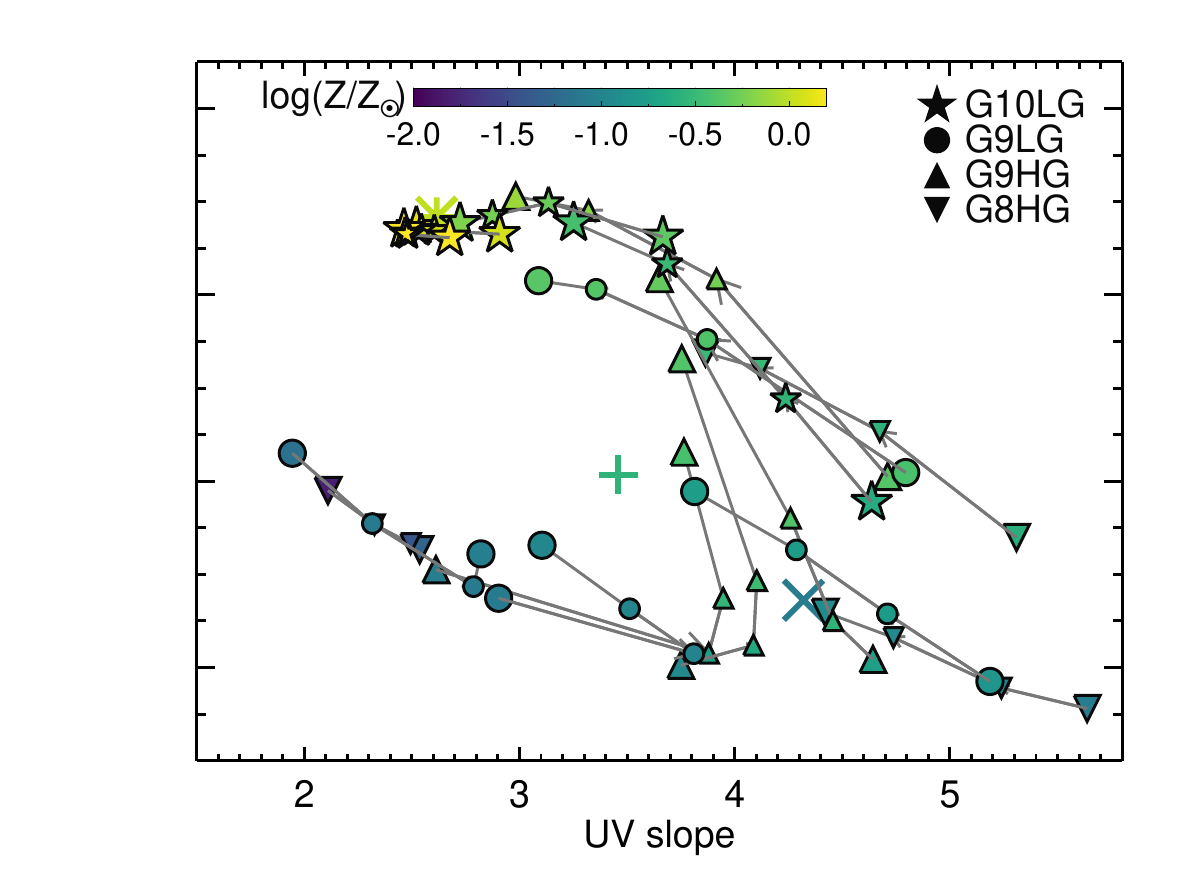}
\caption{Bump strength versus UV-to-optical slope for the different simulations. Arrows give the direction of time sampled at 100, 200, 300 and $400\,\rm Myr$. The big black points are the average values of the observed MW, LMC and SMC from~\cite{pei92} and the smaller grey points are values for sampled sight-lines from~\cite{fitzpatrick07} for MW, and from~\cite{gordon03} for SMC and LMC. The left panel is coded by simulation mass (by their color as indicated in the panel), gas fraction (by their symbol as indicated in the panel), and initial metallicity (by their color shade, from dark to light for increasing initial metallicity $Z_{\rm g,0}$, which corresponding metallicity can be seen in Figs.~\ref{fig:dtgvsooverh} or ~\ref{fig:depvsooverh}), while the right panel is color-coded by the gas metallicity at that given time. Milky Way-like galaxies (G10LG) are naturally attracted to extinction curve features that closely resemble that of the Milky Way, with the closest they get to the true metallicity of the Milky Way the closer they are to the extinction properties of the Milky Way. Lower mass galaxies (G8 and G9) explore a much larger span of the extinction curve features' space, and can be attributed to their lower metallicity that correlates with the features of the extinction curves.}
\label{fig:bumpslope}
\end{figure*}

In Fig.~\ref{fig:dtgvsooverh}, we show the DTG and the DTM as a function of the gas-phase metallicity of the galaxy measured in a cylinder of radius $2\rm kpc$ for G8 and G9, and $4\,\rm kpc$ for G10, with semi-height of $0.2\,\rm kpc$.
This is shown for the different simulated galaxy masses, metallicities, and gas fractions, at 4 different time ($t=100$, 200, 300, and $400\, \rm Myr$).
It is compared with the observational relation obtained by~\cite{remy14} where they fit the data with a broken-power (we note that the data from~\citealp{devis17} and also~\citealp{devis19} show a similar break in DTM with metallicity although with a larger scatter at low metallicities).
The values of DTG increase with the gas phase metallicity across all mass ranges with a significant departure from the linear relation corresponding to constant DTM, in agreement with predictions from cosmological simulations~\citep[e.g.][]{popping17,hou19,li19,graziani20,parente22,lewis23}.
The relation is reasonably well reproduced with a nearly linear relation (i.e. constant DTM) around solar metallicities and with a departure from linear around a gas phase metallicity of a few tenths of solar.
Galaxies with different initial gas fractions also show different behaviour in the DTM-metallicity relation.
For G9HG galaxies (purple squares in Fig.~\ref{fig:dtgvsooverh}), their  DTM is nearly constant after $100\,\rm Myr$, except for the lowest metallicity case, and their gas is continuously enriched as a result of important levels of SFR ($\simeq 0.5\,\rm M_\odot\,yr^{-1}$).
Gas-poor intermediate mass galaxies G9LG have little evolution in metallicity
(low ${\rm SFR}\simeq 0.03-0.04\,\rm M_\odot\, yr^{-1}$) -- except during the first $100\,\rm Myr$ for the two lower metallicity cases $0.01\,\rm Z_\odot$ and $0.03\,\rm Z_\odot$ due to a short initial burst in SFR --, but they strongly increase their DTM.
For the lowest mass galaxies G8HG (gas-rich), the two lowest metallicity cases with initial $0.01\,\rm Z_\odot$ and $0.03\,\rm Z_\odot$ show both a a strong evolution in metallicity and in DTM, and the higher metallicity cases with initial value of $0.1$ and $0.3\,\rm Z_\odot$ has only an important evolution in DTM.
Therefore the various evolution histories due to varying initial metallicities, galaxy mass, or gas fractions contribute to the break in the DTM/DTG-metallicity relation at $0.1\,\rm Z_\odot$ and to its large dispersion at around this value.

The G9 galaxy cases show a different DTM(DTG)-metallicity relation (normalisation) for the gas-poor and the gas-rich galaxies, but also G9HG galaxies can have different DTM at a given metallicity depending on their initial metallicity value.
The reason is that for gas-rich G9HG galaxies, there is a strong evolution in metallicity.
We assumed Milky Way chemical compositions, while due to the significant release of fresh metals over a few 100 Myr, there is an increase of $\alpha$ elements (Mg, Si, O which are the other silicate-bearing elements) over Fe (see the first and second panel of Fig.~\ref{fig:depvsooverh} for resp. [O/Fe] and [C/Fe]).
We recall that Fe was the limiting element of silicate growth for the Milky Way analogue G10LG (see section~\ref{section:depletion}), which is also the case here, as can be seen in Fig.~\ref{fig:depvsooverh} (compare the third and fourth panels) due to the increase of $\rm [\alpha/Fe]$.
Therefore, there is a larger reservoir of $\alpha$ elements that are not accreted due to Fe being strongly depleted in the gas phase.
The mass of elements locked into dust $1-f_{\rm dep}$ (see third, fourth and fifth panel of Fig.~\ref{fig:depvsooverh}) increases with metallicity (in qualitative agreement with the trends observed in SMC, LMC, and MW, see~\citealp{roman22}), as a result of increased accretion rates of refractory elements with metallicity of the ISM.
A crucial insight is that the break in the depletion of Fe as a function of metallicity happens at a different value than for the depletion of C~\citep[see also][]{choban24}, i.e. $f_{\rm dep}=0.5$ is passed at $Z\simeq 0.1\,\rm Z_\odot$ and $Z\simeq \rm Z_\odot$ for Fe and C, respectively.
This disparity might be key to explain variations in extinction curves~\citep{li21} from one simulation to another as we will see in the following section.

\subsection{Extinction curves}

We now characterise the extinction curves by measuring the UV-to-optical slope and the bump strength at $2175\, \AA$.
We follow~\cite{salim20} and~\cite{li21}, where the slope is given by the ratio of extinction at $1500\,\AA$ ($A_{1500}$) over $A_{\rm V}$, and the bump strength by $\Delta A_{2175}/\tilde A_{2175}$, where $\tilde A_{2175}=A_{1500}/3+2A_{3000}/3$, and $\Delta A_{2175}=A_{2175}-\tilde A_{2175}$.
The bump versus slope relation is shown in Fig.~\ref{fig:bumpslope} with the mean values for the global extinction curves of MW, LMC and SMC inferred from the~\cite{pei92} extinction curves, and some sampled lines-of-sights MW from~\cite{fitzpatrick07}, and of the LMC and SMC from~\cite{gordon03}.
As already shown in section~\ref{section:resultsmilkyway}, the Milky Way analogue G10LG exhibits excellent agreement with the extinction features of the Milk Way.
This is true at all times, in particular for the G10LG simulations starting with metallicities at solar values or above (i.e. for $Z_{\rm g,0}=1$ and $2\,\rm Z_\odot$), and only at the latest  times ($t=400\,\rm Myr$) for simulations starting with metallicities below solar value (i.e. for $Z_{\rm g,0}=0.1$ and $0.3\,\rm Z_\odot$), indicating that the gas phase metallicity is a key property to obtain particular features in the extinction curves of a galaxy.

\begin{figure*}
\centering \includegraphics[width=0.38\textwidth]{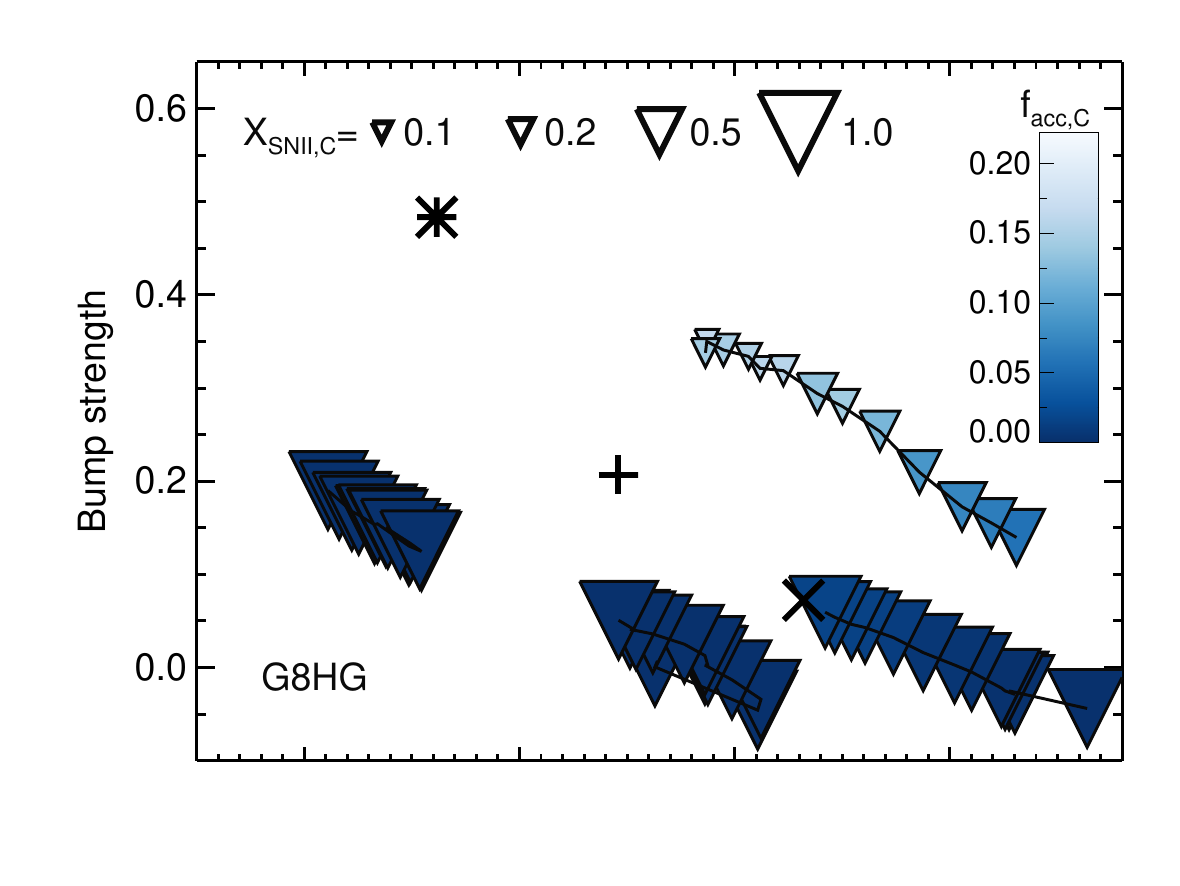}\hspace{-1.6cm}
\centering \includegraphics[width=0.38\textwidth]{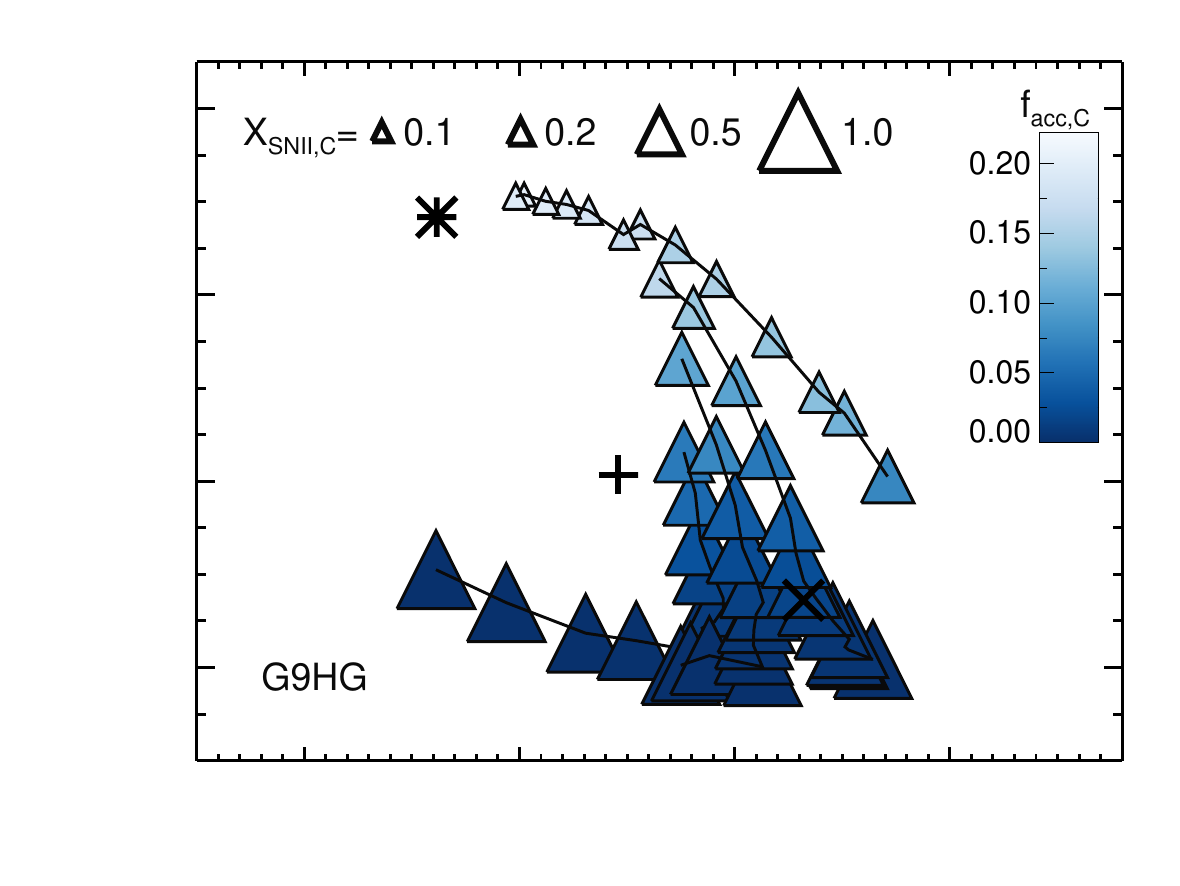}\hspace{-1.6cm}
\centering \includegraphics[width=0.38\textwidth]{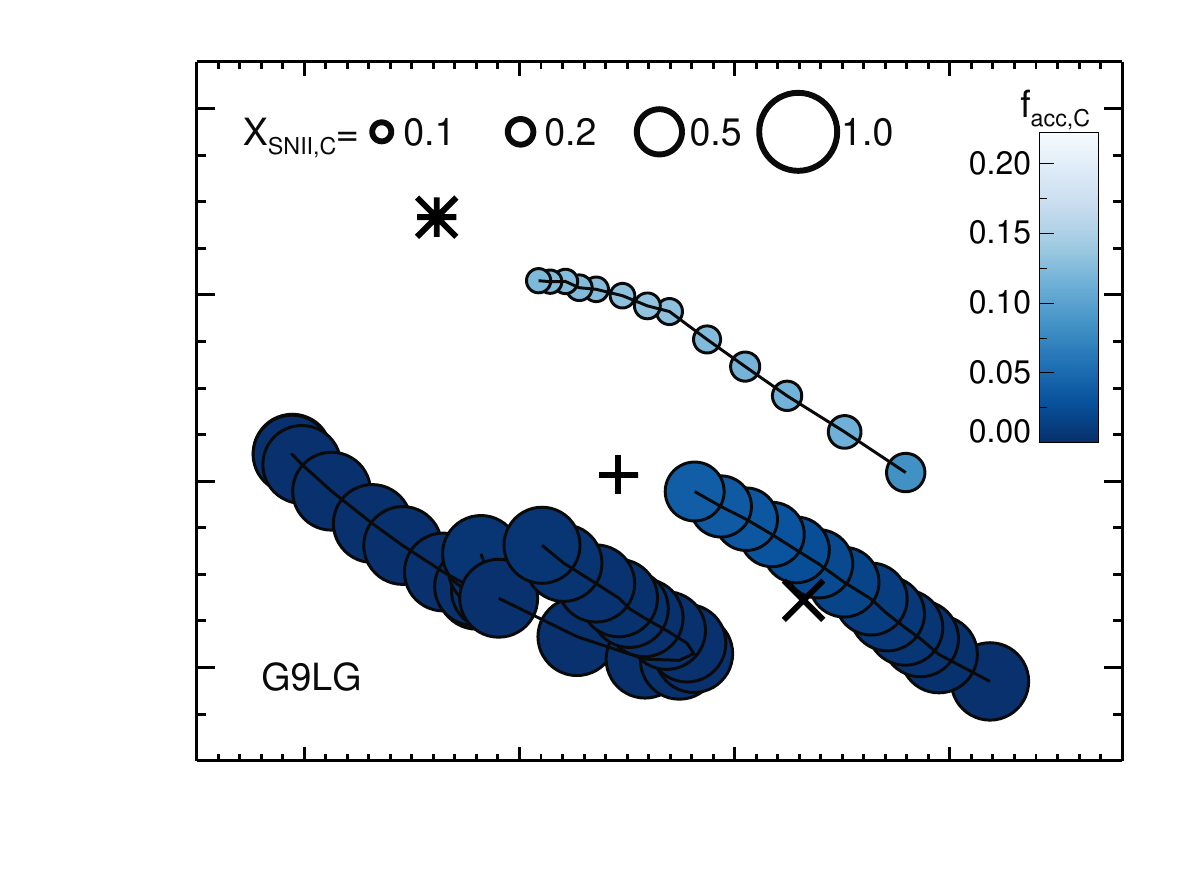}\vspace{-1.15cm} \\
\centering \includegraphics[width=0.38\textwidth]{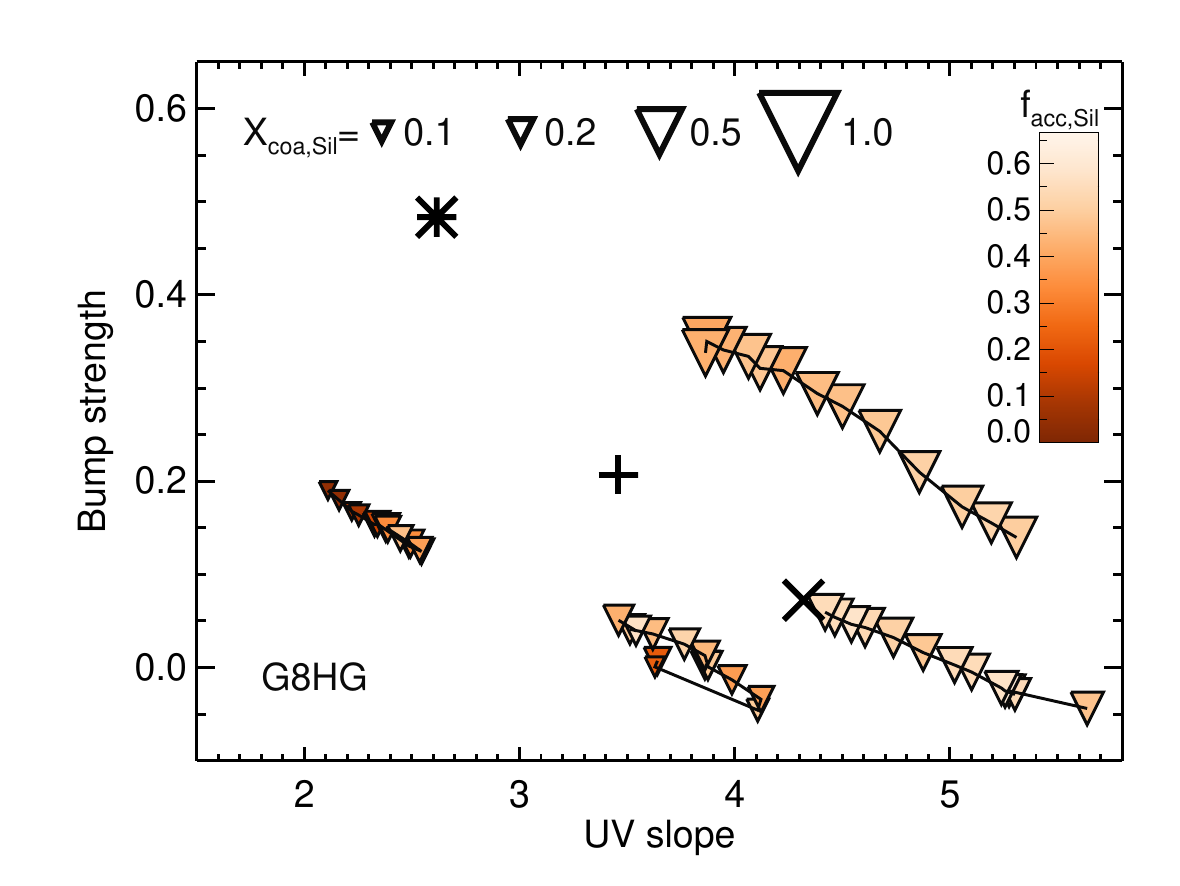}\hspace{-1.6cm}
\centering \includegraphics[width=0.38\textwidth]{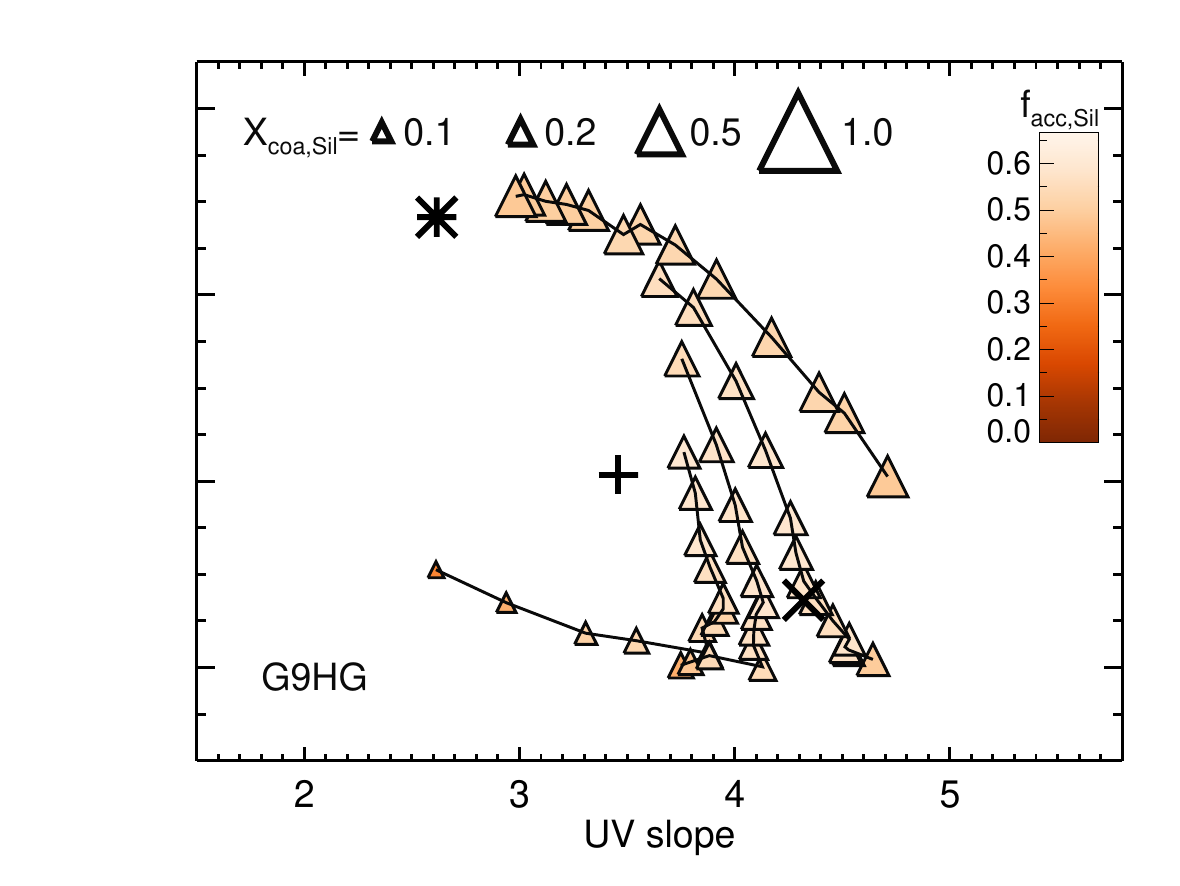}\hspace{-1.6cm}
\centering \includegraphics[width=0.38\textwidth]{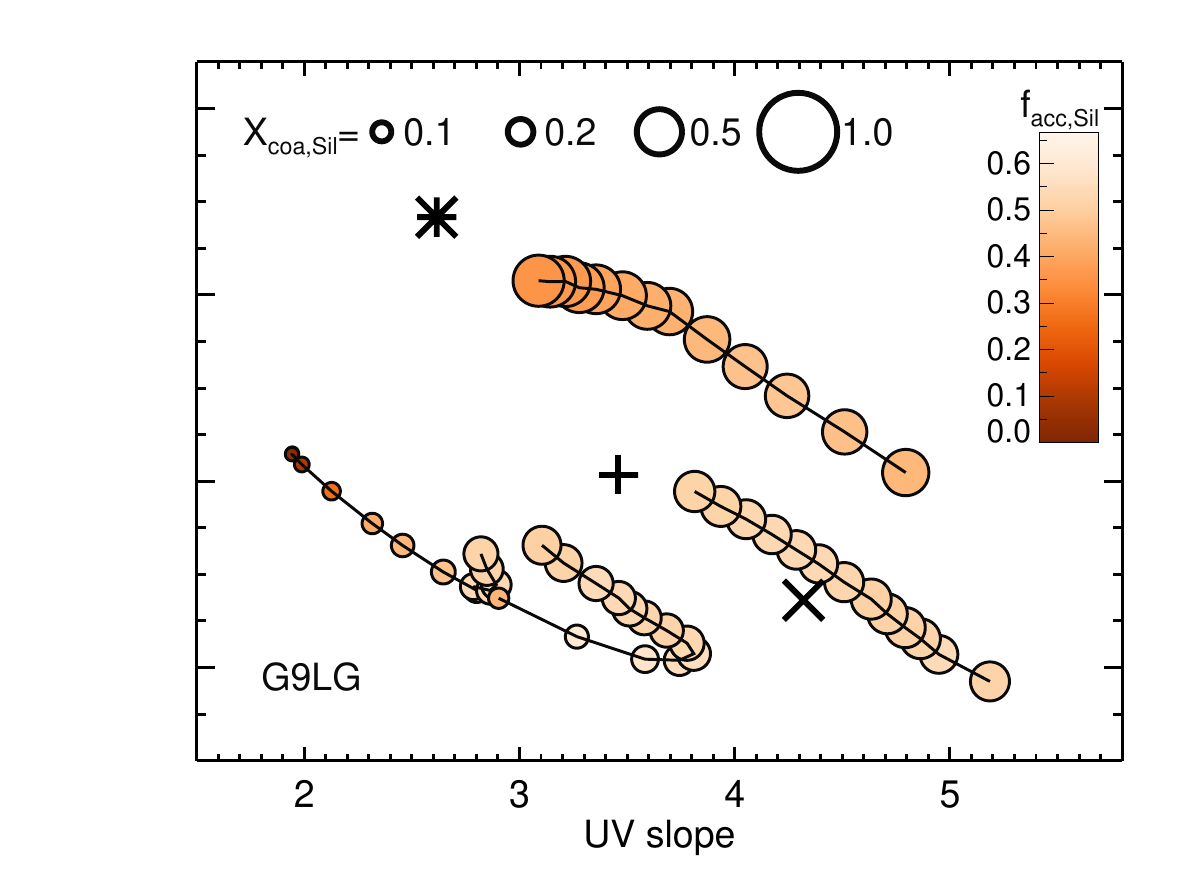}
\caption{Similar to Fig.~\ref{fig:bumpslope} where values of the extinction curve features are size-coded either by the SN ejecta $X_{\rm SNII,C}$ growth strength for the case of carbonaceous grains, or by the coagulation growth strength $X_{\rm coa,Sil}$ for the case of silicate grains, and are color-coded by the accretion fractions $f_{\rm acc,C}$ in blue or $f_{\rm acc,Sil}$ in red (see text for details). Values are shown every $25\,\rm Myr$ from $100\,\rm Myr$ to $400\,\rm Myr$. Extinction curves features for different galaxy type (G8HG, G9HG, G9LG) are shown in different panels (respectively left, middle and right), and each panel contains simulations with different initial metallicities $Z_{\rm g,0}$, from low to high initial metallicity moving from the bottom left of each panel to the top/top-right of the panel in anti-clock wise manner.}
\label{fig:bumpslope_maindmdt_car_sil}
\end{figure*}

It is tempting to compare the results of the simulated G9 low mass galaxies to the extinction curves of the Magellanic Clouds. 
Indeed, the LMC stellar mass and gas mass are estimated at respectively $2\times 10^9 \,\rm M_\odot$ and $5\times 10^8 \,\rm M_\odot$~\citep{kim98}, similar to the
SMC stellar and gas mass of $2\times 10^9 \,\rm M_\odot$ and $5\times 10^8 \,\rm M_\odot$~\citep{stanimirovic04}.
Hence, the stellar mass of G9HG is close to those of the Magellanic Clouds, while our adopted gas mass is larger.
SMC and LMC share extended streams of gas of $\sim 5\times 10^8\,\rm M_\odot$~\citep{bruns05}, which indicates that they have endured important processing of their baryonic content in an interaction.
Finally the stellar metallicities in SMC and LMC are respectively $[{\rm Fe/H}]\simeq-1$ and $\simeq -0.45$~\citep{cole05,choudhury18,choudhury21}.
Additionally, it is important to note that there is a large scatter (and uncertainty) in the UV-to-optical slope of sampled lines-of-sights for SMC and LMC, but very little uncertainty in the bump strength (compare with the distribution of points for the MW in Fig.~\ref{fig:bumpslope}).
In conclusion, the comparison with the extinction curves of SMC and LMC should be taken with extreme care given the dynamical interaction between the two systems, and the large dispersion in the UV-to-optical slope.

With this important warning in mind, we show the extinction curve features (bump and slope) of the simulated lower mass galaxies, G9HG and G8HG for different initial gas metallicities in Fig.~\ref{fig:bumpslope} together with the values for the actual Magellanic Clouds (and Milky Way).
Those two galaxies show a clear departure from the Milky Way extinction curve with i) a lower bump strength, and ii) a steeper slope.
For G9HG, the extinction curve features are attracted to values in the broad range of SMC and LMC, in particular for initially low metallicities.
Nonetheless, they strongly evolve in time mostly due to the strong evolution in metallicity they have (see the right panel of Fig.~\ref{fig:bumpslope}).

This metallicity effect has some important consequences in terms of the grain size distribution of dust.
We recall that SN ejecta only releases large grains, while dust growth by accretion strongly favours the growth of small grains ($t_{\rm acc}\propto a^{-1}$).
Consequently, the extinction curve should be flat in the UV-to-optical regime and the bump should be erased in galaxies whenever the growth by stellar ejecta becomes important with respect to the growth by accretion.
This effect is illustrated in Fig.~\ref{fig:bumpslope_maindmdt_car_sil}: the slope and bump for the low mass galaxies are color-coded by $f_{{\rm acc},i}=\dot M_{{\rm acc}, i}/(\dot M_{{\rm acc,}i}+\dot M_{\rm ej}+\dot M_{{\rm coa,}i^*})$ ($i^*=$ C or Sil, when $i=$ Sil or C respectively) and are size-coded either by $X_{\rm SNII,C}=\dot M_{\rm SNII, C}/(\dot M_{\rm SNII, C}+\dot M_{\rm acc, C})$ or by $X_{\rm coa,Sil}=\dot M_{\rm coa, Sil}/(\dot M_{\rm coa, Sil}+\dot M_{\rm acc, Sil})$.
The corresponding growth rates are computed by taking the average value from the previous $30 \,\rm Myr$ of evolution of the given time to smooth out the high-frequency variations (in particular for the ejection rates).
The reason we employ $M_{{\rm coa,}i^*}$ and not $M_{{\rm acc,}i^*}$ in the calculation $f_{{\rm acc,}i}$ is because the damping of the bump (produced by small carbonaceous grains) or of the UV-to-optical slope (produced by small silicate grains) is due mostly to large grains.

It appears that simulated galaxies have a low bump strength when both $X_{\rm SNII,C}$ are high and $f_{\rm acc,C}$ are low (dark colours), as a natural consequence of the larger fraction of large carbonaceous grains released in stellar ejecta over accretion in the ISM.
When the accretion of C atoms onto carbonaceous dust grains is large with respect to the release by ejecta the bump strength becomes larger, and the extinction curves also evolve towards shallower UV-to-optical slopes.
This increase in bump strength with $X_{\rm SNII,C}$ also corresponds to an increase in $f_{\rm acc,C}$, i.e. in the fraction of accretion to the total growth of dust (carbonaceous and silicate grains).
The increase in the UV-to-optical slope also follows an increase in accretion ($f_{\rm acc,Sil}$), which also correspond to an increase in metallicity (compare with Fig.~\ref{fig:bumpslope}) until they get shallower at the highest metallicities.
Importantly, one can see that these trajectories in extinction curve features bifurcate for some of these runs (e.g. G8HG\_VVVLZ, G8HG\_VVLZ, G9HG\_VVVLZ, G9HG\_VVLZ, G9LG\_VVVLZ or G9LG\_VVLZ) which corresponds to a significant increase in $X_{\rm coa,S}$.
For the largest metallicity runs (e.g. G8HG\_LZ, G9HG\_LZ, or G9LG\_LZ) $f_{\rm acc,Sil}$ is lower compared to lower metallicities, i.e. the relative growth of carbonaceous grains increases (see top panels of Fig.~\ref{fig:bumpslope_maindmdt_car_sil}), and they also present large relative contribution from coagulation $X_{\rm coa, Sil}$, both conspiring to reduce the steepness of the UV-to-optical slope.

\begin{figure}
\centering \includegraphics[width=0.45\textwidth]{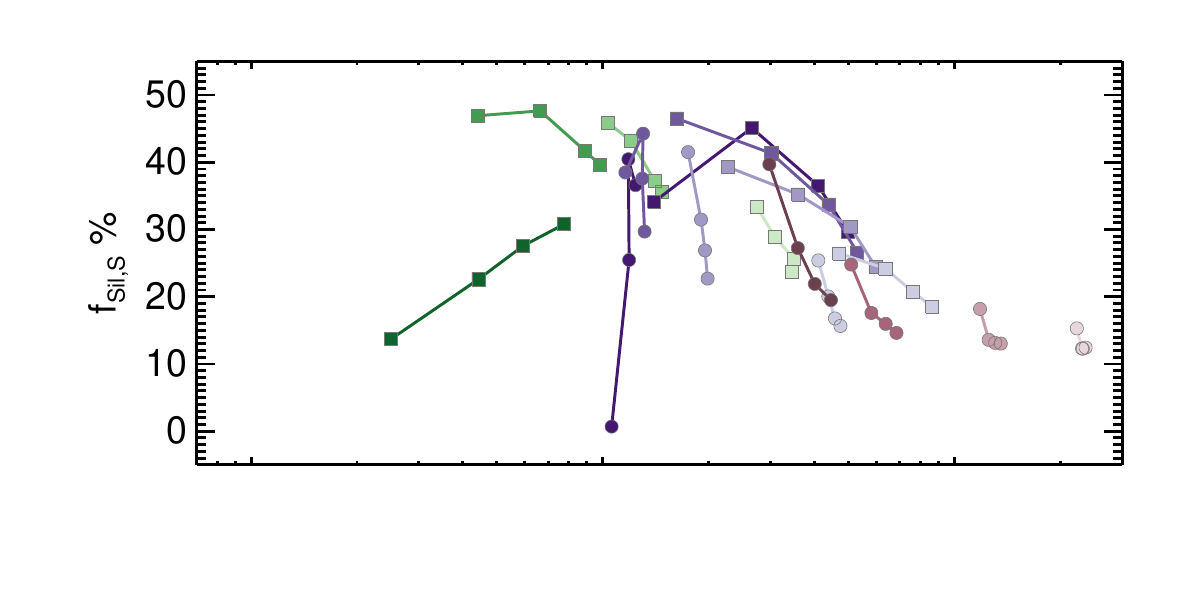}\vspace{-1.35cm}
\centering \includegraphics[width=0.45\textwidth]{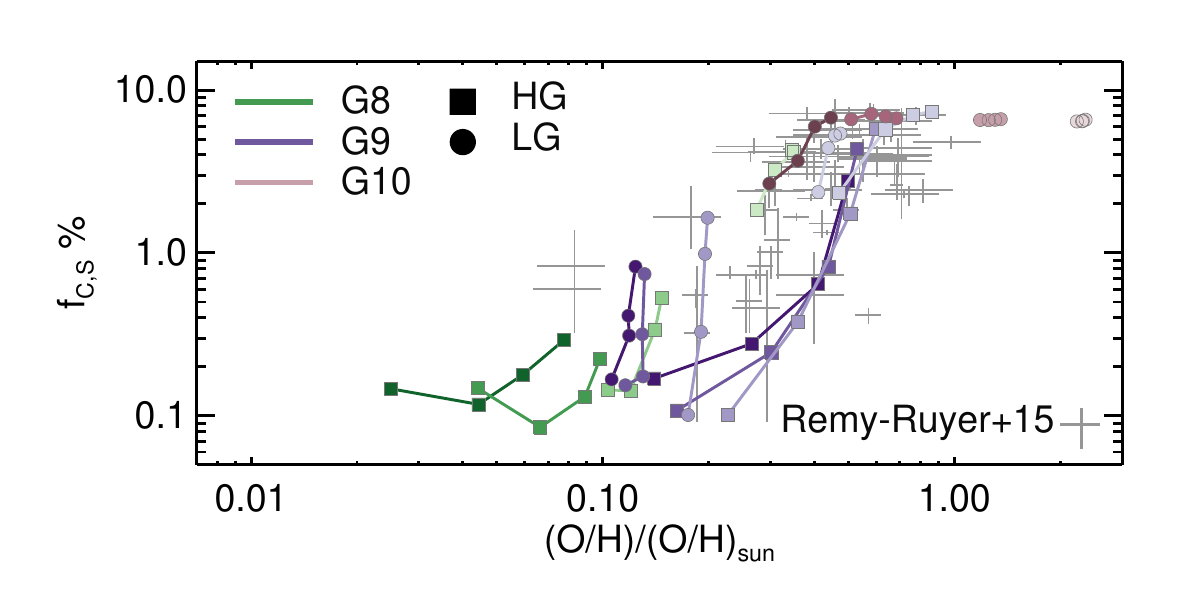}
\caption{Fraction of small silicate (top) and of small carbonaceous (bottom) grains in the overall dust mass. Similar symbol and color coding as in Fig.~\ref{fig:dtgvsooverh}. The grey crosses are the observational data from~\cite{remy15} with error bars for the fraction of dust mass in PAHs.}.
\label{fig:fsvsooverh}
\end{figure}

The consequences of the relative contribution of ejecta, accretion rates, and coagulation to the dust growth of the different size components can be seen in the fraction of small grains that constitute the dust population.
Figure~\ref{fig:fsvsooverh} shows the fraction of small silicate $f_{\rm Sil, S}$ and small carbonaceous $f_{\rm C, S}$ (relative to the total mass of dust) as a function of the gas metallicity.
At very low metallicities ($Z\lesssim 0.1\,Z_\odot$), there is an increase of $f_{\rm Sil, S}$, that stems from the increased contribution in silicate accretion over ejecta.
It is followed by a decrease of $f_{\rm Sil, S}$ ($Z\gtrsim 0.1\,Z_\odot$) caused by the increased coagulation rates.
Carbonaceous grains have a significantly different behaviour.
In qualitative agreement with the increased contribution from accretion (over mass released by ejecta) to the growth of carbonaceous grains (Fig.~\ref{fig:bumpslope_maindmdt_car_sil}, top panels), the fraction of small carbonaceous grains $f_{\rm C,S}$ increases by $\sim 2\,\rm dex$ from $\sim 0.1\,\%$ up to $8\,\%$ from the lowest metallicities $\sim 0.1\,\rm Z_\odot$ up to $\sim 0.5\,\rm Z_\odot$.
Above $Z\gtrsim 0.5\,\rm Z_\odot$ $f_{\rm C,S}$ starts to saturate with the increasing contribution from coagulation.
We also show the mass fraction of polycyclic aromatic hydrocarbons (PAHs) from the observations of nearby galaxies~\citep{remy15}, along with the results from our simulations for $f_{\rm C,S}$, which show excellent agreement. 
Although this is reassuring in the ability of our model to capture a population of small carbonaceous structures in agreement with the data, we must stress that PAHs have additional growth and destruction channels to that of small carbonaceous grains (although they might well co-evolve).
We will discuss further this aspect in section~\ref{section:caveats}.

In summary for low metallicity galaxies, ejecta dominates the growth of dust until they reach sufficiently high metallicity and that the accretion of silicates dominates its growth, increasing the UV-to-optical slope.
The behaviour is similar for carbonaceous grains but it happens at higher metallicities (and this can be traced back to the depletion of Fe and C in the gas phase, see Fig.~\ref{fig:depvsooverh}), which explains the increase in the bump strength.
Finally, the UV-to-optical slope gets shallower at the largest metallicities when both the amount of small carbonaceous grains (grown by accretion) and the coagulation of silicate grains increase.

\begin{figure}
\centering \includegraphics[width=0.45\textwidth]{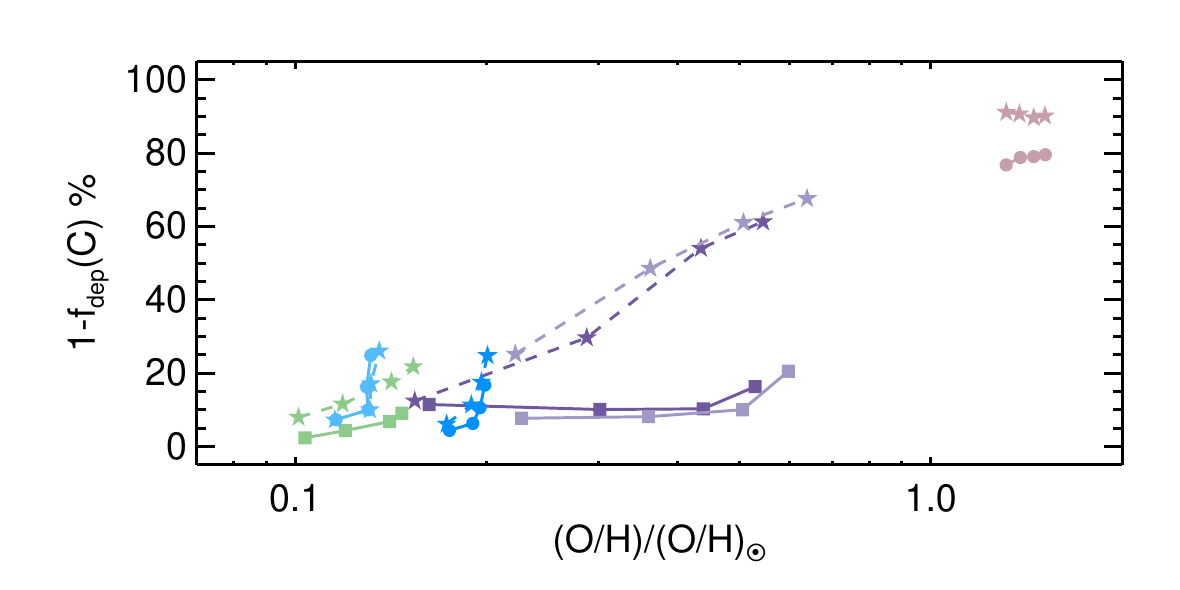}\vspace{-0.35cm}
\centering \includegraphics[width=0.45\textwidth]{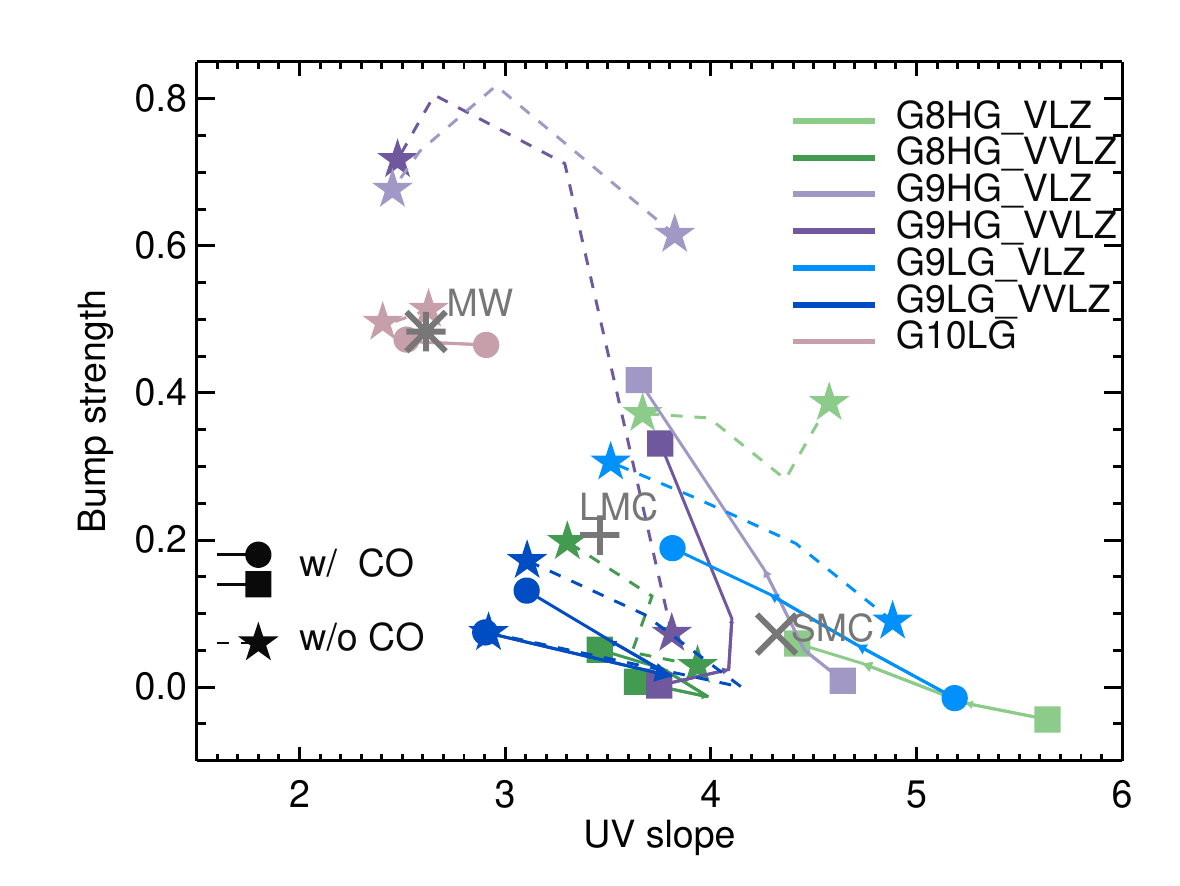}
\caption{Bump strength versus UV-to-optical slope of the extinction curve for some selected simulations considering or not the formation of CO as a limiter to carbonaceous grain growth (as indicated in the panel). Assuming capture of C atoms in molecules which are not available for grain growth has the effect of reducing the amount of small carbonaceous grains in the ISM and, hence, the bump strength of the extinction curve.
}
\label{fig:bumpslope_noco}
\end{figure}

\begin{figure}
\centering \includegraphics[width=0.50\textwidth]{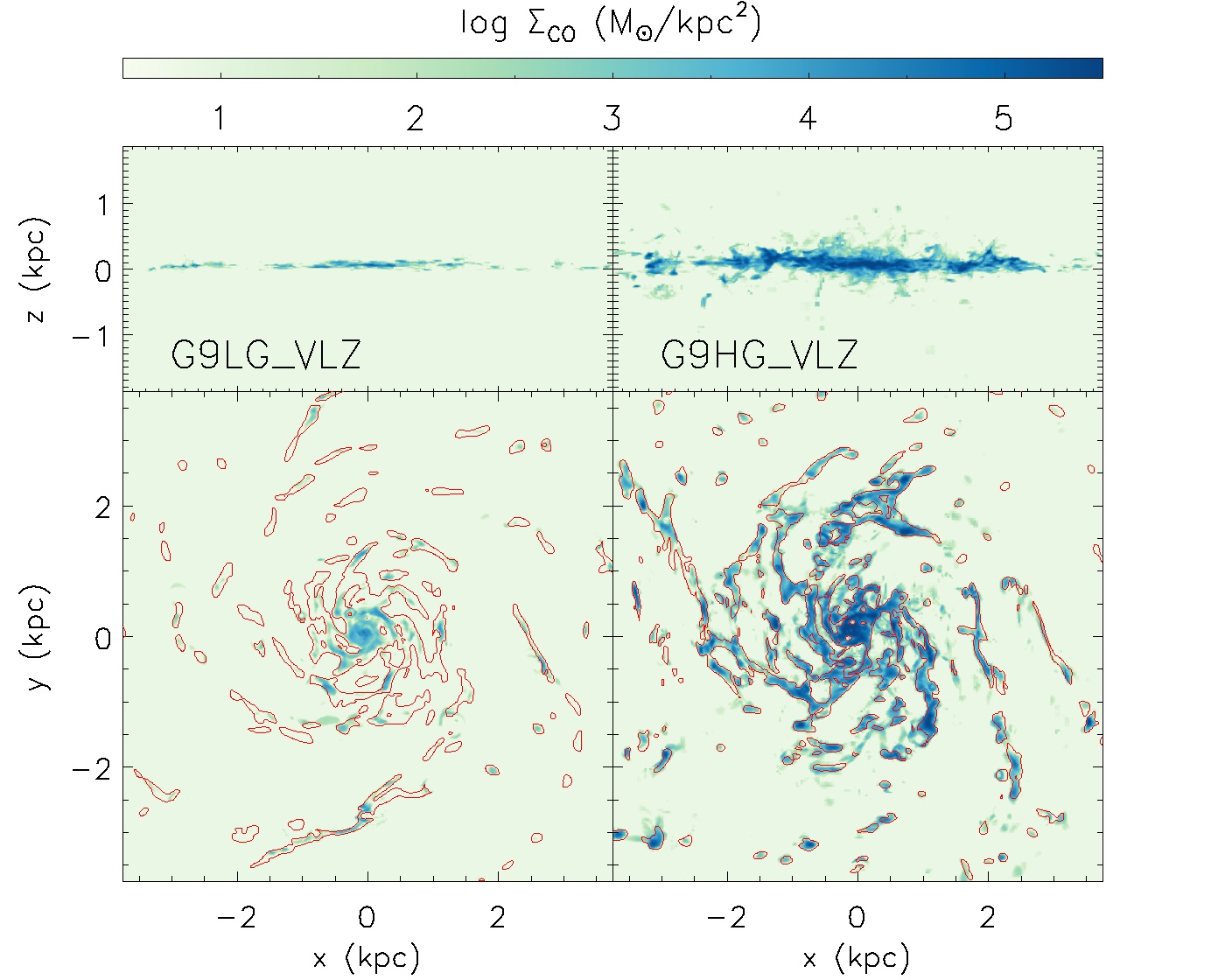}
\caption{Estimated CO gas surface densities $\Sigma_{\rm CO}$ for the gas-poor G9LG\_VLZ galaxy (left panels) and the gas-rich G9HG\_VLZ galaxy (right panels) at $t=400\,\rm Myr$. Red are for the isocontours of projected gas density of $10\, \rm H\,cm^{-3}$. The gas-rich galaxy produce much more molecular gas mass than the gas-poor galaxy.}
\label{fig:nice_co_maps}
\end{figure}

\subsection{The role of CO in the $2175\, \AA$ bump}

In the dust accretion rate we introduced a limit to the growth of carbonaceous grains 
argued on the basis of rapid formation of CO molecules at high gas densities (for $n>10^3\,\rm H\,cm^{-3}$, see section~\ref{section:accretion}).
To explore the impact of CO, we also tested a range of models where we neglect the role of the removal of C atoms into CO molecules: i.e. both carbonaceous and silicate grains stoped their growth by accretion of refractory material at densities $n>10^4\,\rm H\,cm^{-3}$ due to icy mantles coating the grain's surface.

The effect on the depletion of C and on the extinction curve is shown in Fig.~\ref{fig:bumpslope_noco}, resp. top and bottom panels, for several galaxy masses and for different gas fractions.
Removing the consideration on CO formation for the subgrid model of carbonaceous grain growth results in a larger depletion of C, i.e. more carbonaceous grains are grown from the gas phase. 
Consequently, the $2175\,\AA$ bump strength  of the extinction curve is more prominent as a result of a large amount of small carbonaceous grains due to their stronger growth by accretion, which favours the growth of small grains.
In addition, without considering CO formation, the extinction curves are also shallower in their UV-to-optical slope due to the larger ratio of carbonaceous to silicate grains, since there is, proportionally, a lower amount of small silicate grains that are responsible for the steepness of the slope.
A striking behaviour with CO modelling and its impact on carbonaceous dust growth is that galaxies with the low gas fractions exhibit weak variation of the amount of (small) carbonaceous grains with respect to the CO modelling, as opposed to simulated galaxies with a high fraction of gas.
Indeed galaxies with high gas fractions have much more CO formed compared to galaxies with low gas fractions as they reach high gas surface densities, are more turbulent and, hence, get more gas that can reach typical threshold of CO formation at $10^{3}\,\rm H\,cm^{-3}$.
We can also see that for the simulations with the largest difference in bump strength due to the absence of CO corresponds also a significant shift in the UV-to-optical slope: there is overall more carbonaceous grains with respect to silicate grains, which produces a shallower slope of the extinction curve.

Figure~\ref{fig:nice_co_maps} illustrates this last point by showing the estimated CO gas surface densities $\Sigma_{\rm CO}$ for two simulated galaxies of the same mass and with the same initial gas metallicity but which differ in initial gas fraction (G9LG\_VLZ versus G9HG\_VLZ).
Due its larger gas fraction, the G9HG\_VLZ run produces much more CO than the G9LG\_VLZ run, with CO concentrated in regions of high gas densities.
G9HG\_VLZ contains $\simeq 60$ times more CO than G9LG\_VLZ, while the ratio of their total gas mass is only $\simeq 3$.
Since gas accretion primarily proceeds in dense gas, the effect of C capture in CO molecular formation is a strong limiter to growth of carbonaceous grains, and is, hence, the stronger in galaxies with large gas densities.
The estimated CO masses are naive as we only assume instantaneous processing of the entire C gas mass into molecules at a threshold density: it does not take into account the time it takes to grow the CO or destroy it, nor the effect of destruction by UV radiation field and by cosmic rays. 
We defer such investigation and its impact on carbonaceous grain growth to future studies.

\section{Caveats}
\label{section:caveats}

Despite the apparent success in reproducing several key observables of dust properties in our simulated galaxies, our model for dust evolution has made a few important assumptions that we will discuss.

\subsection{Polycyclic aromatic hydrocarbons}

The presence of multiple emission bands in the mid-IR is well attested and the vibrational excitation of small organic molecules, made of PAHs, is the favourite picture to explain these emission features~\citep{leger84,desert90,draineli07,tielens13}.
Observations of the 3.3/11.3~$\mu$m band ratio near photo-dissociation regions suggests that, within the Milky Way ISM, PAHs are primarily composed from molecules with~$\sim 50-100$ C atoms~\citep{croiset16}, well below the centroid of our small grain bin.
Despite their small size, PAHs can lock up to~$\sim 10$\% of C in the ISM. They are hypothesised to play a key role in the thermo-chemistry of the ISM, by heating HI gas and the warm phase via the photo-ejection of electrons \citep[e.g.][]{bakes94,weingartner01}, catalysing the production of molecular hydrogen, or acting as the formation sites of small hydrocarbons \citep[see the review by][and references therein]{kaiser15}.

Theoretical modelling of PAH survival (subject to the diffuse ISM radiation field, SN shocks and thermal sputtering) implies that efficient mechanisms for their replenishment are necessary to account for their abundance \citep{micelotta09}, may this be either through a `top-bottom' picture via the shattering of small and/or large carbonaceous grains \citep{seok14,hirashita&murga20,naranayan23}, or from combustion-like reaction in the ISM giving rise to the most fundamental aromatic species \citep[e.g. benzene, naphthalene,][]{reizer22}.
Observations imply a significantly more linear relation of PAH mass with galaxy metallicity~\citep[e.g.][]{madden06,galliano21,shivai24}, and a strong correlation of PAH band emission with the interstellar radiation field~\citep[e.g.][]{pilleri12,egorov23}, supporting that there are ISM processes that alter differently the PAH and the dust grain abundances.
Of fundamental relevance for the present work, PAHs are also candidate sources of the $2175\, \AA$ feature.
However, in our extinction curve modelling, the origin of this feature comes from the graphitic properties assumed for our small carbonaceous grains ($5\, \rm nm$) with the \cite{weingartner&draine01} optical constants.
Therefore, a model in which this feature is originated instead by smaller, less resilient particles (e.g. PAHs), could result in a larger variation (compared to the one presented here) of the $2175\, \AA$ bump strength with ISM conditions and galaxy metallicity.

Although the extremely high radiation fields (i.e. intensity in the Habing band of $G_0 \gtrsim 10^{8}$) necessary for the thermal sublimation of dust grains are only attained near extreme sources like active galactic nuclei or $\gamma$-ray bursts~\citep{guhathakurta89}, radiation can have a direct effect on the content and properties of the dust grains considered in this work. In particular, if amorphous grains deviate from pure spherical geometry, they present differential scattering and absorption of photons. This means that under an anisotropic radiation field, they will experience an effective radiation torque, capable of accelerating them to the point that their cohesive forces cannot prevent the fragmentation of the grain by centrifugal forces \citep[see][and references therein]{hoang20}. Given that small grains are more easily slowed down due to IR emission and drag from the gas atoms, this radiative-torque disruption (RATD) preferentially fragments large grains, thus altering the shape of the grain size distribution. This may play an important role within low-metallicity, highly star-forming systems, competing with SN shattering/destruction, allowing for steeper UV-slopes than the ones achieved in this work~\citep{fudamoto20}. We will further explore the effect of RATD in the dust extinction curve in future work (Rodr\'iguez Montero et al., in prep.).

\subsection{Dust in stellar ejecta}

We have seen that for low metallicity galaxies ($\gtrsim 0.1\,\rm Z_\odot$), their extinction curves are reflective of their inability to grow carbonaceous dust by gas accretion, hence, their dust grain composition is skewed towards silicates, and carbonaceous dust is only composed of big grains shaped by the SNII ejecta.
For extremely low metallicities galaxies ($\lesssim 0.1\,\rm Z_\odot$), the gas accretion onto silicate grains is also severely quenched, and SNII ejecta become the only source of both carbonaceous and silicate dust growth in the ISM.
Dust condensation efficiencies for SNe are highly uncertain~\citep{schneider23}.
Dust in SNe condenses in the shell of the young ejecta, with dust masses ranging $[0.03-1]\,\rm M_\odot$ for $20\,\rm M_\odot$ progenitor~\citep[e.g.][]{sarangi&cherchneff15,marasi19,brooker22}, before the destruction of dust takes on by the ignition of the SN reverse shock through sputtering and grain-grain collisions~\citep[see e.g.][]{kirchschlager19}, which settles the final amount of dust that is going to be mixed with the ISM.
How much of the dust survives in this final phase depends on the magnetic field, initial dust size distribution, and clumpiness of the ejecta.
Therefore, final dust masses remain largely uncertain, with dust survival rates ranging from a few \% to a few 10 \%, and with large uncertainties in the final dust size distributions~\citep[e.g.][]{silvia10,micelotta16,kirchschlager19,kirchschlager23,slavin20}.
On the observational side, dust mass estimates are also affected by strong uncertainties related to dust composition, size distribution and temperature, with best estimates for Cassiopeia A and Crab Nebula that are in the range $[0.1-1]\,\rm M_\odot$ and $[0.025-0.05]\,\rm M_\odot$, respectively~\citep[see the compilation of data and the discussion in][]{schneider23}.

Given all those uncertainties in dust yields, it is difficult to predict accurately the dust composition and, hence, the extinction curves from low mass and low metallicity galaxies.
As we have seen in our simulations, the contribution from ejecta  is of critical importance in $Z\gtrsim 0.1\,\rm Z_\odot$, as opposed that of silicate grains in that same metallicity range. 
To gauge the impact of the uncertainties on dust ejecta, we ran some of the G8HG low mass galaxies with different fraction of small grain size fraction in stellar ejecta $f_{\rm ej,S}$ in appendix~\ref{appendix:fejsmall}.
Values of the extinction curve UV-to-optical slope and bump strength are significantly increased with higher $f_{\rm ej,S}$ as long as the grain growth of a given species is dominated by the contribution from stellar ejecta.
If carbonaceous grains are the direct product of SN ejecta in low mass galaxies as suggested by these simulations, the extinction curves of the Magellanic Clouds, and young galaxies of the early Universe~\citep{witstok23,markov24} through their attenuation curves, could serve as indirect constraints on mean values of condensation efficiencies in SNII ejecta.

\subsection{Approximations in the grain properties}

An important foundation of the model at the heart of the obtained grain size distribution, especially for the MRN-like size distribution in the Milky Way analogue, is the velocity dispersion of grains. 
We assumed, as in other work~\citep[e.g.][]{aoyama17,gjergo18,granato21}, a unique mean grain velocity dispersion for coagulation and a unique $\sigma_{\rm gr}$-$n$ scaling relation for shattering, although different in magnitude for the small and big grains (for respectively coagulation and shattering rates).
However, it is clear that the grain velocity dispersion should vary with the grain properties, such as its charge, and with the ISM properties, such as its density, ionisation, turbulence, and magnetisation~\citep{yan04,moseley23}.
Additionally, the velocity dispersion of grains should imply a grain-gas decoupling for the largest grain sizes at the molecular cloud scale, with a typical diffusion length of $L\simeq 5\,a_{0.1}(n/1\,{\rm cm^{-3}})^{-1} (\sigma_{\rm gr}/c_{\rm s})\,\rm pc$, with the consequence that global shattering rates should increase. 
We defer such investigations with, using e.g. a diffusion-like approach~\citep{lebreuilly19}, to future work.

We adopted a coarse two-size grain decomposition, as opposed to decompose fully the size spectrum of grains based on the model of~\cite{hirashita15}~\citep[a similar approach that is employed in][]{hou17,aoyama17,gjergo18,granato21}.
This approach significantly reduces the memory and computational requirements of the model, as it only necessitates 4 bins (2 sizes, 2 chemical compositions) instead of the $2N_{\rm bin}$, where $N_{\rm bin}\gtrsim 10$ used in practice for full size spectrum decomposition~\citep{mckinnon18,aoyama19,li21}.
Furthermore, the two-size bin approximation greatly simplifies the treatment of coagulation and shattering that otherwise would have needed cross-coupling between size bins, which makes the computations of coagulation and shattering rates to scale like $N_{\rm bin}^2$.
While a detailed description of fine-sampled grain size distribution associated with accurate models of grain size dispersion could be considered a gold standard, this might come at the expense of spatial resolution, which is crucial for capturing the multiphase ISM responsible for phase transitions between accretion/coagulation-dominated regions (in dense neutral or molecular gas) and shattering-dominated regions (in diffuse ionised gas).

\section{Conclusion}
\label{section:conclusion}

We introduced a novel model of dust growth for galaxy simulations in the {\sc ramses} code that was coupled to the chemical enrichment of the gas.
This model decomposed dust into carbonaceous and silicate grains with two grain sizes of $5\, \rm nm$  and $0.1\, \rm\mu m$.
It included the modelling of dust growth by accretion, stellar ejecta, dust destruction by SNe, thermal sputtering and astration, and also included coagulation and sputtering to allow dust to cycle between small and large grain sizes.
Using hydrodynamical simulations of idealised disc galaxies with a multiphase ISM structure captured at a resolution of $20\,\rm pc$, we explored various galaxy masses, metallicities, gas fractions, and their effect on the galactic dust properties and extinction curves.

These simulations are able to reproduce the main dust properties of the Milky Way, and they produce dust-metallicity scaling relations in good agreement with observational constraints.
They draw a scenario for the physical origin of the observed differences between the extinction curve of the Milky Way and those of the Magellanic Clouds, that stem in the emergence of different dominant processes for dust growth in the ISM (ejecta, accretion and coagulation) driven by the metallicity of the gas.  
In more details, our main results are the following:
\begin{itemize}
    \item The numerical analogue of the Milky Way shows excellent agreement with the DTM, the MRN size distribution, the extinction curve, and the depletion factors of various elements of the Milky Way.
    \item Accretion is the leading growth mechanism in Milky Way-like of galaxies, and the DTM is sensitive to the gas metallicity which controls the amount of accretion onto dust grains.
    \item In Milky Way-like galaxies, coagulation and shattering play an important role to shape the most prominent features of the extinction curve: its bump and UV-to-optical slope.
    \item DTM is sensitive to spatial resolution effects but, as introduced here, a subgrid model that captures the unresolved density contrasts due to turbulence greatly improves the convergence properties of the accretion scheme.
    \item Varying the galaxy mass, gas fraction and metallicity has some impact on the DTM/DTG, and the DTM/DTG-$Z$ relation is naturally reproduced in our model with a rapid decrease in DTM below $0.1\,\rm Z_\odot$ as suggested by observations. This transition is due to a dominant growth mode by ejecta, at low metallicities, to growth by accretion, at high metallicities.
    \item A crucial insight is that the break in the depletion of silicate-bearing elements as a function of metallicity happens at a different value than that of C: at $Z\simeq 0.1\,\rm Z_\odot$ and $Z\simeq \rm Z_\odot$ respectively. This offset, driven by the different accretion efficiencies of silicate and carbonaceous grains, is key to explain variations in extinction curves for low mass and low metallicity galaxies, such as for the Magellanic Clouds.
    \item By exploring different galaxy masses, gas fractions and metallicities, we obtain diversity in the bump strength and UV-to-optical slope of extinction curves. In particular, these features are strongly correlated with the metallicity of the gas, as it controls the relative amount of growth by accretion (small grains) relative to ejecta (large grains) and to coagulation.
    \item Due to their lower metallicity, we can interpret the steeper slopes and the weaker bumps of the extinction curves of the Magellanic Clouds with respect to that of the Milky Way as the result of a lower accretion of metals onto dust relative to ejecta (especially for C), of a more efficient accretion of silicate elements with respect to C, as well as of a less efficient coagulation of grains.
    \item Because of the capture of C atoms in CO molecules, there is a strong suppression of small carbonaceous grains in gas-rich galaxies, which guarantees the good agreement with the extinction curves of the Magellanic Clouds. 
\end{itemize}

Those simulations were carried out in an idealised context without considering the effect of fresh gas replenishment by halo-scale gas accretion.
A natural direct extension of this work is to test the validity of the dust model in fully cosmological hydrodynamical simulations to understand the influence of cosmic infall and mergers on the properties of dust.
In particular, such cosmological simulations could be used to understand the cosmological build up of dust in galaxies, and test the fraction of galaxies with obscured star formation~\citep{magnelli20,pozzi21,fudamoto21,inami22}.
In this work, we limited ourselves to testing extinction curves of simulated galaxies, which gives a direct insight on dust properties.
Although attenuation curves are the result of the complex interplay between the dust properties, and the ISM and stellar distributions~\citep{narayanan18}, the flurry of observational evidence might serve as an important testbed for the cosmic evolution of dust properties.

The dust model introduced in this paper is currently extended to track PAHs, as well as to include a self-consistent coupling of the dust and ISM non-equilibrium thermo-chemistry in {\sc ramses-rtz}~\citep{katz22a} with radiation-hydrodynamics (Rodr\'iguez Montero, et al., in prep.).
This `Dusty-{\sc PRISM}' model would be capable of exploring the influence of radiation, chemistry, gas properties and dust grains on the properties of PAHs, shedding light on the origin of the correlations between mid-IR bands and the properties of the ISM across cosmic time.

\section*{Acknowledgments}

We thank Hakim Atek, Benoît Commer\c con, Pierre Cox, Frédéric Galliano, Pierre Guillard, Patrick Hennebelle, Harley Katz, Guillaume Laibe, Ugo Lebreuilly, Guillaume Pineau des For\^ets, and Nicolas Prantzos for useful discussions.
This work has made use of the Infinity cluster on which the simulation was run and post-processed, hosted by the Institut d'Astrophysique de Paris.
We warmly thank S.~Rouberol for  running it smoothly.
FRM is supported by the Wolfson Harrison UK Research Council Physics Scholarship.
This project has received funding from the European Research Council (ERC) under the European Union’s Horizon 2020 research and innovation programme (grant agreement No 693024).
MT acknowledges support from the NWO grant 016.VIDI.189.162 ("ODIN").
SKY acknowledges support from the Korean National Research Foundation (2020R1A2C3003769, 2022R1A6A1A03053472).
Part of the interpretation was  based on the calculations using KIST's Nurion (KSC-2022-CRE-0088, KSC-2022-CRE-0344, KSC-2023-CRE-0343).
\bibliographystyle{aa}
\bibliography{author}

\appendix

\section{Released metallicities and dust}
\label{appendix:yields}

\begin{figure*}
\centering \includegraphics[width=\textwidth]{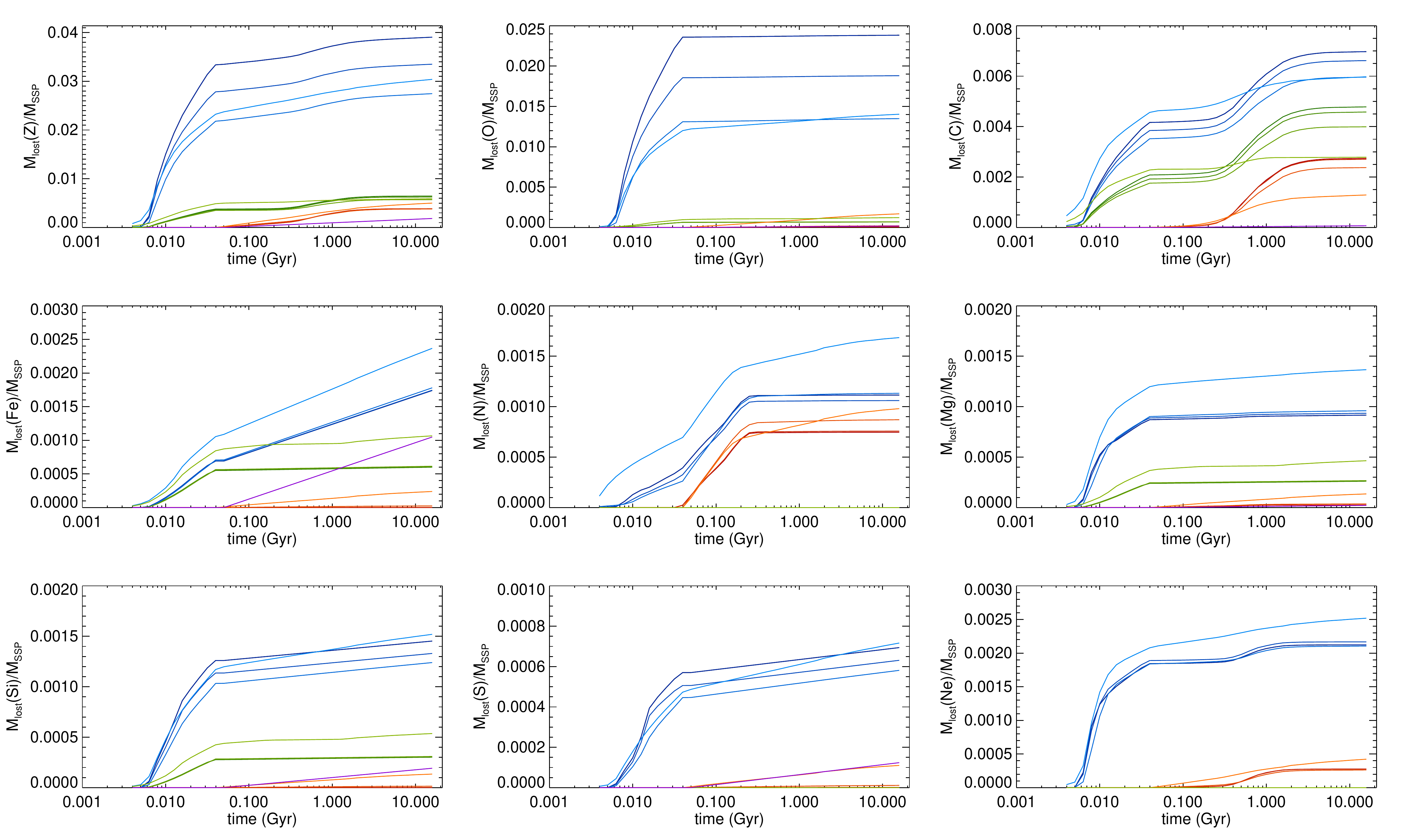}
\caption{Evolution of the SSP yield  adopted in this work for total metallicity and different individual elements. The blue lines are for the total elemental release (gas plus dust). Red lines are the contribution from intermediate stars. Purple lines are the contribution from SNIa. Green lines are for dust. Lines goes from dark to light for an initial SSP metallicity of $Z_{\rm ZAS}=10^{-3},10^{-2},10^{-1},1\,\rm Z_\odot$.}
\label{fig:yield_ej}
\end{figure*}

The time evolution of the resulting chemical and dust yields for a simple stellar population of different initial metallicities $Z_{\rm ZAS}=10^{-3},10^{-2},10^{-1},1\,\rm Z_\odot$ are shown in Fig.~\ref{fig:yield_ej} with assumptions for stellar tracks, IMF, stellar rotations, and dust condensation efficiencies as described in sections~\ref{section:stellar_yields} and~\ref{section:SSPdust}.
It shows that the total mass return, and that of O, Mg, Si, S and Ne are overall largely dominated by SNII. 
N mass release is largely dominated by AGB winds.
C has an important contribution from AGB while still dominated by SNII, and, finally, Fe has an equal mass contribution from SNIa and SNII.
For dust release, we see as much carbonaceous dust released by AGN as by SNII, while silicate dust is mostly produced by SNII.

\section{Ejecta sizes in low mass galaxies}
\label{appendix:fejsmall}

\begin{figure}
\centering \includegraphics[width=0.45\textwidth]{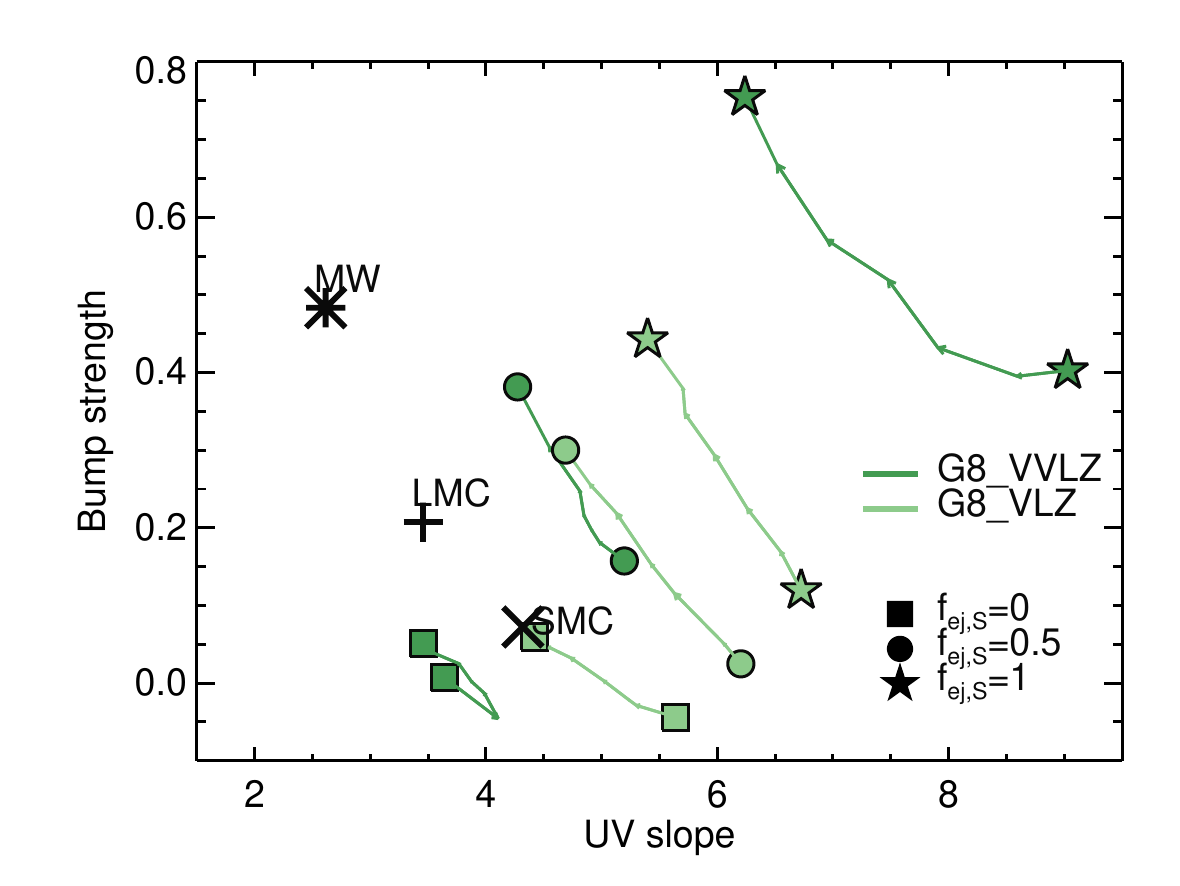}
\caption{Bump strength and UV-to-optical slope of the extinction curve for the low mass galaxies with initial gas metallicity $0.03\,\rm Z_\odot$ (G8\_VVLZ) and $0.1\,\rm Z_\odot$ (G8\_VLZ) for different fractions of small grain mass in stellar ejecta $f_{\rm ej, S}$ as indicated in the panel. Values of the extinction curve features for the Milky Way (`MW'), the Large Magellanic Cloud (`LMC'), and the Small Magellanic Cloud (`SMC') are also shown.}
\label{fig:bumpslope_fej}
\end{figure}

We show in Fig.~\ref{fig:bumpslope_fej}, the changes in the UV-to-optical slope and bump strength of the extinction curves of the G8HG\_VVLZ and G8HG\_VLZ for different values of small grain size fraction in stellar ejecta $f_{\rm ej, S}=0$ (default value), $0.5$, and $1$.
G8HG\_VVLZ has been selected for its important contribution of SNII ejecta component (as opposed to accretion) for both carbonaceous and silicate grains in the fiducial model, and G8HG\_VLZ because of its significant SNII ejecta component for carbonaceous grains (only).
As expected, these two low mass and low metallicity galaxies are very sensitive to the grain size mixture of their stellar ejecta for both the UV-to-optical slope and bump strength for G8\_VVLZ, but only significantly for the bump strength for G8\_VLZ since G8\_VLZ has silicate grain (mainly responsible for the UV-to-optical slope) growth dominated by ISM accretion.

\section{Sputtering times}
\label{appendix:sputtering}

\begin{table*}
    \centering
    \caption{Coefficients of the fifth-order polynomial fit for the sputtering rate ($i=$ silicate or carbonaceous grains) $\log(Y_i/(\mu {\rm m} \, {\rm yr^{-1}}\, {\rm  cm}^3))=\sum_{n=0}^5 a_n (\log (T/{\rm K}))^n$ of~\cite{hu19} to the values predicted by~\cite{nozawa06}.}
    \begin{tabular}{c c c c c c c}
grain type & $a_0$ & $a_1$ & $a_2$ & $a_3$ & $a_4$ & $a_5$ \\
        \hline
        \hline
silicate & -234.791 & 133.209 & -31.3027 & 3.71345 & -0.221824 & 0.00531746 \\
carbonaceous & -234.334 & 138.486 & -33.9022 & 4.17705 & -0.258281 & 0.00638828 \\
        \hline
    \end{tabular}
    \label{tab:hu19}
\end{table*}

\begin{figure}
\centering \includegraphics[width=0.45\textwidth]{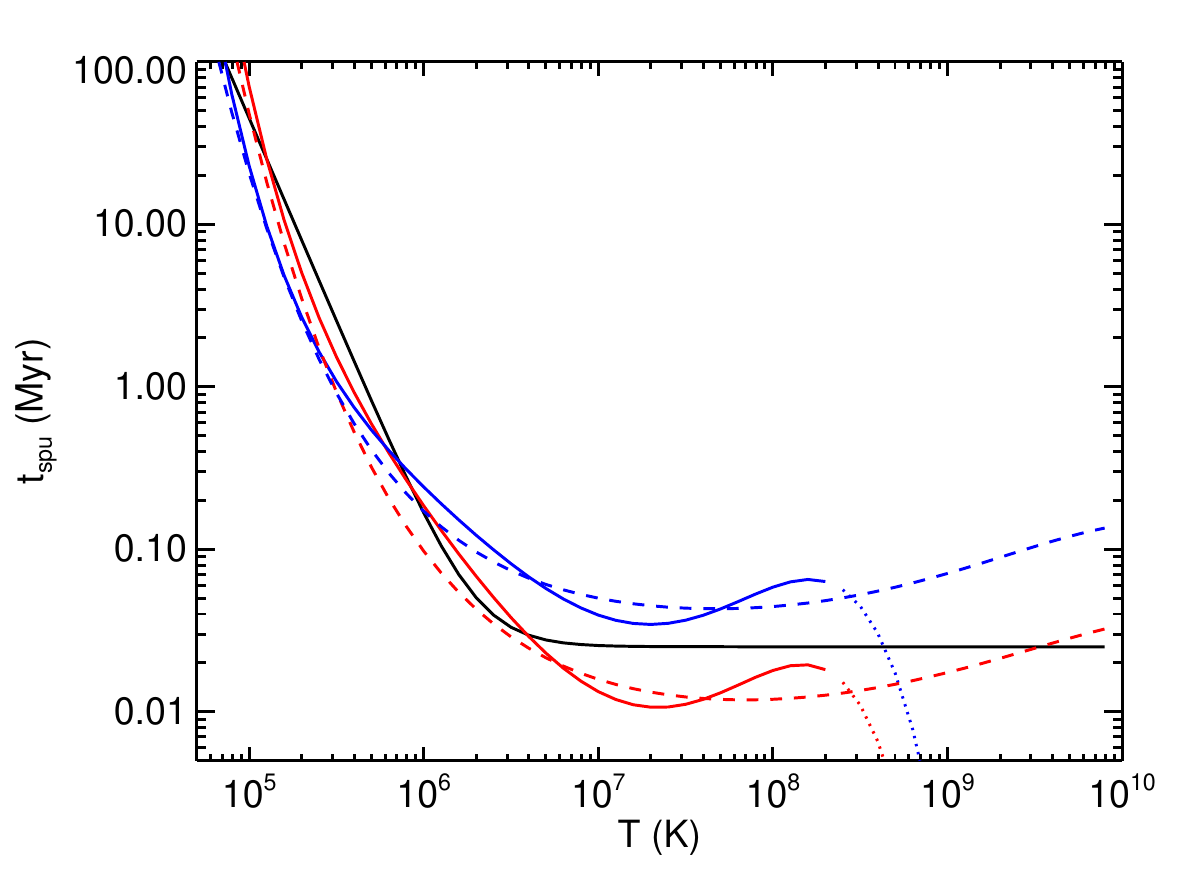}
\caption{Sputtering time as a function of temperature for a gas density of $n=1\,\rm H\,cm^{-3}$ and a grain size of $a=0.1\, \mu\rm m$. In red and blue solid and dotted are the results of~\cite{tielens94} for respectively Si and C grains using their fitting functions valid up to $T\simeq 2\times 10^8\,\rm K$. The black line is the fitting formula from~\cite{tsai&mathews95}. The red and blue dashed lines are the fitting formulas for respectively Si and C grains from~\cite{hu19}.}
\label{fig:sputtering_time}
\end{figure}

Figure~\ref{fig:sputtering_time} compares the sputtering times as a function of the gas temperature for a grain size of $a=0.1\, \mu\rm m$ and a gas density of $n=1\,\rm H\,cm^{-3}$ from~\cite{tielens94}, \cite{tsai&mathews95}, and ~\cite{hu19}.
\cite{hu19} provides a convenient fifth-order polynomial fit to the predictions of the sputtering rates $Y(T)$ from~\cite{nozawa06}.
The values of the \cite{hu19} fit with sufficient digits are reported in table~\ref{tab:hu19} (private communication with C.-Y. Hu).
The formula from~\cite{tsai&mathews95} is intermediate to the sputtering times of carbonaceous and silicate grains from~\cite{tielens94} and~\cite{hu19}, which both show a general good agreement up to a temperature of $T\sim10^8\,\rm K$.
We note that these calculations are all for a semi-infinite target, however, the finite size of the grain modifies the sputtering rates for grains with size smaller than the penetration depth of the impinging ion~\citep[e.g.][]{kirchschlager19}.

\section{Sticking coefficient}
\label{appendix:sticking_coefficient}

\begin{figure}
\centering \includegraphics[width=0.45\textwidth]{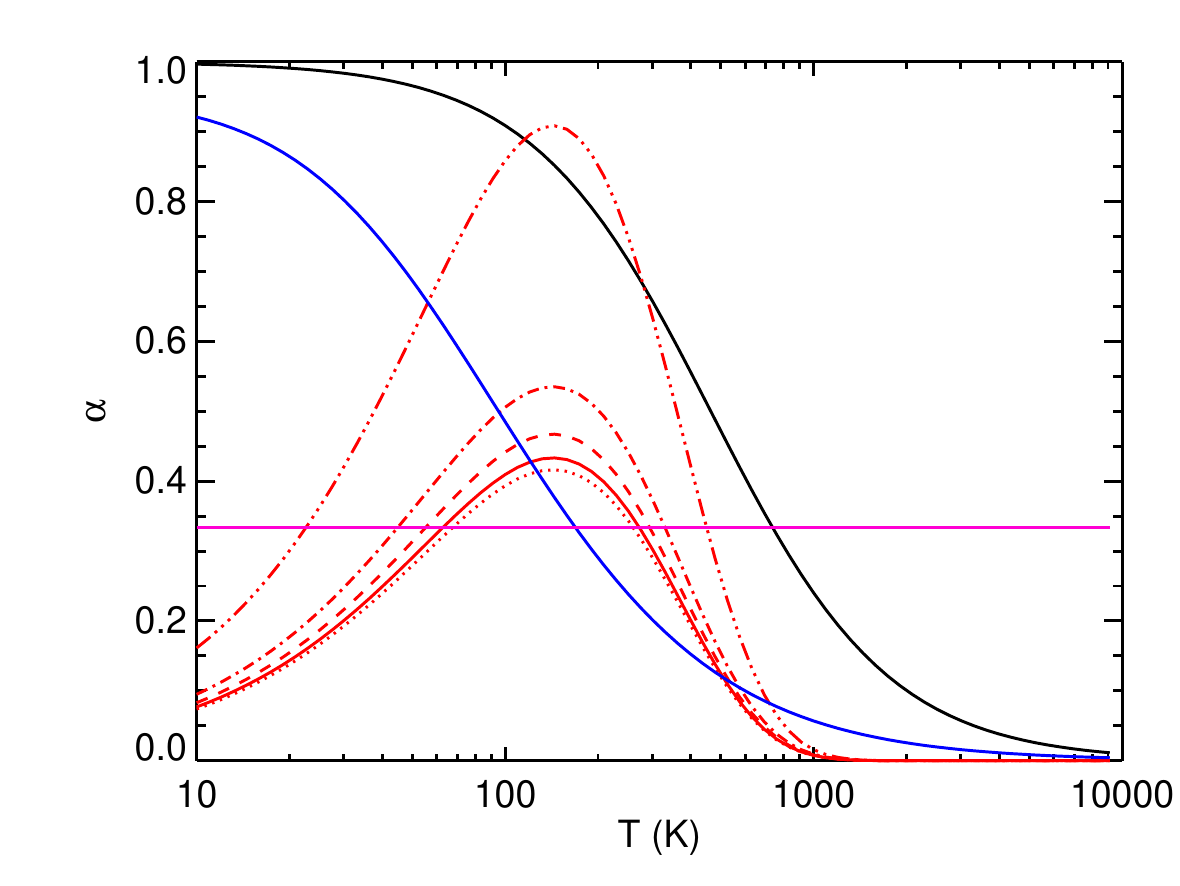}
\caption{Sticking coefficient for gas accretion onto dust as a function of the gas temperature $T$ for different models: \cite{lebourlot12} in black, \cite{chaabouni12} in blue, and \cite{leitch85} in red for different dust temperatures $T_{\rm D}=10,20, 40, 80, 300\,\rm K$ for the dotted, solid, dashed, dot-dashed, and triple dot-dashed red lines respectively. The magenta line at $\alpha=1/3$ is our preferred value for unresolved gas densities (see Section~\ref{section:accretion}).}
\label{fig:stickcoef}
\end{figure}

There are various predictions for the sticking coefficient $\alpha$, which value depends on the temperature of the gas (and grains), e.g.:
\begin{enumerate}
 \item $\alpha_{\rm C}(T)=0.95(1+\beta T/T_0)/(1+T/T_0)^\beta$, $\beta=2.22$ and $T_0=56$~K measured from laboratory experiments for chemisorption of H$_2$ molecules on silicate surfaces~\citep{chaabouni12}.
 \item $\alpha_{\rm LB}(T)=(1+10^{-4}T^{1.5})^{-1}$ from theoretical considerations from~\cite{lebourlot12}.
 \item  $\alpha_{\rm L}(T)=1.9\times 10^{-2} T (1.7\times 10^{-3}T_{\rm D}+0.4)\exp(-7\times 10^{-3}T)$ taken from the fit to the data from quantum calculations of~\cite{leitch85} in~\cite{grassi14}, and where $T_{\rm D}$ is the dust temperature. The dust temperature for big grains is at local thermodynamical equilibrium around 20 K for local galaxies~\citep[e.g.][]{schreiber18}, however very small grains can be stochastically heated~\citep{draine03} to large temperatures of about 1000 K.
 \end{enumerate}
We present in Fig.~\ref{fig:stickcoef} the behaviour of each of these predictions of the behaviour of sticking coefficient with $T$.
They all show an increase of $\alpha$ with a decrease in temperature down to $T\sim 200\,\rm K$, where $\alpha$ ranges from 0.3 to 0.9 at this temperature. 
Below this temperature, the values from \cite{leitch85} predicts a decrease of $\alpha$ with a decrease of $T$, while the values from \cite{lebourlot12} and \cite{chaabouni12} show an increase of $\alpha$ with a decrease of $T$ up to $\alpha=1$.

Since the exact value and rate of change with temperature of the sticking coefficient $\alpha$ remains highly uncertain~\citep[see the discussion in][]{zhukovska16}, we adopted a constant value of $\alpha=1/3$ for our unresolved high-gas density regime driven by turbulence (see Section~\ref{section:accretion}).
For comparison, the turbulence-driven subgrid model for dust accretion is activated for $\lambda_{\rm J}<4\Delta x$ and the effective temperature-density relation in our simulations is $T\simeq10^4 (n/0.1\,\rm H\,cm^{-3})^{-2/3}$ (see Fig.~\ref{fig:histo_dust_g10lg}) for $n>0.1\,\rm H\,cm^{-3}$ and $T<10^4\,\rm K$, hence, the temperature at which the subgrid model is triggered for a fiducial resolution of $\Delta x=20 \,\rm pc$ is $T\simeq 750\,\rm K$.
This temperature gives a value of $\alpha$ from \cite{lebourlot12} that is close to our adopted constant value of $\alpha$ in the subgrid model.
We ran a G10LG simulation with an increased value of $\alpha=1$ in that regime: the DTM increases by $5\%$, mostly driven by an increase in the fraction of carbonaceous grains, which goes from a fiducial value of $f_{\rm C}=0.34$ to $f_{\rm C}=0.38$, and produces a significant decrease of the depletion factor of C by a factor of 2 at $n\simeq 10\,\rm H\,cm^{-3}$.

\begin{figure}
\centering \includegraphics[width=0.45\textwidth]{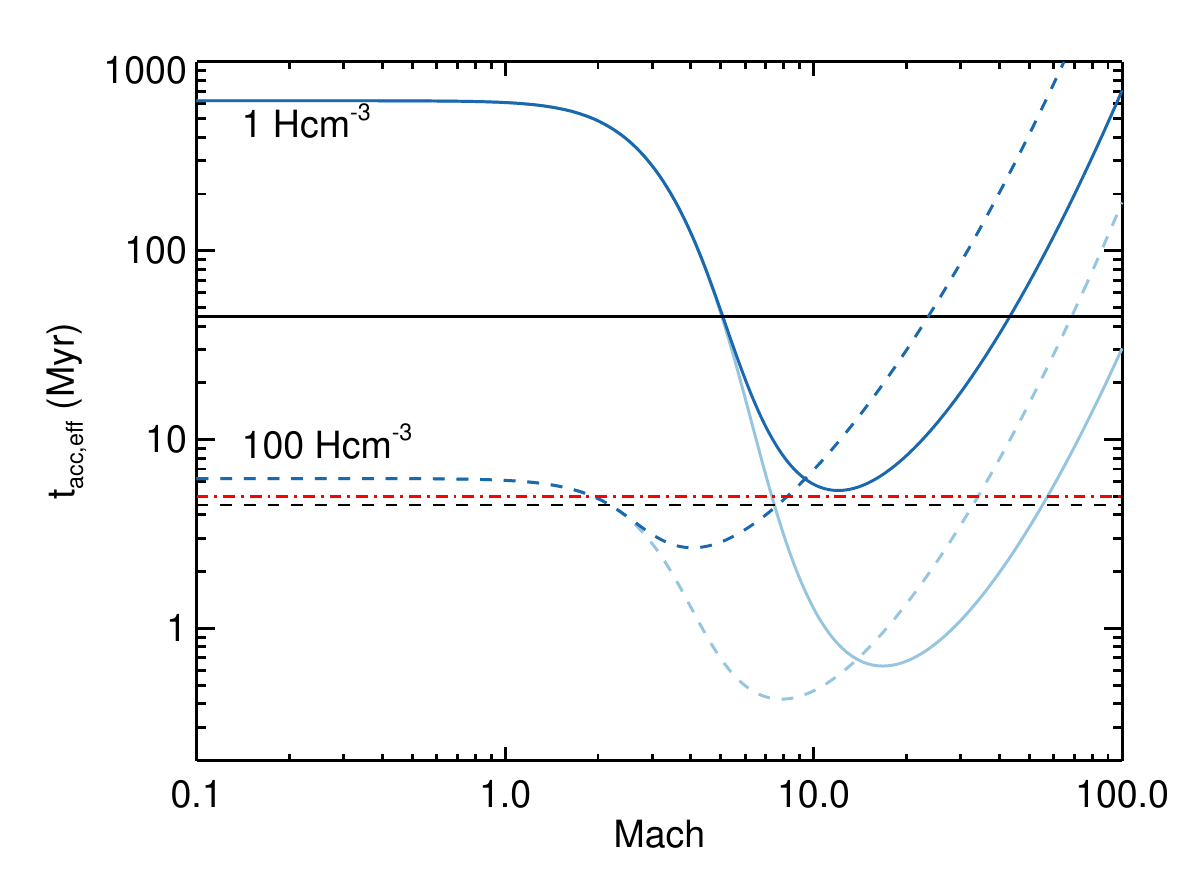}
\caption{Effective accretion rate of refractory materials onto grains (here for carbonaceous grains) as a function of the turbulent Mach number for two different mean density $\bar n=1$ and $100\,\rm H\, cm^{-3}$ (respectively in solid blue and in dashed blue). The various colors stand for different density cut-offs $n_{\rm max}=10^3$ and $10^4\,\rm H\, cm^{-3}$ (respectively in dark and light blue). For comparison we show the free-fall time corresponding to the quoted gas density $\bar n$ in black, and the SN time delay in red dot-dashed.
}
\label{fig:boost_acc}
\end{figure}

The turbulence-driven subgrid model for gas accretion onto dust grains has a maximum density above which the accretion of new refractory elements is impossible (either $n_{\rm max}=10^3$ or $10^4\,\rm H\, cm^{-3}$ (see section~\ref{section:accretion}).
This value modifies the behaviour of the effective accretion timescale as a function of the turbulent Mach number and is shown in Fig.~\ref{fig:boost_acc} for the case of accretion onto carbonaceous dust grains.
The effective timescale for accretion $t_{\rm acc,eff}$ exhibits a decrease with turbulent Mach number above $\mathcal{M}>1$ until it reaches a minimum for strong $\mathcal{M}\simeq 5-10$ after which the effective timescale increases. 
The exact behaviour is driven by the maximum value $n_{\rm max}$.
To guide the eye, we show the corresponding free-fall time of the gas at the given mean gas densities $\bar n$, and the (single) time delay for SNII adopted in this work. 

\section{Sizes and extinction curves for low mass galaxies}

In this appendix we show in Figs.~\ref{fig:size_ext_all_1} and~\ref{fig:size_ext_all_2} the dust grain size distribution and the extinction curves of the G9HG, G9LG, and G8LG listed in table~\ref{tab:simulations}.

\begin{figure*}
\centering \includegraphics[width=0.34\textwidth]{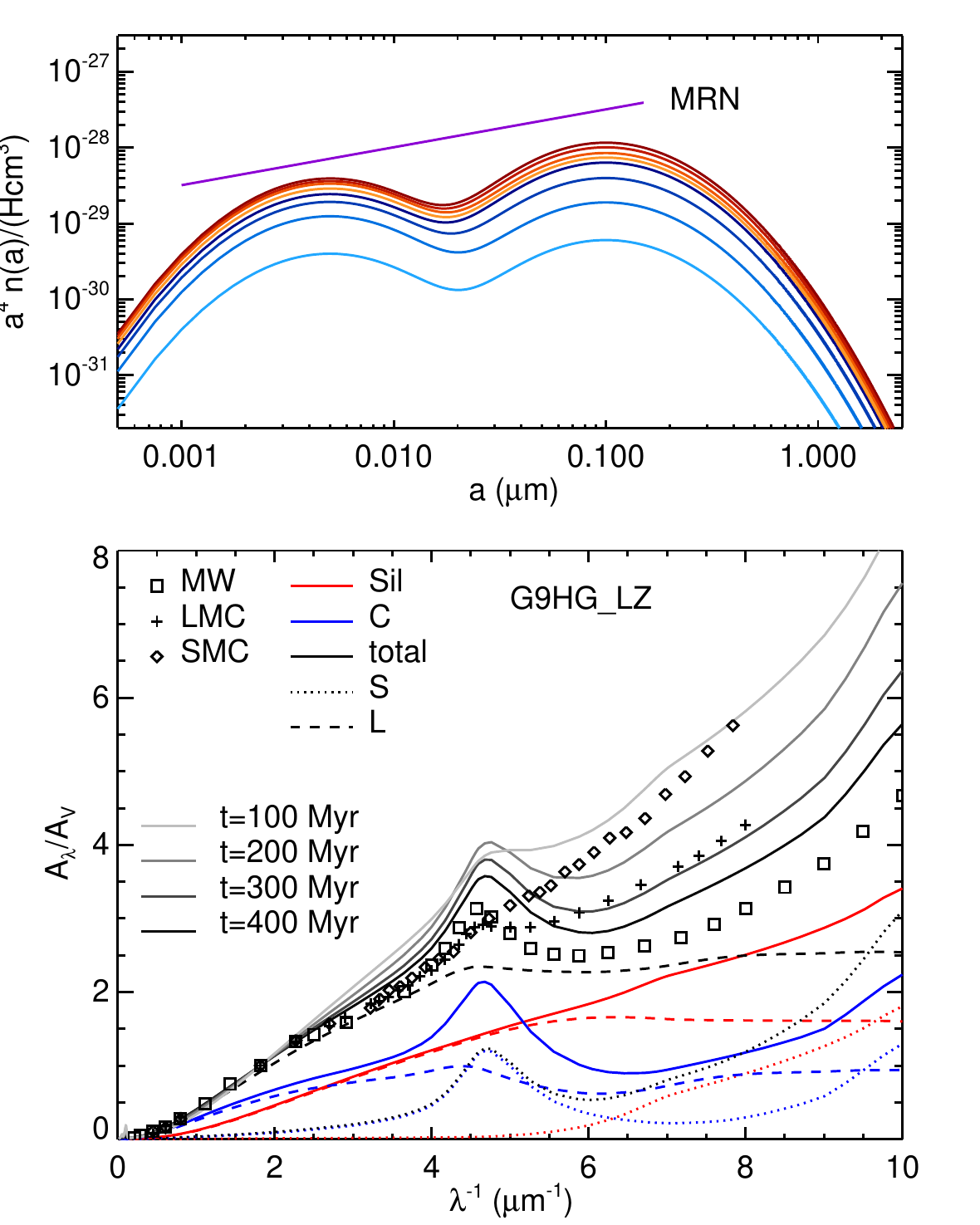}\hspace{-0.5cm}
\centering \includegraphics[width=0.34\textwidth]{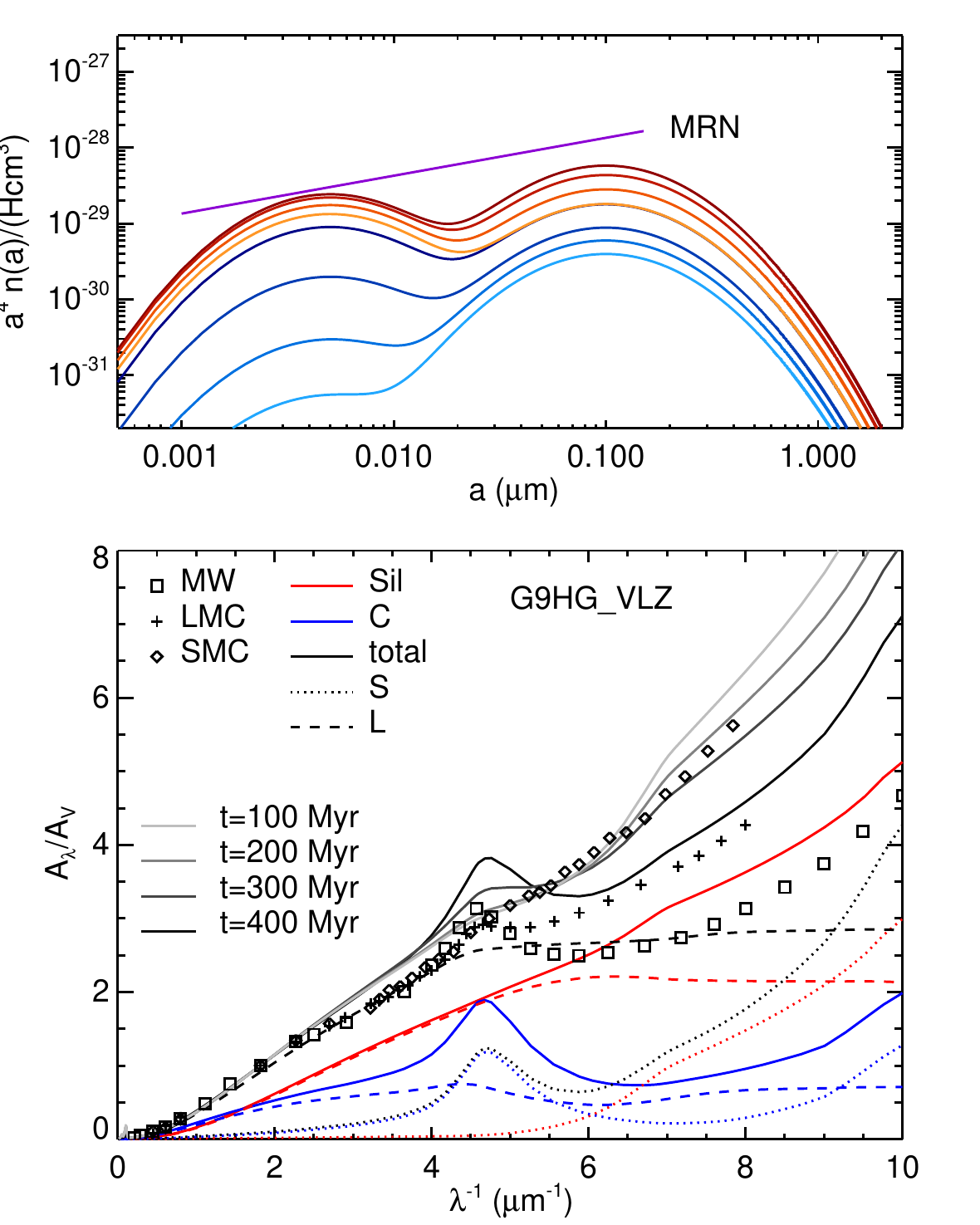}\hspace{-0.5cm}
\centering \includegraphics[width=0.34\textwidth]{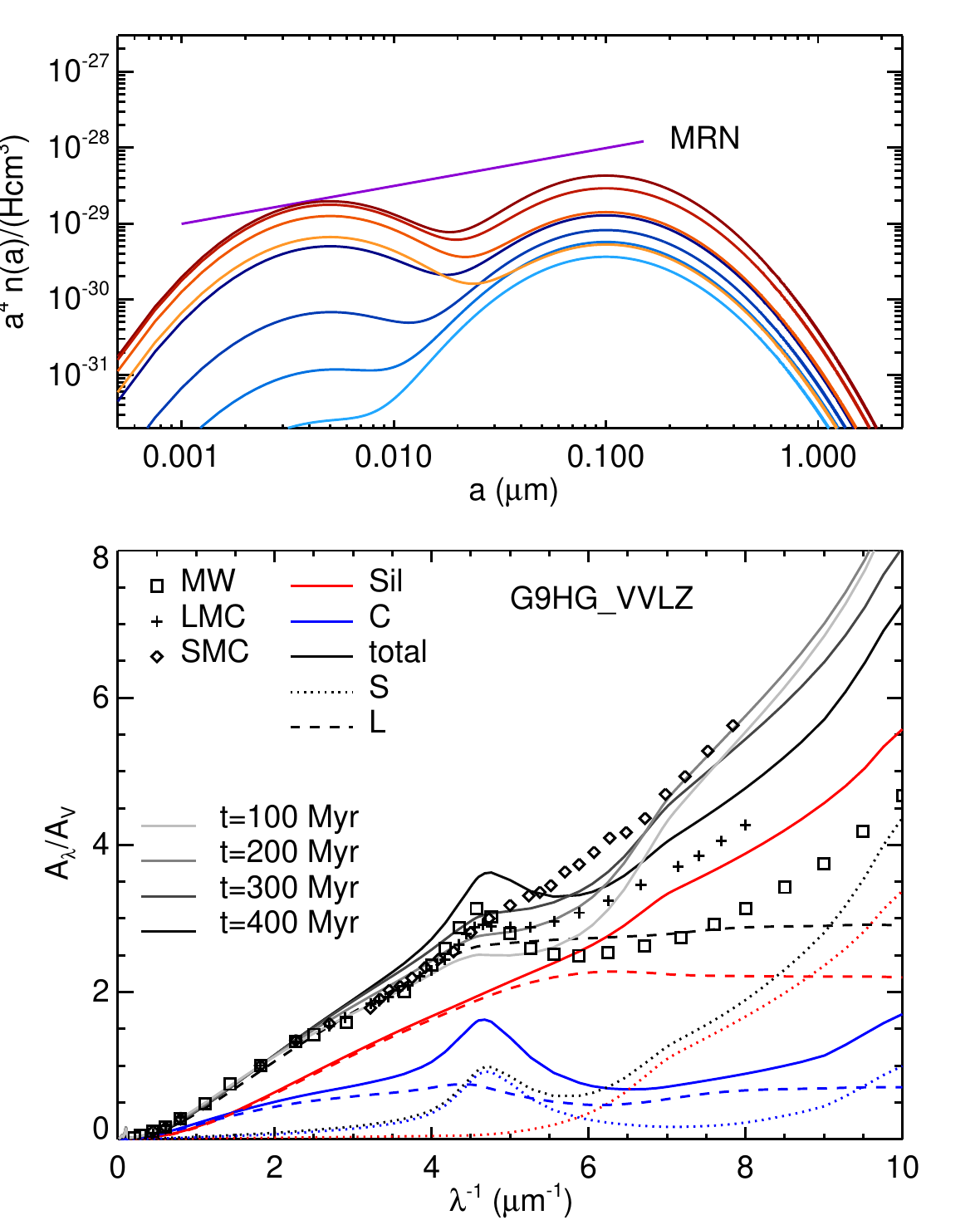}\hspace{-0.5cm}
\centering \includegraphics[width=0.34\textwidth]{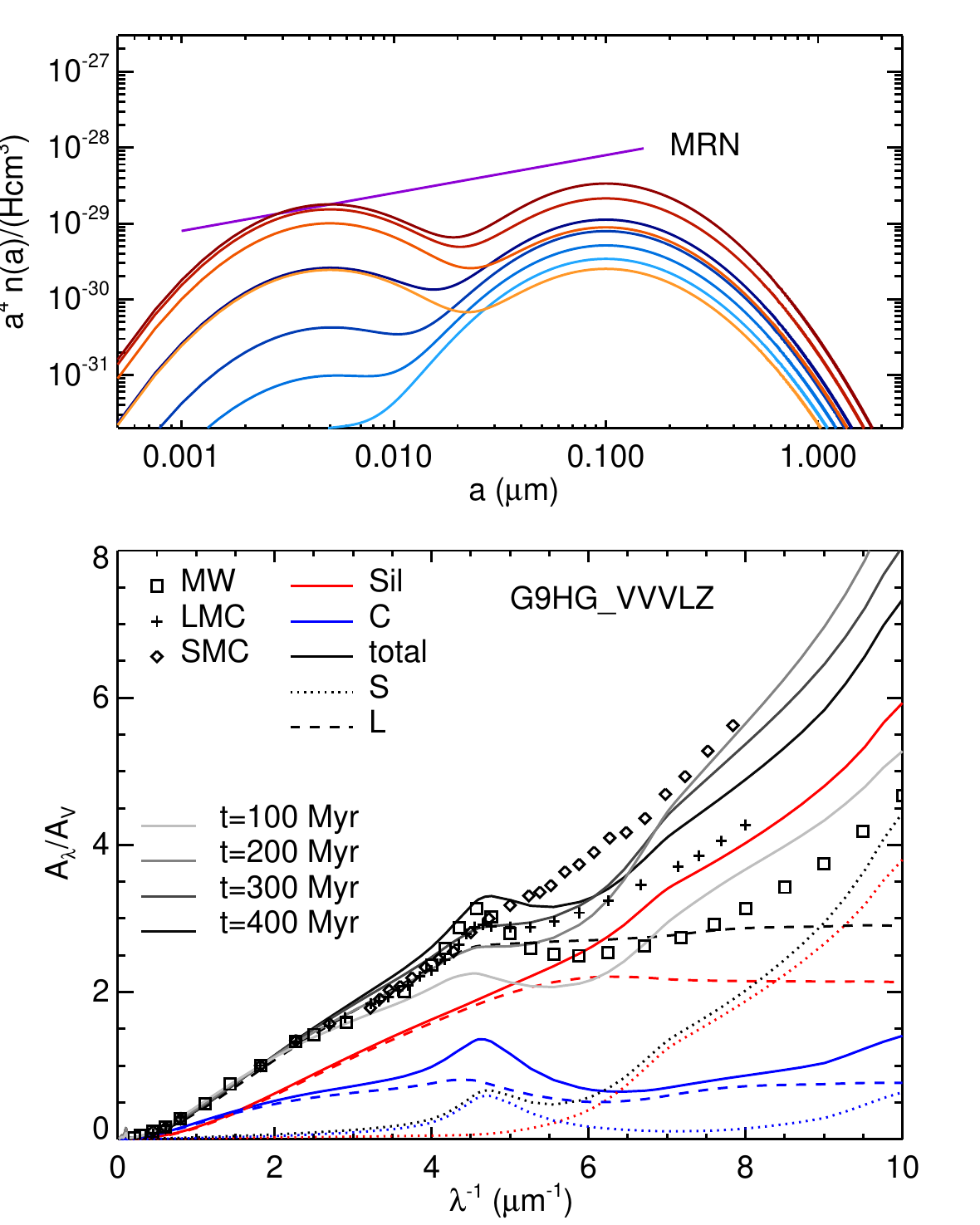}\hspace{-0.5cm}
\centering \includegraphics[width=0.34\textwidth]{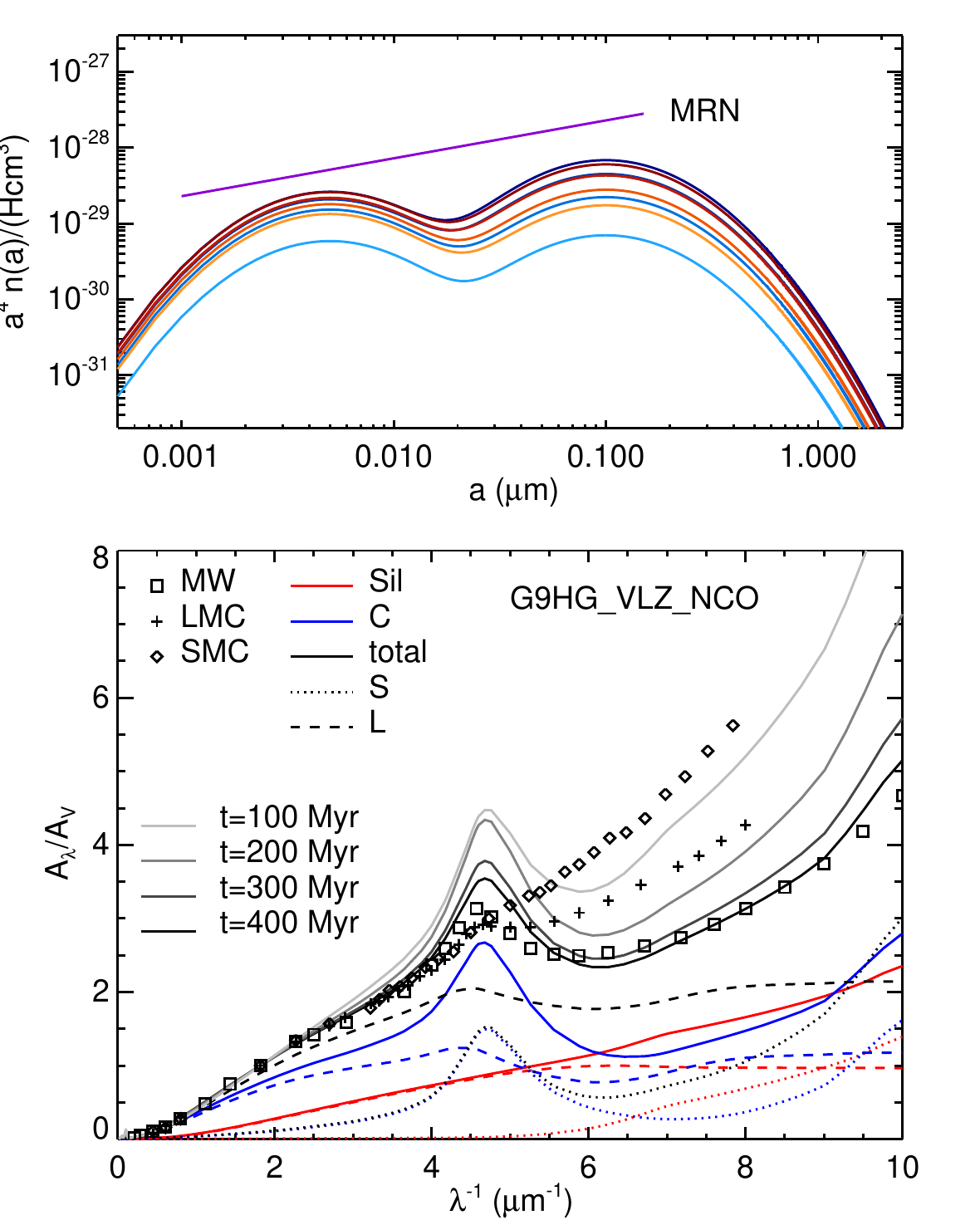}\hspace{-0.5cm}
\centering \includegraphics[width=0.34\textwidth]{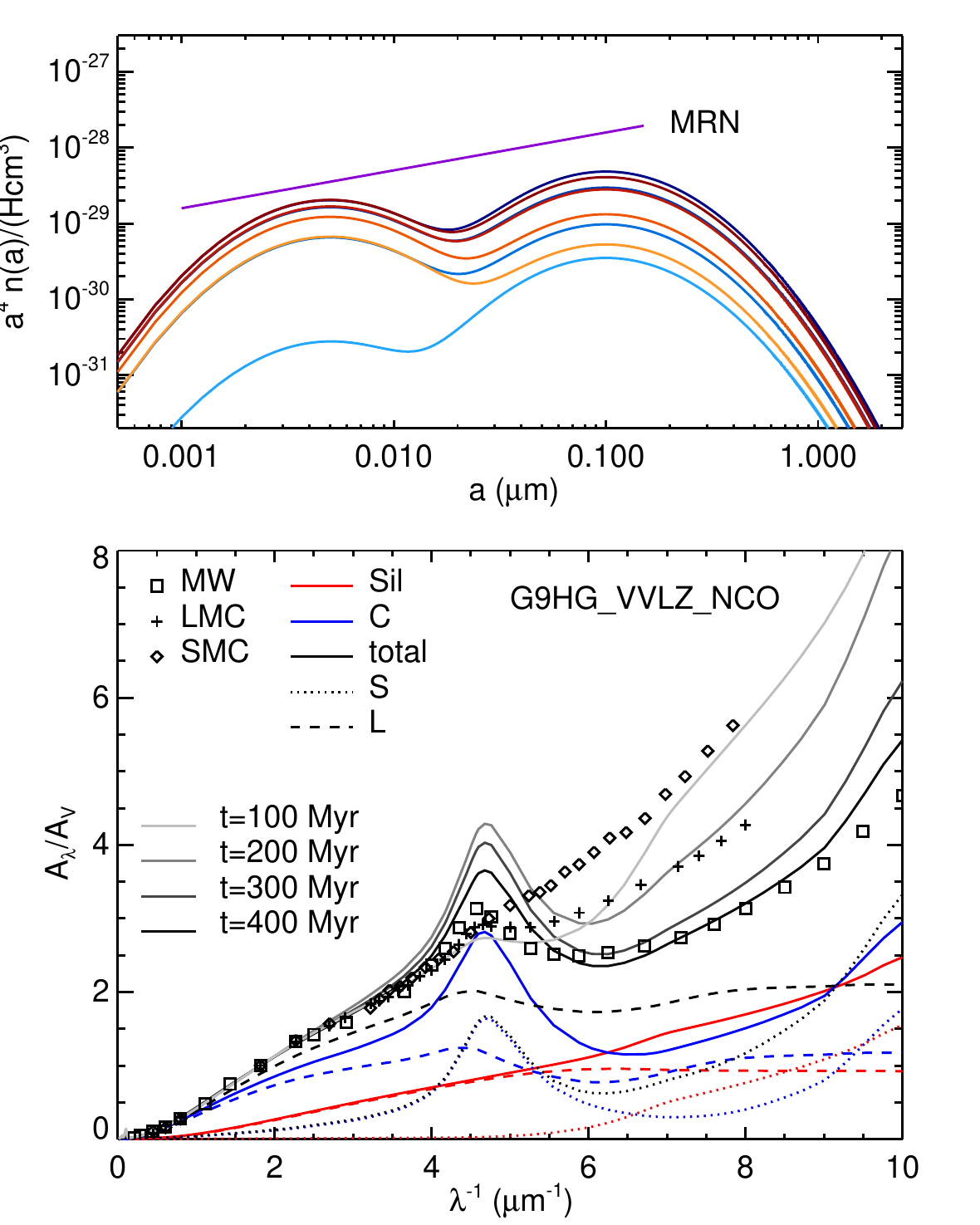}\hspace{-0.5cm}
\centering \includegraphics[width=0.34\textwidth]{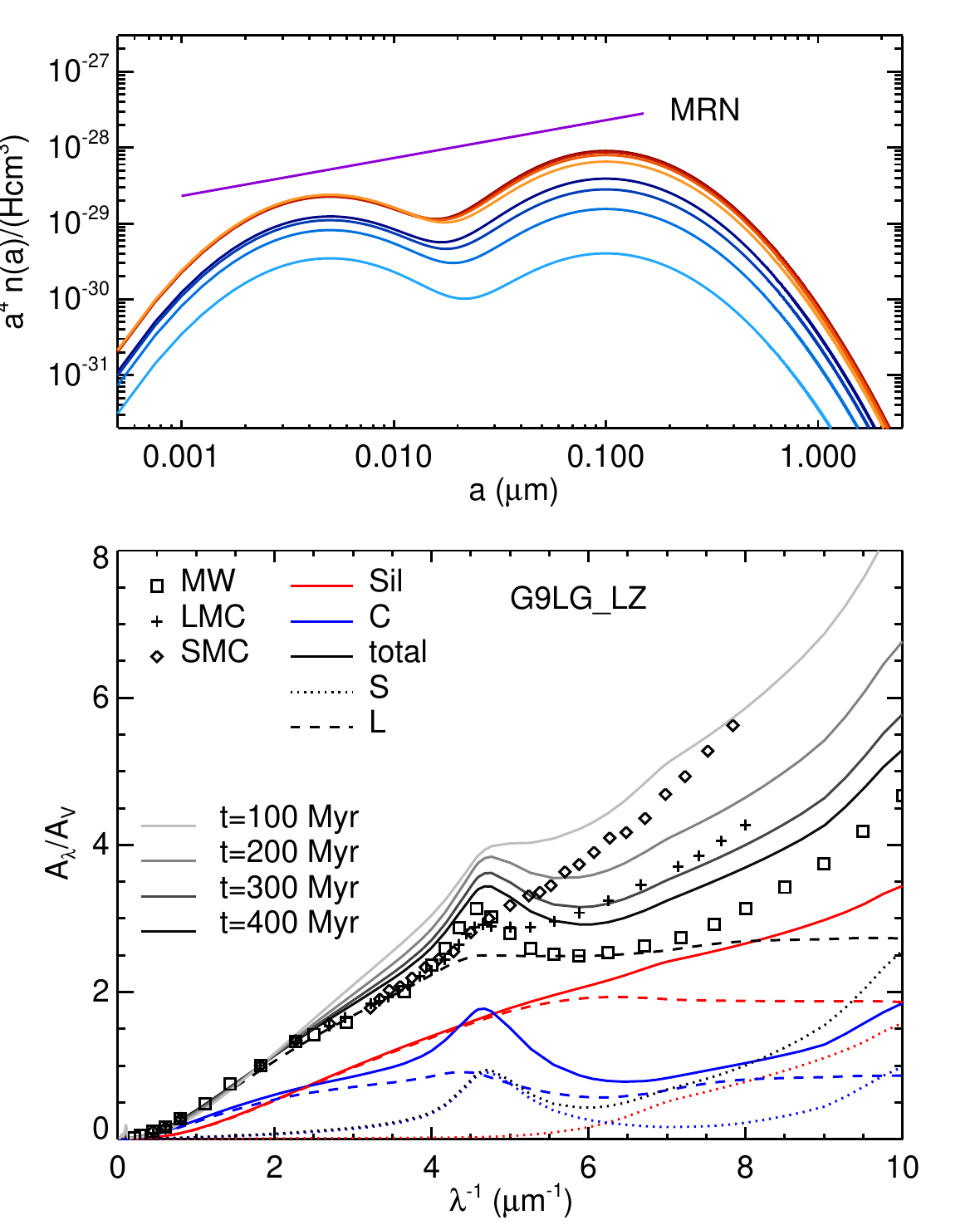}\hspace{-0.5cm}
\centering \includegraphics[width=0.34\textwidth]{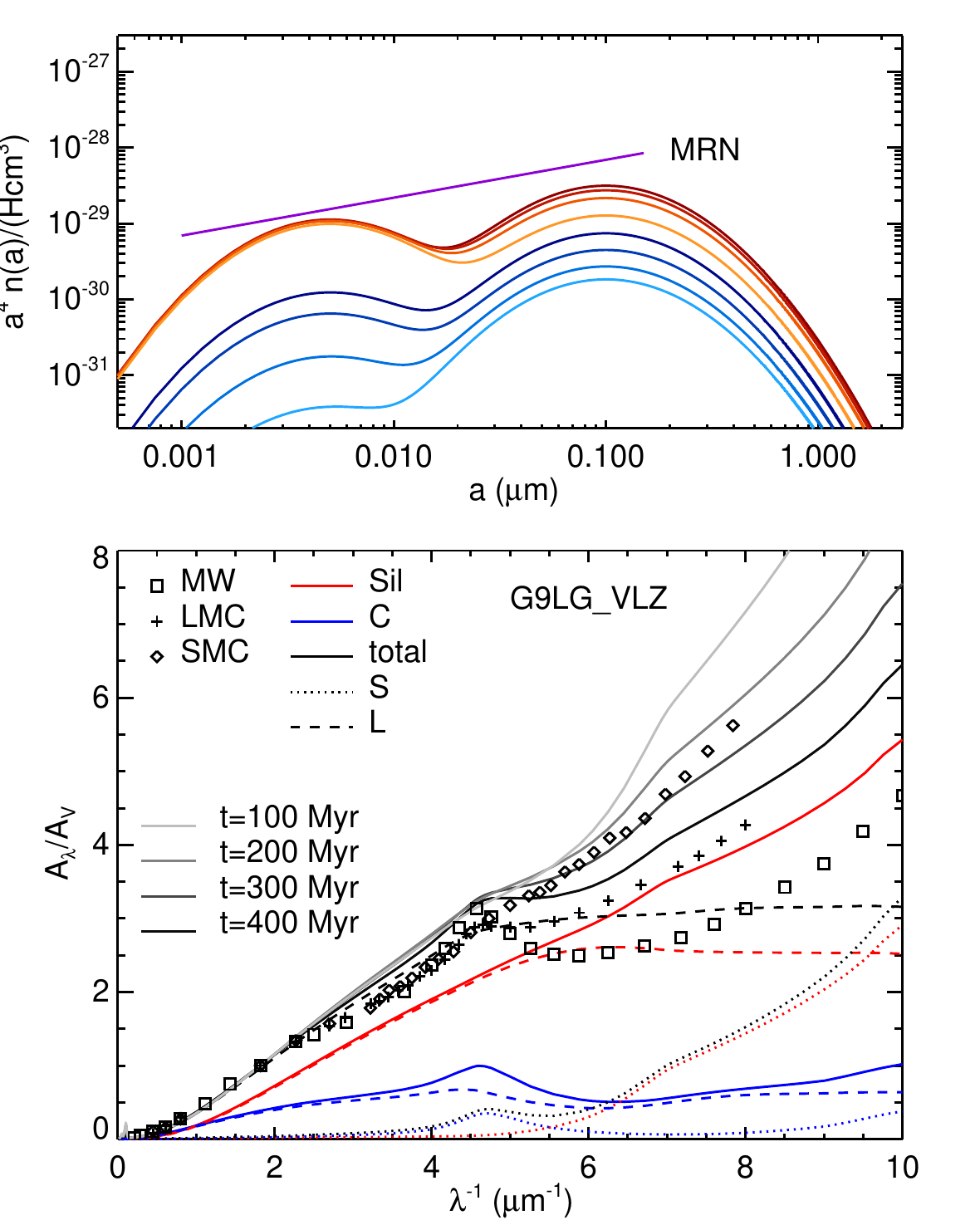}\hspace{-0.5cm}
\centering \includegraphics[width=0.34\textwidth]{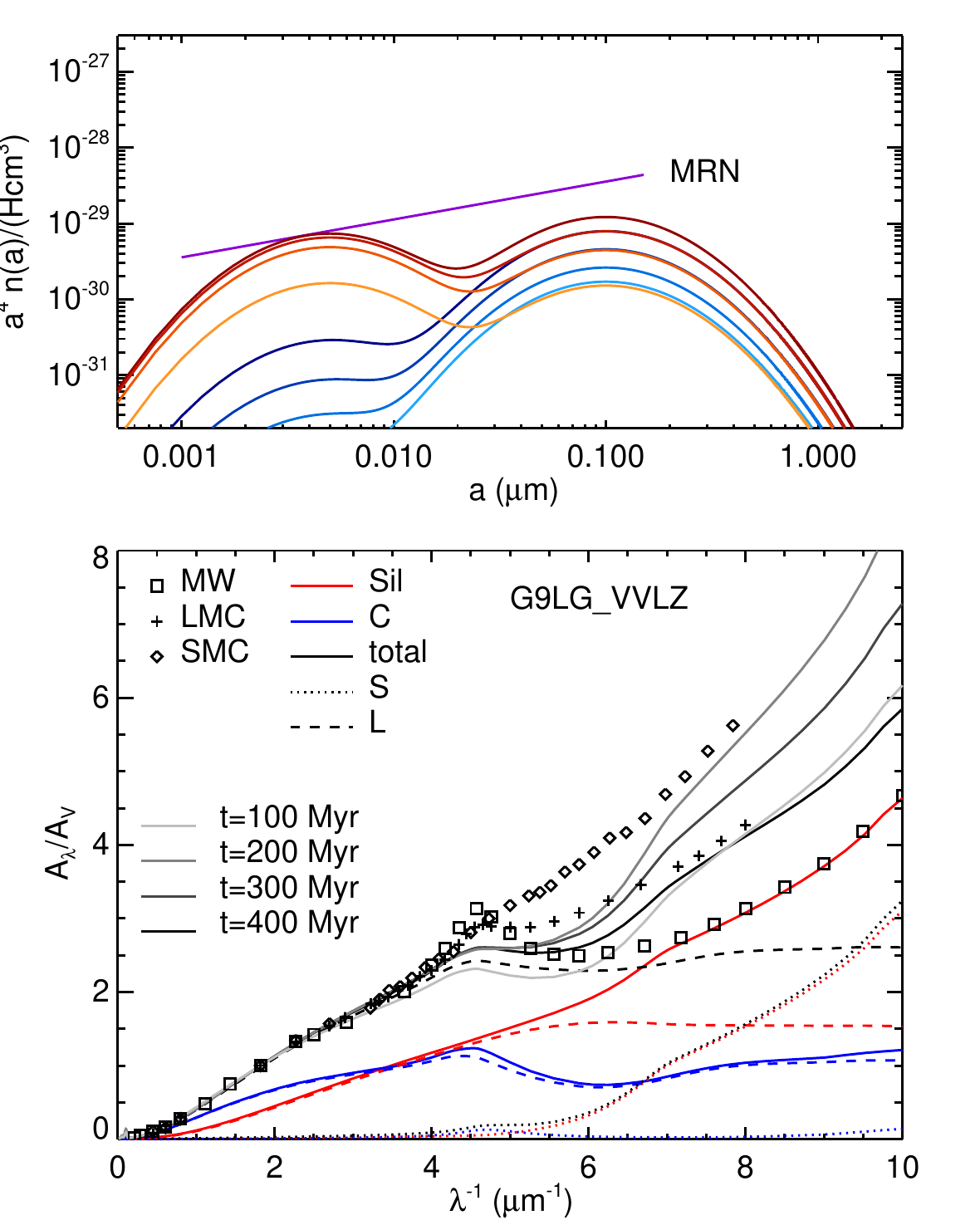}
\caption{Size distribution of grains and extinction curves at different times for different simulations as indicated in the panel. The colors (red and blue) indicate the contribution from different chemical compositions of the dust (resp. silicate and carbonaceous grains), and we only show this decomposition in the extinction curve for the final time of the simulation.}
\label{fig:size_ext_all_1}
\end{figure*}
\begin{figure*}
\centering \includegraphics[width=0.34\textwidth]{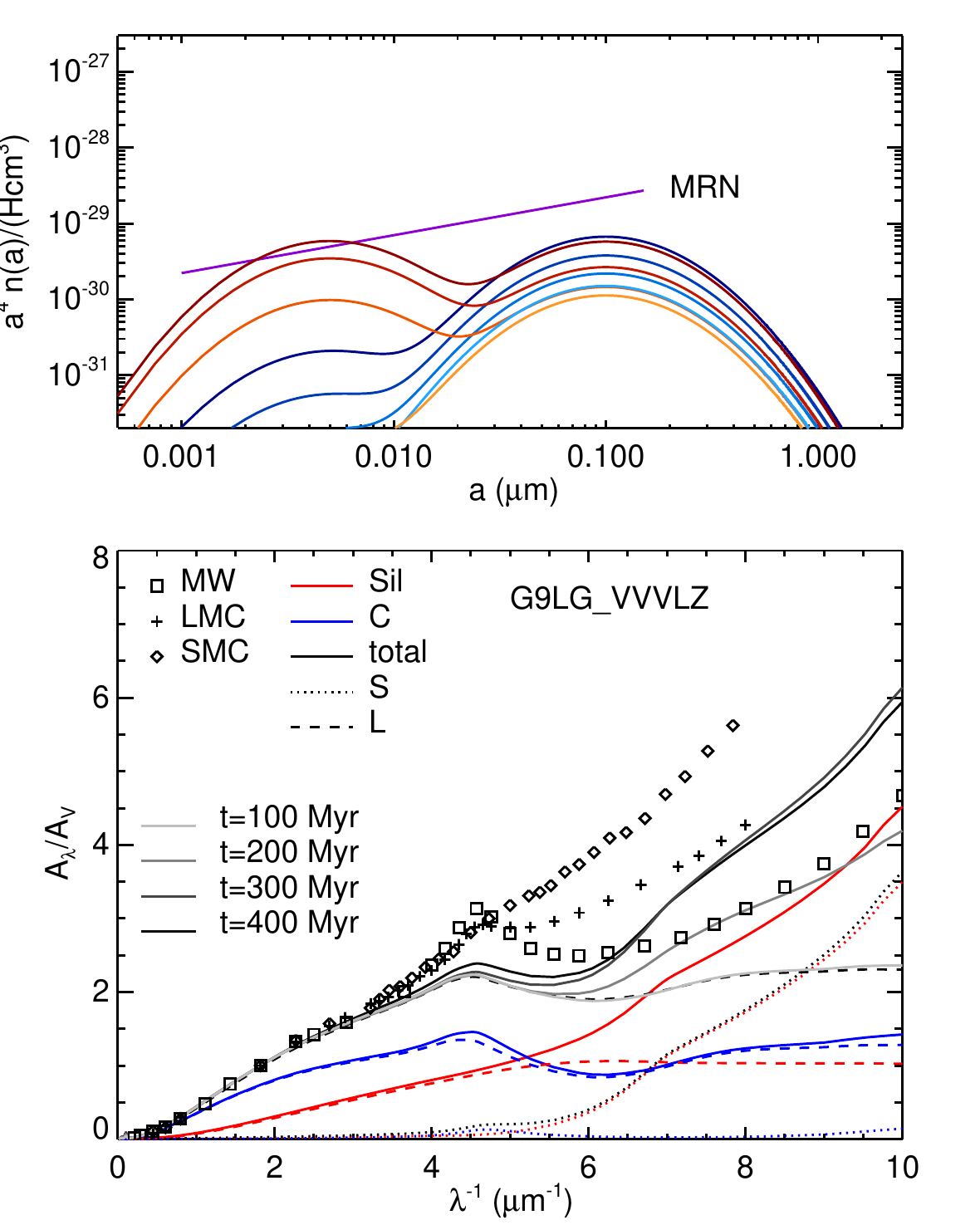}\hspace{-0.5cm}
\centering \includegraphics[width=0.34\textwidth]{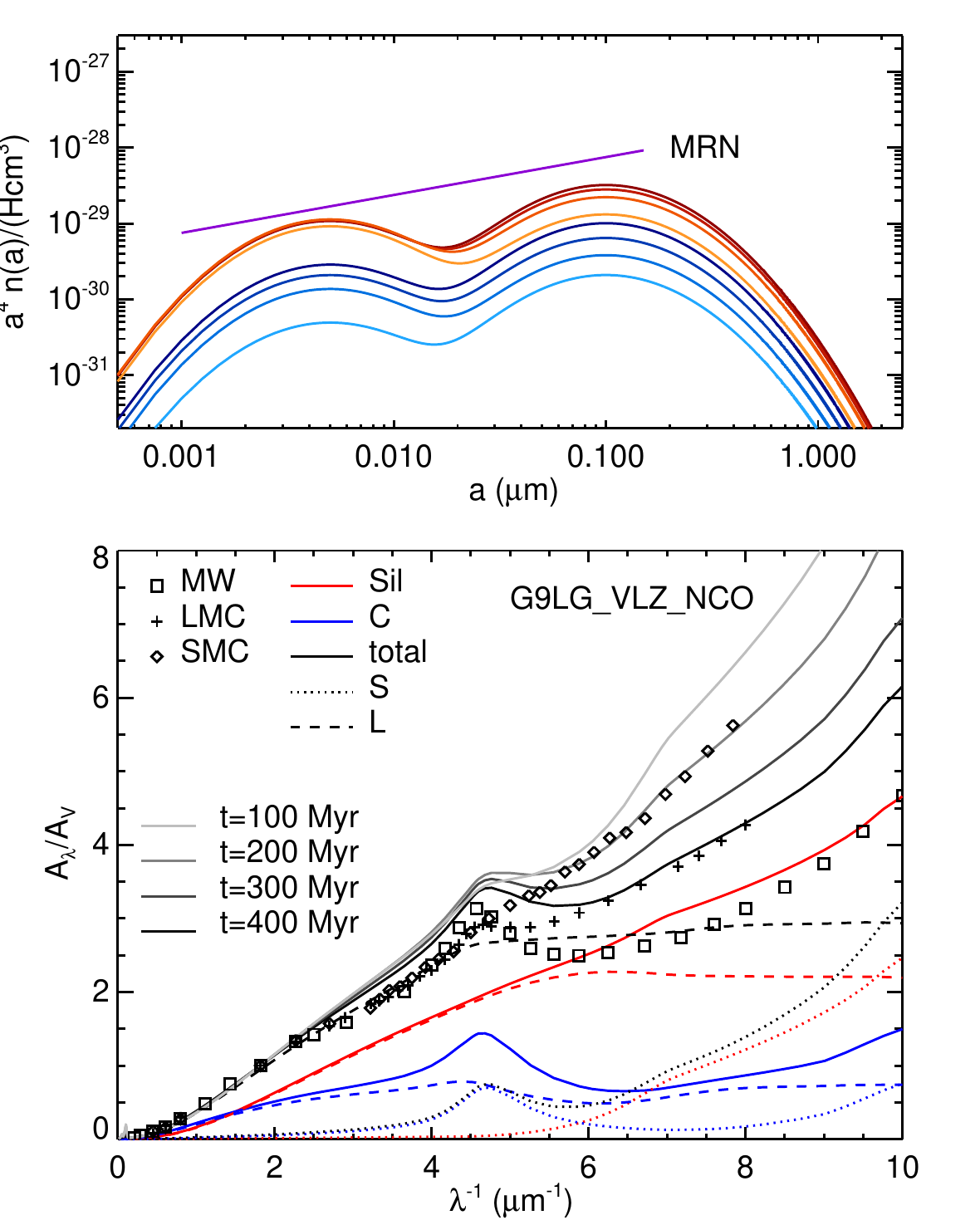}\hspace{-0.5cm}
\centering \includegraphics[width=0.34\textwidth]{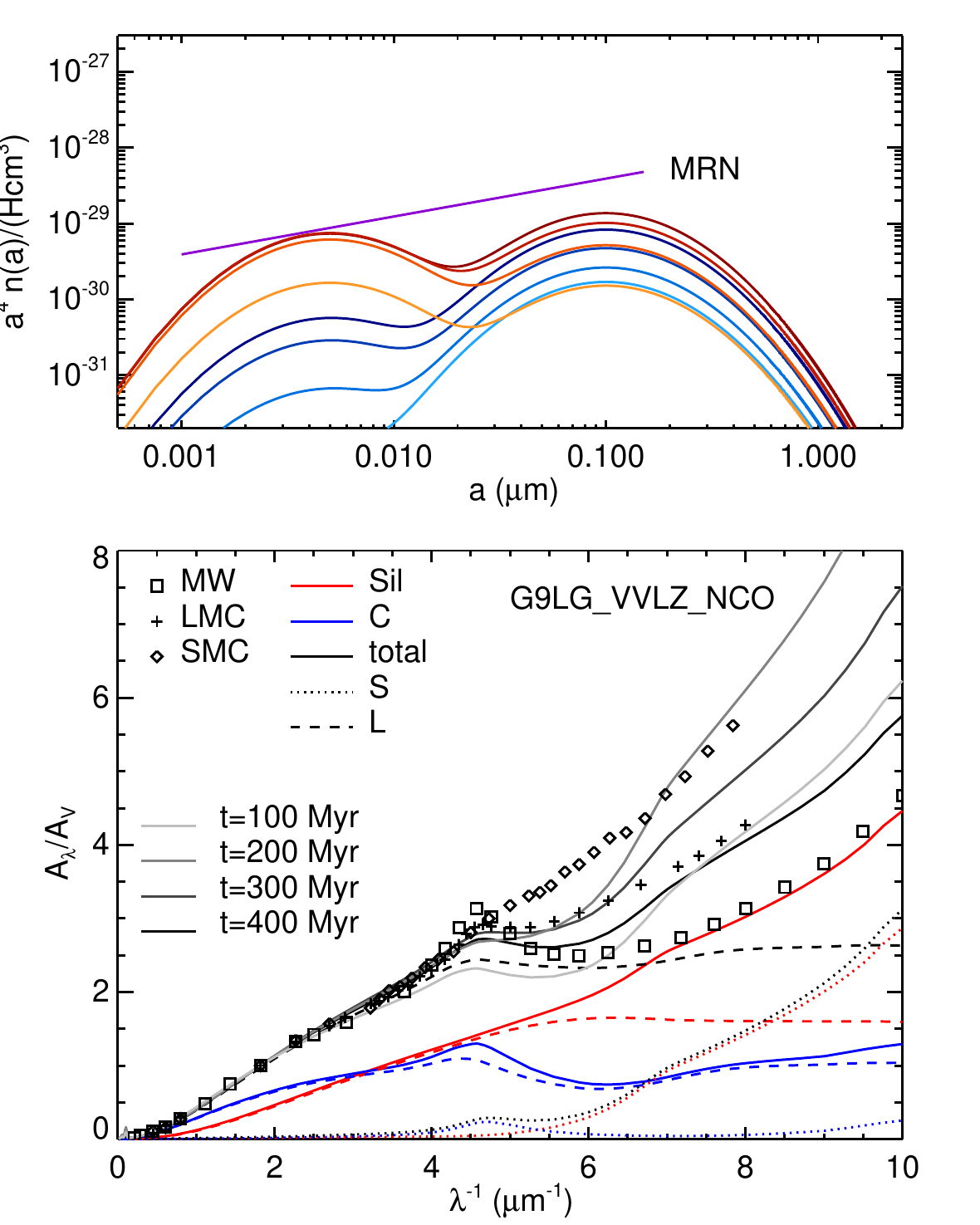}\hspace{-0.5cm}
\centering \includegraphics[width=0.34\textwidth]{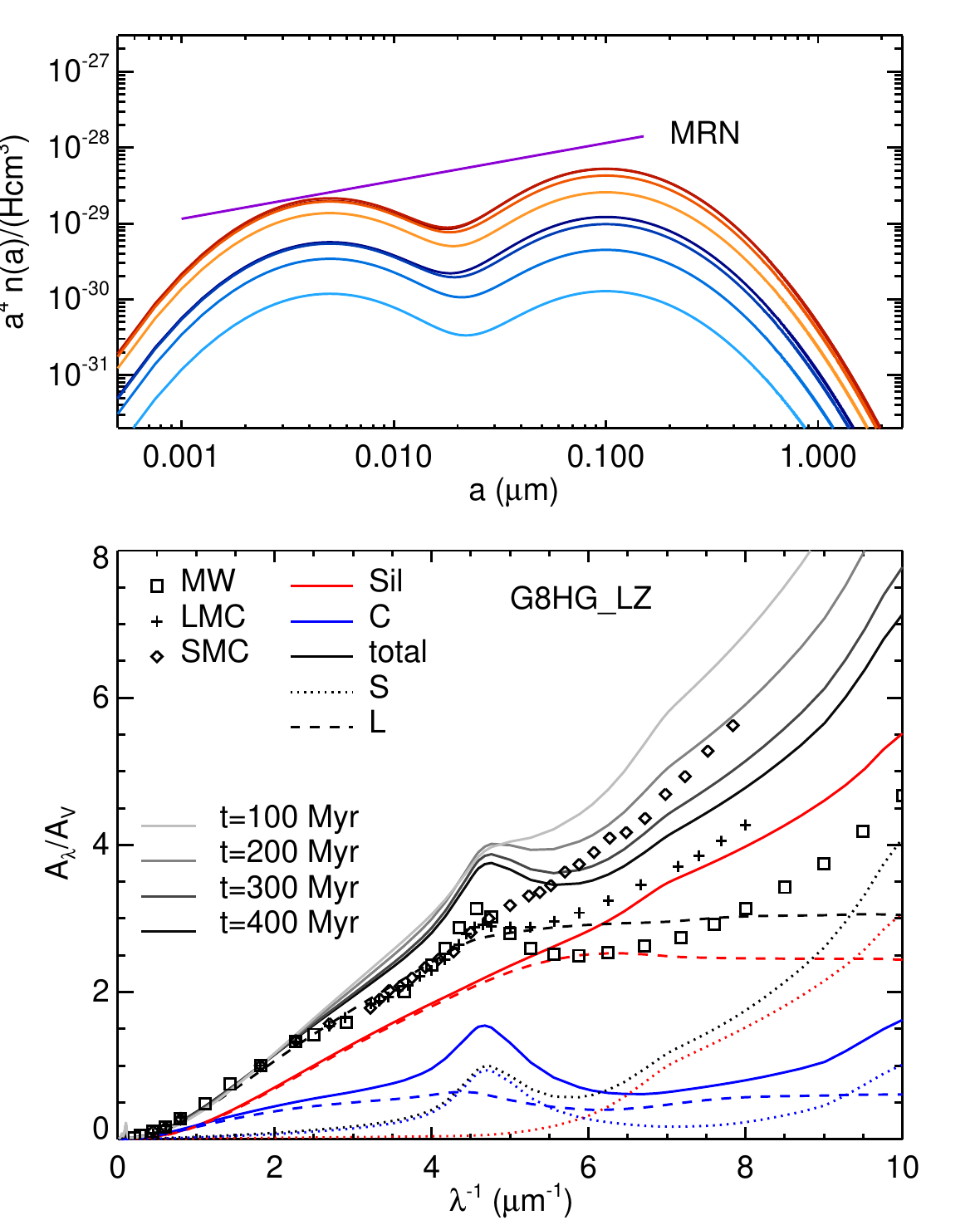}\hspace{-0.5cm}
\centering \includegraphics[width=0.34\textwidth]{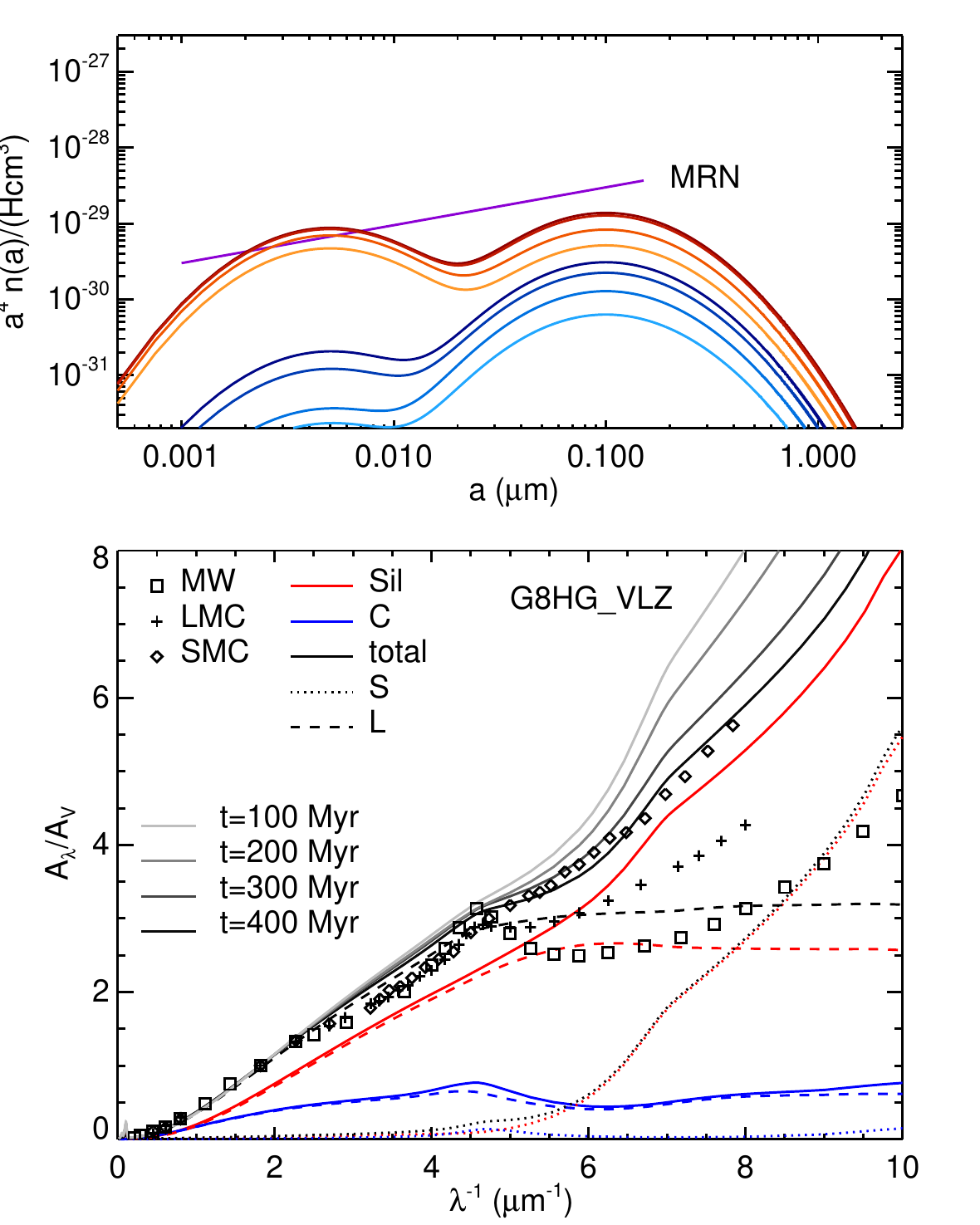}\hspace{-0.5cm}
\centering \includegraphics[width=0.34\textwidth]{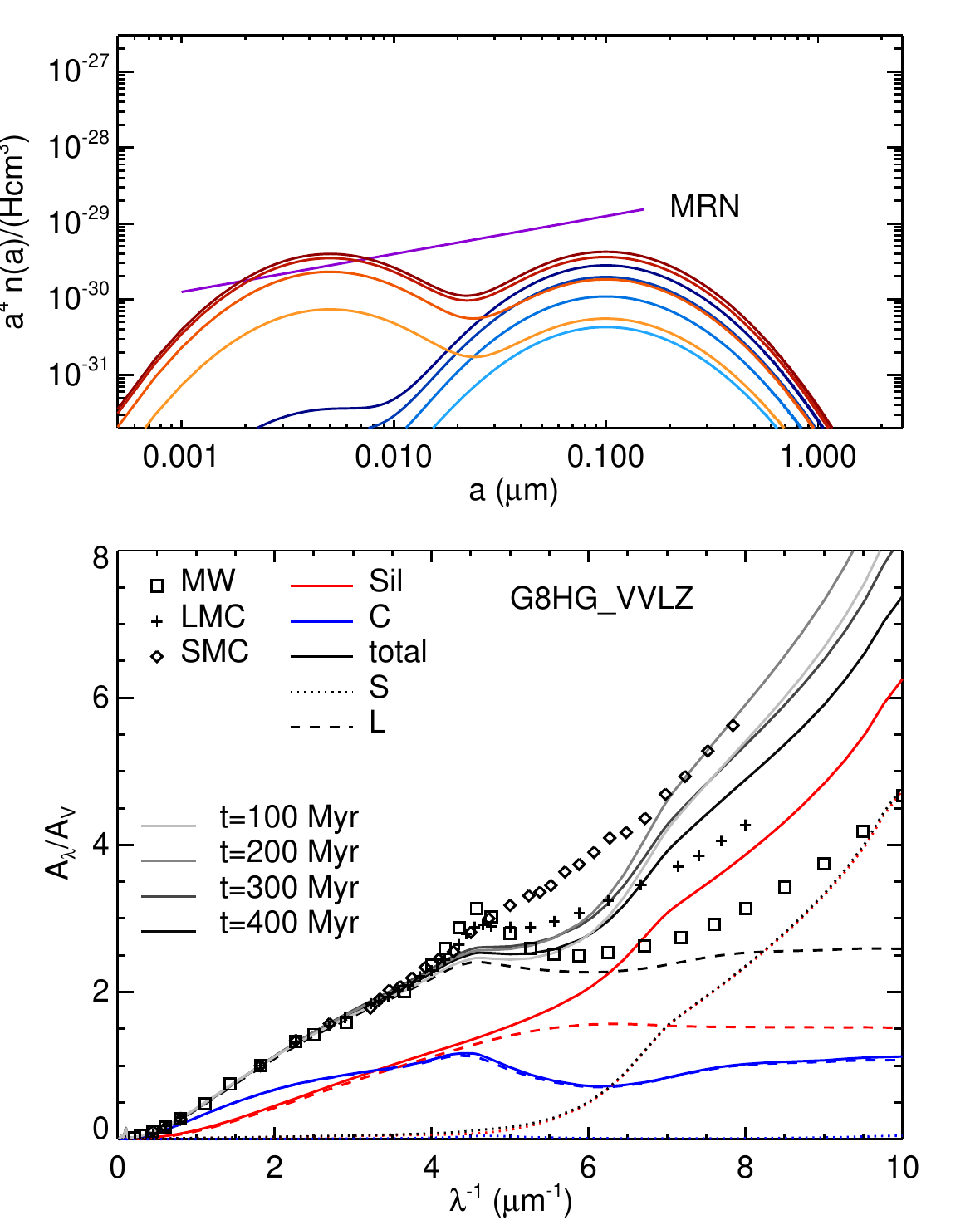}\hspace{-0.5cm}
\centering \includegraphics[width=0.34\textwidth]{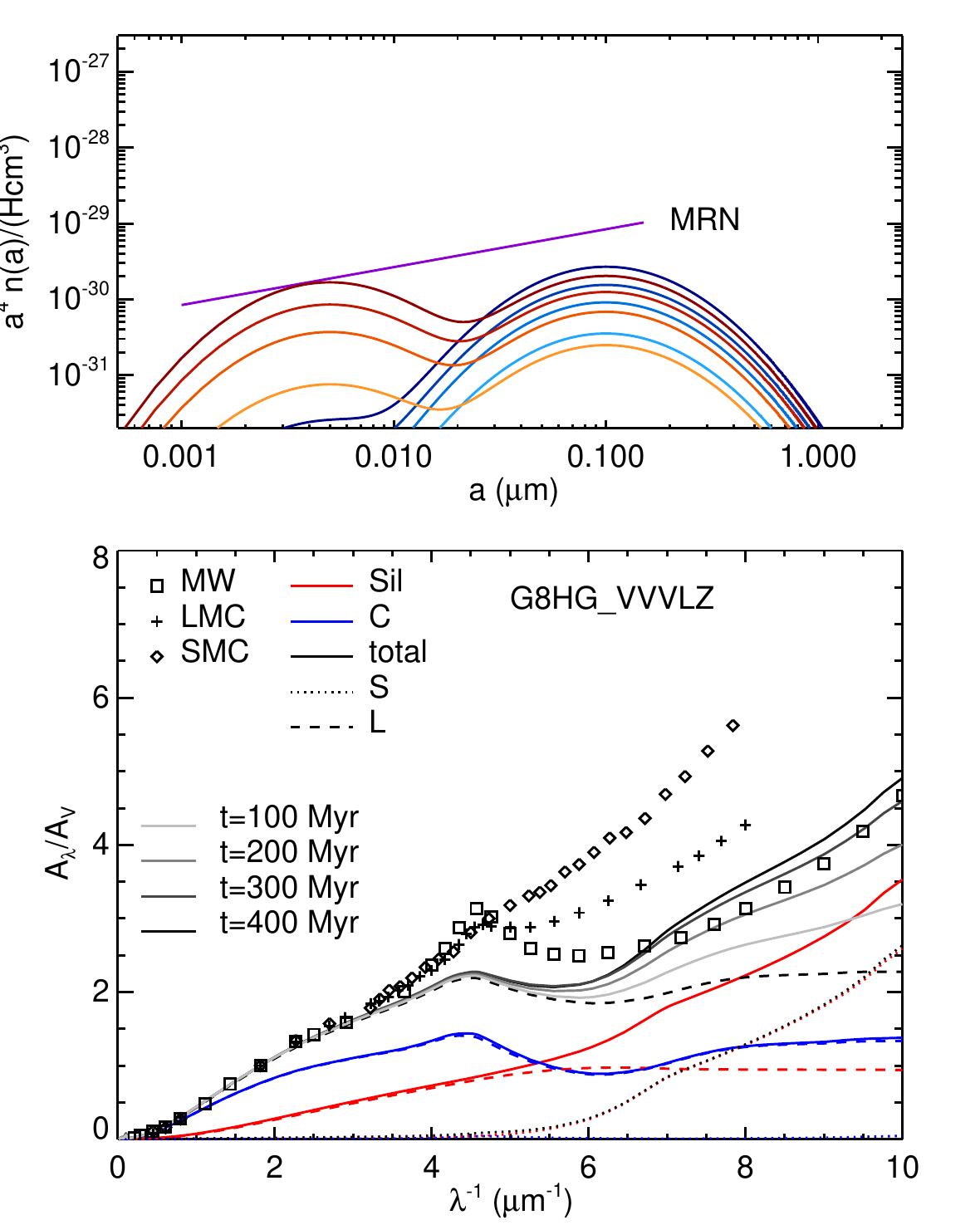}\hspace{-0.5cm}
\centering \includegraphics[width=0.34\textwidth]{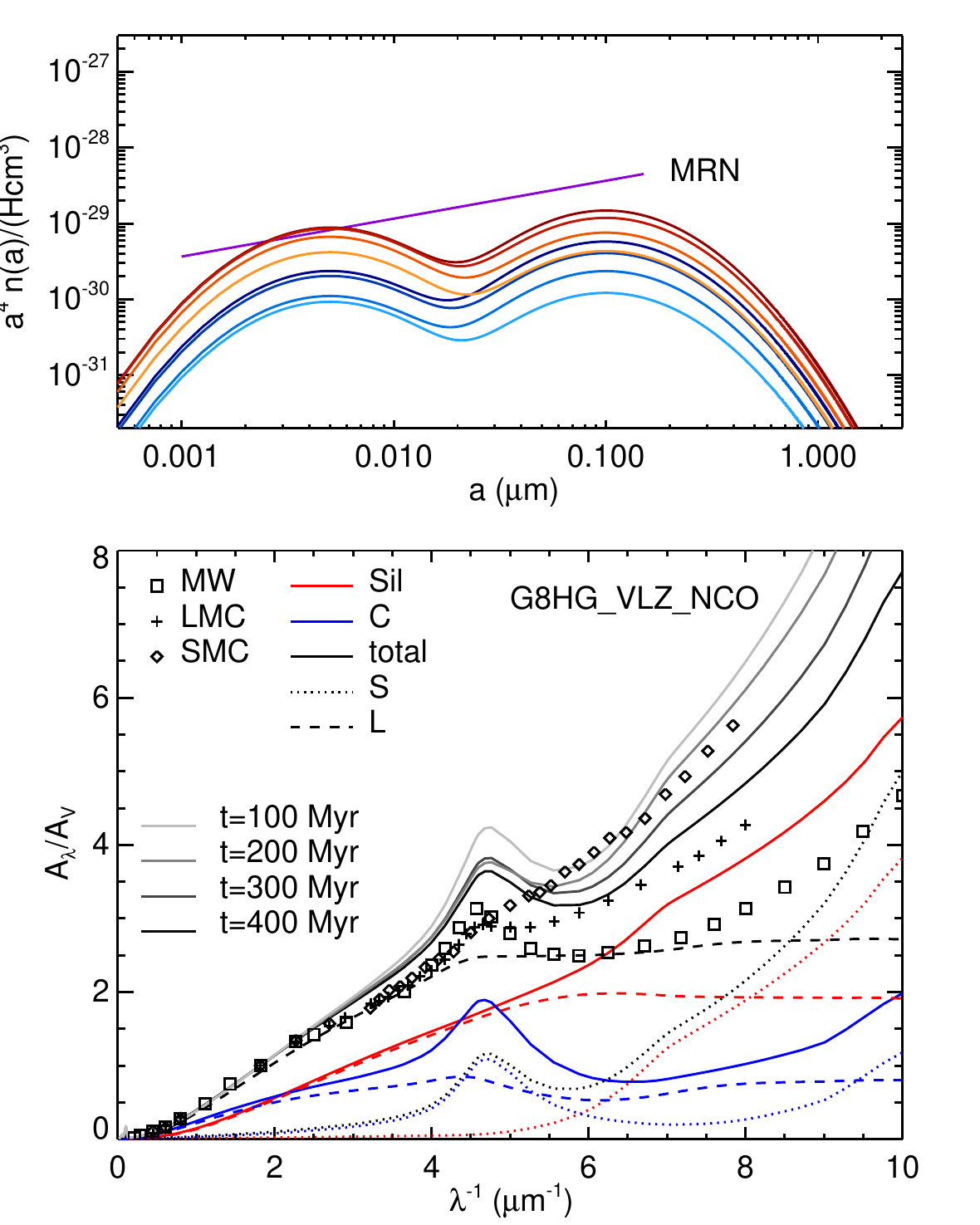}\hspace{-0.5cm}
\centering \includegraphics[width=0.34\textwidth]{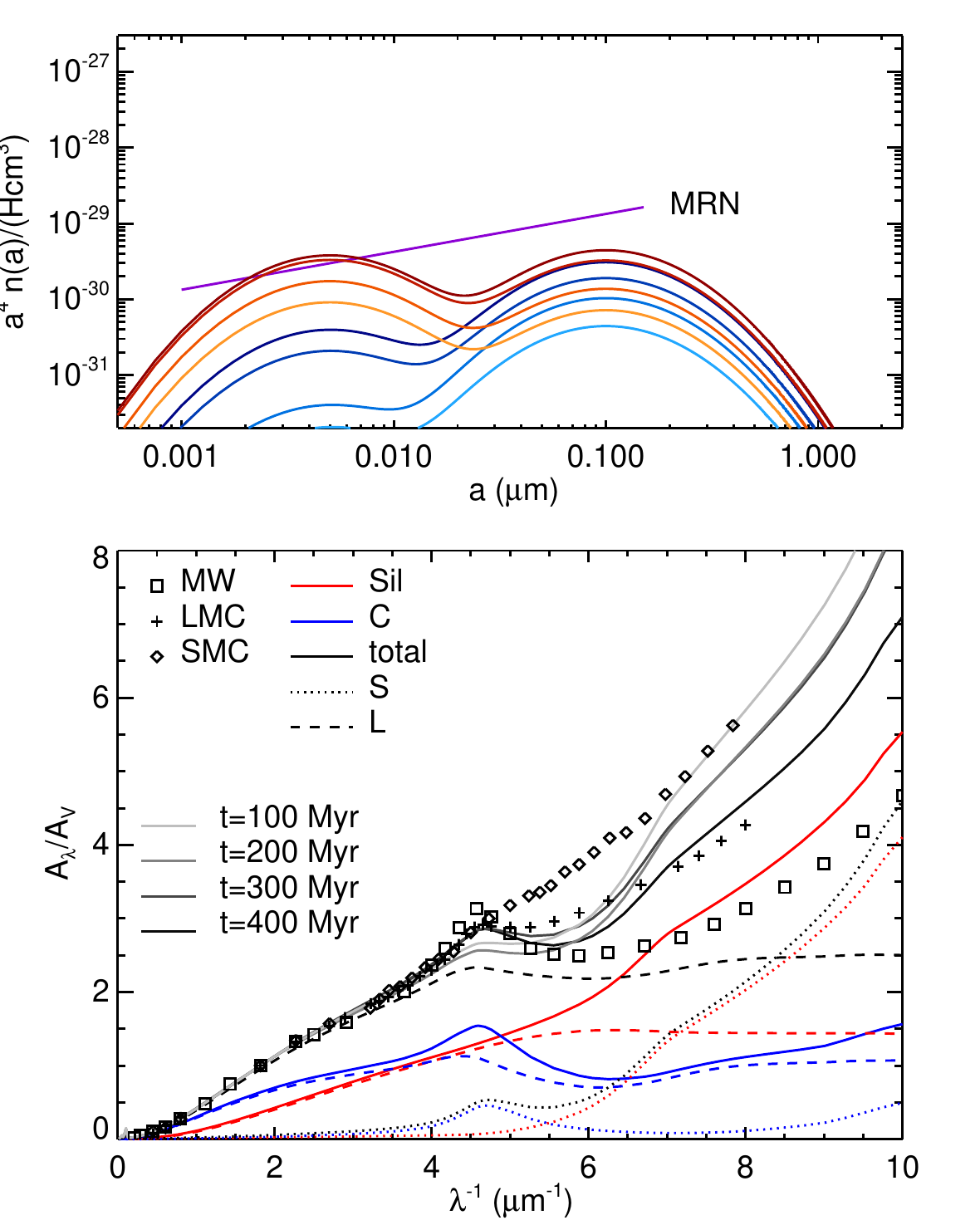}
\caption{Continued from Fig.~\ref{fig:size_ext_all_1}.}
\label{fig:size_ext_all_2}
\end{figure*}

\end{document}